\documentclass[twocolumn,twocolappendix]{aastex631}
\usepackage{amsmath,amssymb}
\usepackage{multirow}
\usepackage{lipsum}

\shorttitle{Radio spectro-polarimetry of CME}
\shortauthors{Kansabanik et al.}

\graphicspath{{./}{figures/}}

\published{June 21, 2023}

\begin{document}

\title{Deciphering Faint Gyrosynchrotron Emission from Coronal Mass Ejection using Spectro-polarimetric Radio Imaging}

\author[0000-0001-8801-9635]{Devojyoti Kansabanik}
\affiliation{National Centre for Radio Astrophysics, Tata Institute of Fundamental Research, S. P. Pune University Campus, Pune 411007, India}

\author[0000-0002-2325-5298]{Surajit Mondal}
\affiliation{Center for Solar-Terrestrial Research, New Jersey Institute of Technology, 323 M L King Jr Boulevard, Newark, NJ 07102-1982, USA}

\author[0000-0002-4768-9058]{Divya Oberoi}
\affiliation{National Centre for Radio Astrophysics, Tata Institute of Fundamental Research, S. P. Pune University Campus, Pune 411007, India}

\correspondingauthor{Devojyoti Kansabanik}
\email{dkansabanik@ncra.tifr.res.in, devojyoti96@gmail.com}

\begin{abstract}
Measurements of the plasma parameters of coronal mass ejections (CMEs), particularly the magnetic field and non-thermal electron population entrained in the CME plasma, are crucial to understand their propagation, evolution, and geo-effectiveness. Spectral modeling of gyrosynchrotron (GS) emission from CME plasma has been regarded as one of the most promising remote sensing technique for estimating spatially resolved CME plasma parameters. Imaging the very low flux density CME GS emission in close proximity to the Sun with orders of magnitude higher flux density, however, has proven to be rather challenging. This challenge has only recently been met using the high dynamic range imaging capability of the Murchison Widefield Array (MWA). Although routine detection of GS is now within reach, the challenge has shifted to constraining the large number of free parameters in GS models, a few of which are degenerate, using the limited number of spectral points at which the observations are typically available. These degeneracies can be broken using polarimetric imaging. For the first time, we demonstrate this using our recently developed capability of high fidelity polarimetric imaging on the data from the MWA. We show that spectro-polarimetric imaging, even when only sensitive upper limits on circularly polarization flux density are available, is not only able to break the degeneracies, but also yields tighter constraints on the plasma parameters of key interest than possible with total intensity spectroscopic imaging alone.
\end{abstract}


\section{Introduction}
Coronal Mass Ejections (CMEs) are large-scale eruptions of magnetized plasma from the solar atmosphere to the heliosphere. CMEs are routinely observed at visible wavelength using ground and space-based coronagraphs. Observation at visible wavelength provide several pieces of crucial information about CMEs -- its large scale three-dimensional structure, velocity, acceleration, electron density \citep[e.g.][]{Webb2012}. 
There are several models available about the origin and evolution of the CMEs \citep[e.g.,][etc.]{Chen2011,Kilpua2021,Sindhuja_2022}, though the exact mechanisms continue to be debated. Nonetheless, it is well established that CME eruption, evolution and geo-effectiveness are all driven by their magnetic fields \citep[e.g,][etc.]{aschwanden2004,Temmer2021,Vourlidas2020,Srivastava2021}. Hence measurements of the magnetic fields both inside the CME plasma and at the shock are essential. 

Routine observations at visible wavelength are not useful for estimating the magnetic field strength of the CME plasma. These observations also cannot be used for measuring the non-thermal electron population either inside the CME plasma or at the shock front. Several techniques have been developed over the last decade or so to measure the average magnetic field strength at the CME shock front \citep[e.g.,][etc.]{Cho2007,Kumari2017typeII_band,Raja2014,Kumari2017_typeIV,Gopalswamy2011,Zhao2019}. 
Although successful, none of these techniques can be used to measure the magnetic field inside the CME plasma and do not provide any information about the distribution of the non-thermal electrons either. 

\begin{figure}[!htbp]
    \centering
     \includegraphics[trim={1.2cm 5cm 1.2cm 4.5cm},clip,scale=0.43]{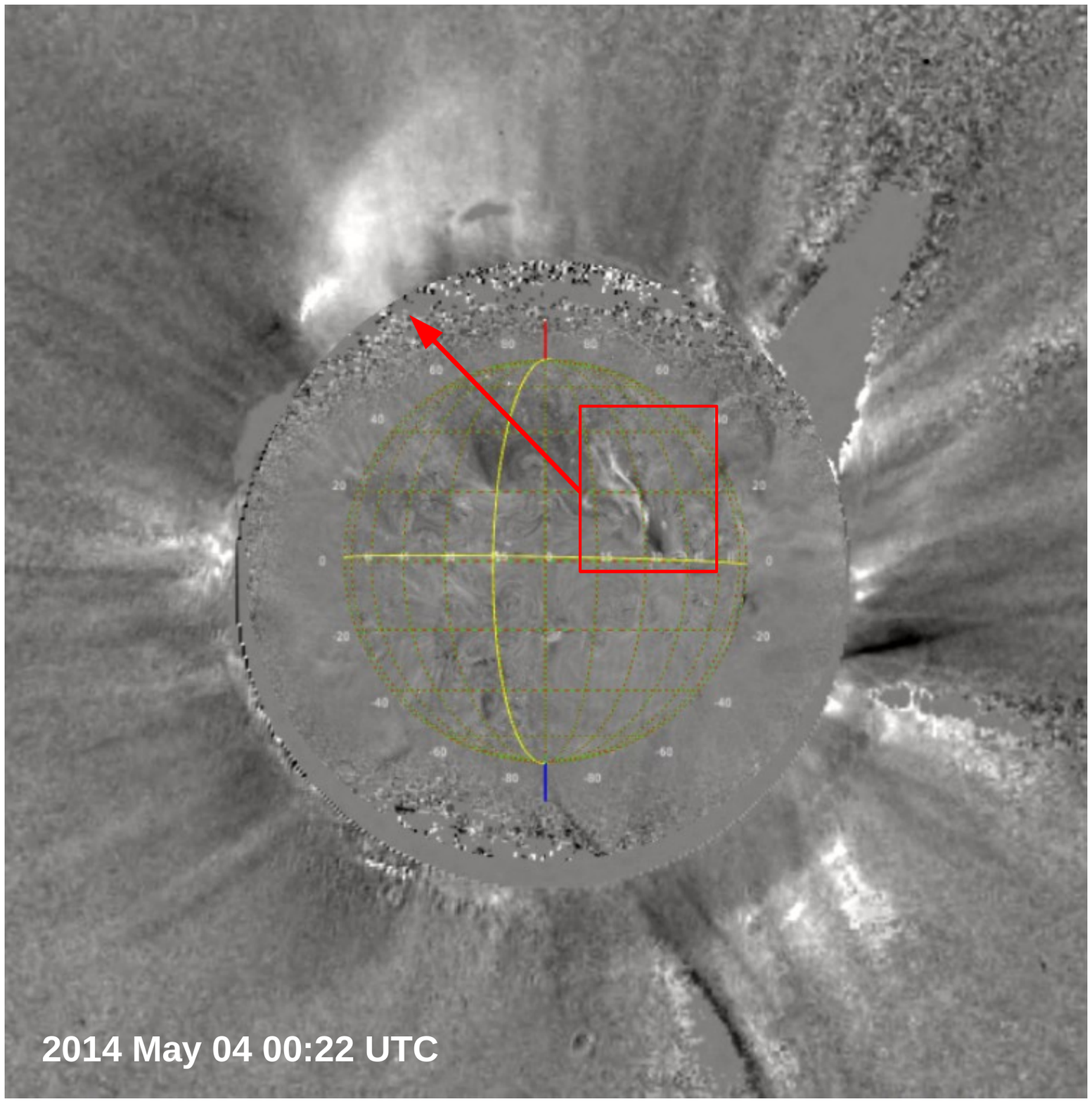}    \caption{\textbf{Eruption of CME-1 as observed using the STEREO-B spacecraft.} CME-1 erupted from behind the visible solar disc. A composite base difference image from the Extreme Ultraviolet Imager (EUVI) and COR-1 coronagraph onboard STEREO-B spacecraft is shown. The red box shows the eruption site and the red arrow shows the propagation direction.}
    \label{fig:north_cme_eruption}
\end{figure}
\begin{figure*}[!htbp]
   \centering
    \includegraphics[trim={0.4cm 0.5cm 1cm 0cm},clip,scale=0.6]{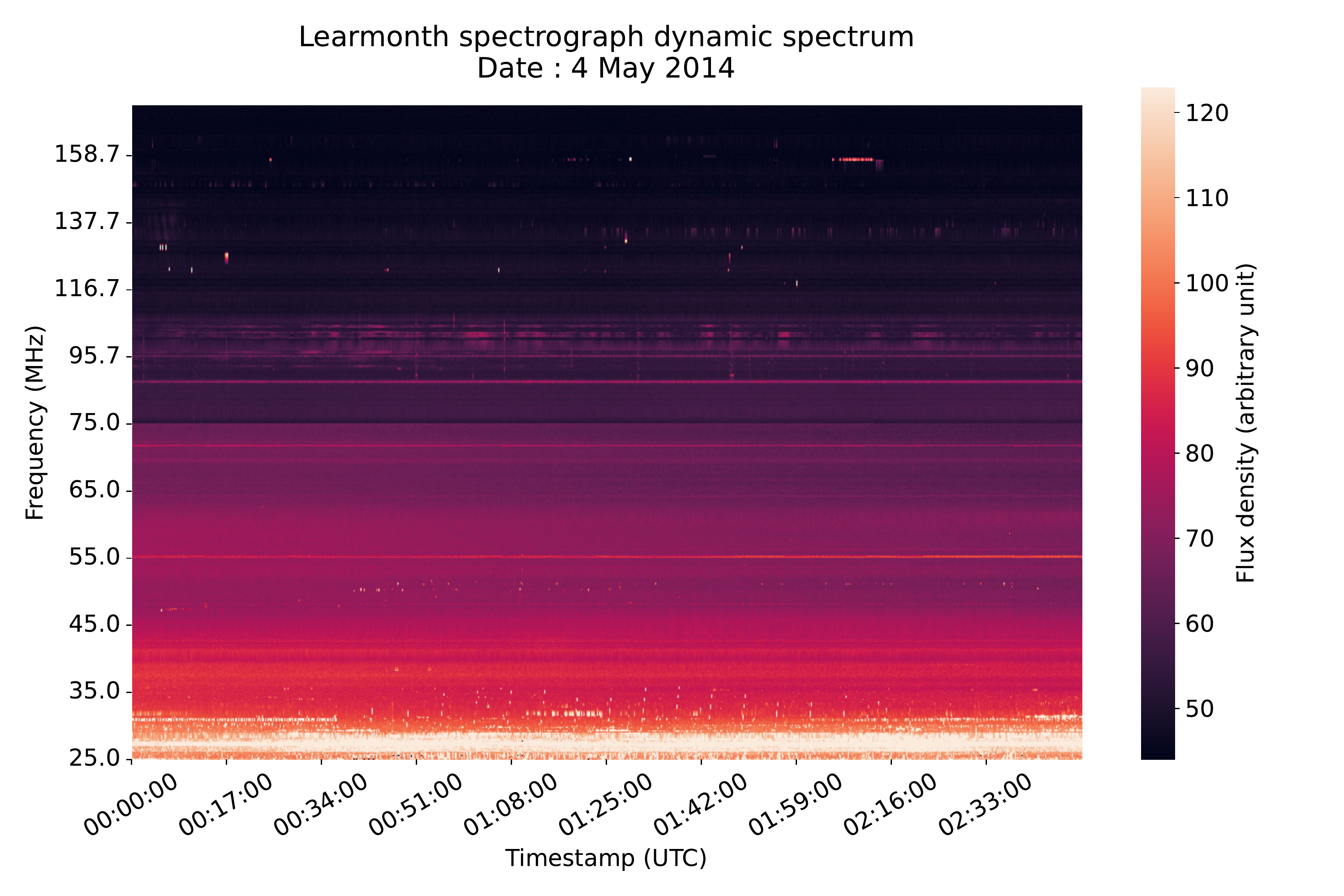}
    \caption{{\bf Dynamic spectrum from the Learmonth radio spectrograph.} No radio bursts are seen from 00:00-03:00 UTC on 04 May 2014. Several bad channels with persistent radio frequency interference (RFI) have been flagged and interpolated across for each time slice independently.}
    \label{fig:learmonth}
\end{figure*}

There exist other remote sensing techniques which can potentially be used to measure the magnetic field entrained in the CME plasma. A recent study by \cite{Ramesh2021} used the induced circular polarization (Stokes V) measurements of thermal emission from CME plasma to estimate CME magnetic field at heliocentric distance $\sim2\mathrm{R_\odot}$. Though this method too can not provide any information about the population of non-thermal electrons. Another method which has been used in past to measure CME magnetic field is the modeling the spectrum of gyrosynchrotron (GS) emission \citep{bastian2001,Maia2007,Tun2013,Bain2014,Carley2017,Mondal2020a,Chhabra_2021}. 
GS emission is produced by the mildly relativistic electrons trapped in the CME plasma. This method can also provide estimates of the non-thermal electron distribution and other plasma parameters. Modeling the GS spectrum is hence regarded to be a very promising method for remote sensing the CME plasma parameters.
Despite the promise it holds and the attention it has commanded, there have been only a handful of successful attempts of detection of GS emissions from CME plasma in the last two decades.

The reason for the limited success of this approach is simply that it is challenging to detect the much fainter GS emission CME plasma (about a few tens to hundreds of $\mathrm{Jy}$) in the vicinity of the much brighter Sun. Even the quiet Sun can be a few SFU (1 SFU = 10$^4 \mathrm{Jy}$, and often, the presence of GS emission overlaps with that of much brighter non-thermal emissions associated with active regions.
To be able to routinely use this method for the estimation of CME magnetic field and other plasma parameters, one routinely needs to be able to achieve high imaging dynamic range over short temporal and spectral spans. This has recently become possible using the state-of-the-art calibration and imaging algorithms \citep{Mondal2019,Kansabanik_principle_AIRCARS} optimized for solar observations with currently perhaps the best-suited radio interferometer for this application, the Murchison Widefield Array \citep[MWA,][]{lonsdale2009,Tingay2013,Wayth2018}.

There are however additional challenges to be overcome beyond routine Stokes I detection of GS emission from CME plasma. The GS model has ten
independent parameters, assuming the non-thermal electron to follow a single power-law distribution \citep{Fleishman_2010,Kuznetsov_2021} and some of them show degeneracies which cannot be broken by Stokes I spectra alone.  Observations of polarized emission can  be used to break this degeneracy. \cite{Tun2013} reported the circular polarization from CME GS emission using the Nan\c{c}ay Radio Heliograph \citep[NRH;][]{bonmartin1983,avignon1989} observations. However, due to calibration uncertainties and low fidelity of the circular polarization image, no spatially resolved study including the constraints from Stokes V measurements could be done. The lack of polarization information and a large number of degrees of freedom of the GS models, in comparison to the available constraints, left no choice for the earlier studies but to rely on several assumptions while fitting the observed Stokes I GS spectrum. Recently a robust polarization calibration and imaging pipeline, {\em Polarimetry using Automated Imaging Routine for the Compact Arrays for the Radio Sun} \citep[P-AIRCARS;][]{Mondal2019,Kansabanik2022,Kansabanik2022_paircarsI,Kansabanik_principle_AIRCARS,Kansabanik_paircars_2} has been developed for the MWA solar observation.  Using the high-fidelity spectro-polarimetric images from P-AIRCARS, this work presents the first spatially resolved estimates of CME GS model parameters using joint constraints from Stokes I spectra and stringent upper limits on Stokes V measurements. This work also presents the first application of Bayesian analysis to this scenario.
Joint constraints from Stokes I and V yields tighter bounds on the distribution of GS model parameters than possible using Stokes I spectra alone.

This paper is organized as follows -- Section \ref{sec:obs_and_data} describes the observation and the data analysis. The imaging results are presented in Section \ref{sec:result}, along with the arguments for the observed emission arising from the GS mechanism. The impact of variations in the different parameters of the GS model on Stokes I and V spectra are presented in Section \ref{sec:spectrum_sensitivity}. Sections \ref{sec:spectrum_modeling} and \ref{subsec:magnetic_green} describe the joint Stokes I and V spectral modeling and the estimates of plasma parameters they lead to. Section \ref{sec:discussion} presents a discussion before presenting the conclusions in Section \ref{sec:conclusion}.

\section{Observation and Data Analysis}\label{sec:obs_and_data}
The observation presented here were made on 04 May 2014. On this day a total six active regions were present on the visible part of the solar disk\footnote{\url{https://www.solarmonitor.org/?date=20140504}}. No large flares (M or X GOES class) were reported. The CME catalogue provided by the Coordinated Data Analysis Workshop (CDAW) reported a total of nine CMEs\footnote{\url{https://cdaw.gsfc.nasa.gov/CME_list/UNIVERSAL/2014_05/univ2014_05.html}}, and most of them are reported as ``poor events". Of these, two have overlapping MWA observations -- one is seen to be propagating towards solar north (CME-1) and the other towards south-west (CME-2). Here we present a detailed spectro-polarimetric imaging analysis of the GS emission from the CME-1.

\subsection{Eruption and Evolution of CME-1}\label{subsec:erup_loc}
The CME-1 first appeared in the FoV of COR1 coronagraph \citep{Thompson2003} onboard STEREO-B spacecraft at 23:52:17 UTC on 03 May 2014, . It did not show any eruptive signature in the Extreme Ultra Violet (EUV) images from the Atmospheric Imaging Assembly \citep[AIA;][]{Lemen2012} onboard the Solar Dynamics Observatory \citep[SDO;][]{Pesnell2012}. This suggests that CME likely erupted from the backside of the Sun. Examining the EUV image from the Extreme Ultraviolet Imager \citep[EUVI;][]{Wuelser2004} onboard STEREO-B, we could identify the filament eruption, which gave rise to CME-1. A composite base difference image from EUVI at 195$\mathrm{\AA}$ and COR1 coronagraph at visible wavelength is shown in Figure \ref{fig:north_cme_eruption}. 
CME-1 first appeared at 00:12 UTC on 2014 May 04 in the FoV of C2 coronagraph of the Large Angle Spectroscopic Coronagraph \citep[LASCO;][]{Brueckner1995} onboard the Solar and Heliospheric Observatory \citep[SOHO;][]{Domingo1995} and was visible till 02:48 UTC in visible in the C2 FoV. 

\begin{figure}
    \centering
    \includegraphics[trim={1.7cm 0.3cm 1cm 0cm},clip,scale=0.7]{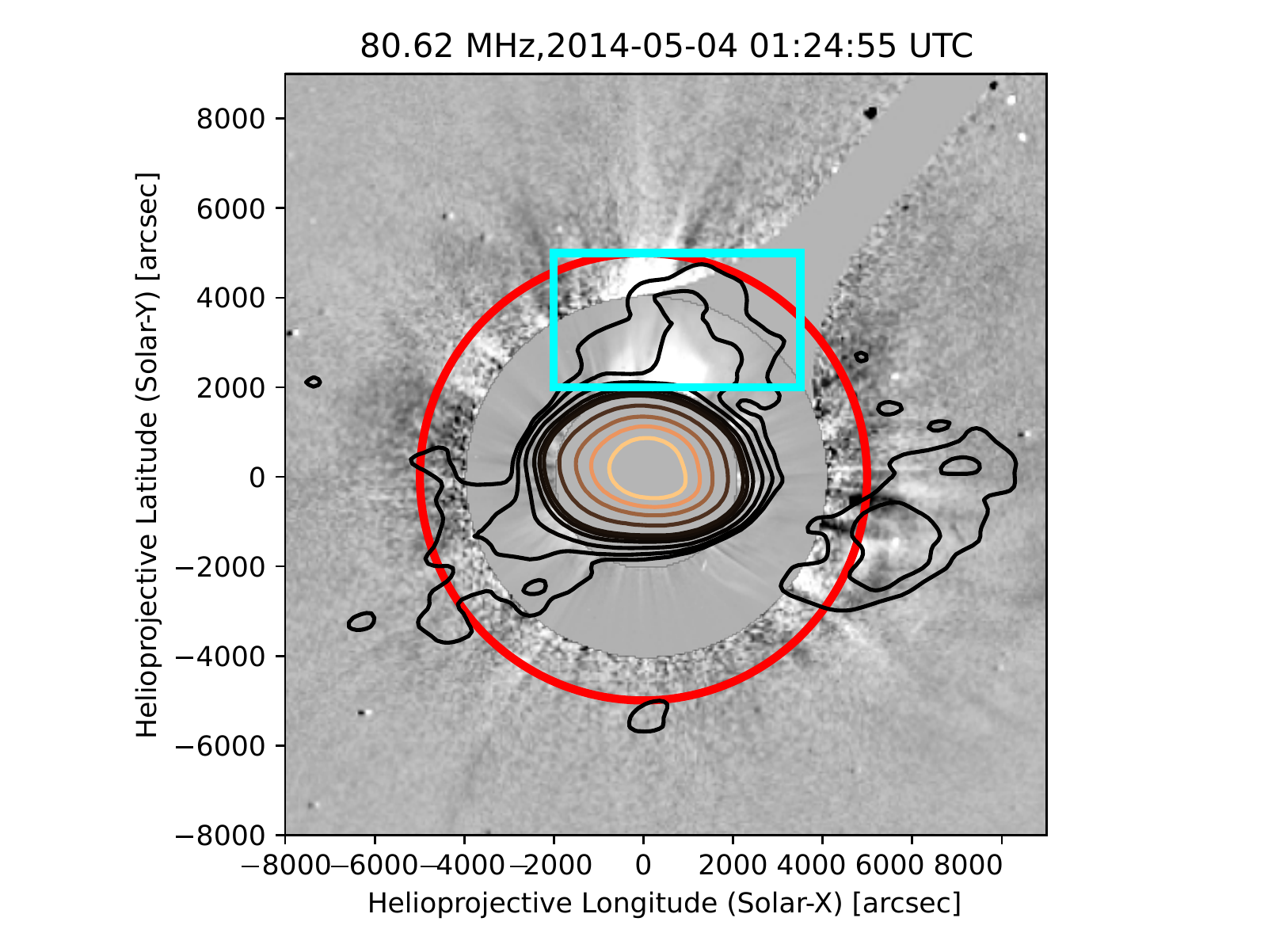}
    \caption{\textbf{Radio emission from CME-1 at 80 MHz.} Stokes I emissions at 80 MHz are shown by the contours overlaid on the base difference coronagraph images. The background shows the LASCO C2 and C3 coronagraph images from the nearest available timestamps. The inner white-light image is from C2 coronagraph and the outer image is from C3 coronagraph. Radio image is at 01:24:55 UTC. Contours levels are at  0.5, 1, 2, 4, 6, 8, 20, 40, 60, 80 \% of the peak flux density. Radio emission marked by the cyan box is from CME-1, which is detected on the sky plane out to 5.2 $R_\odot$ shown by the red circle.}
    \label{fig:north_cme}
\end{figure}

\begin{figure}
   \centering
    \includegraphics[trim={0.4cm 0.2cm 0.3cm 0.3cm},clip,scale=0.53]{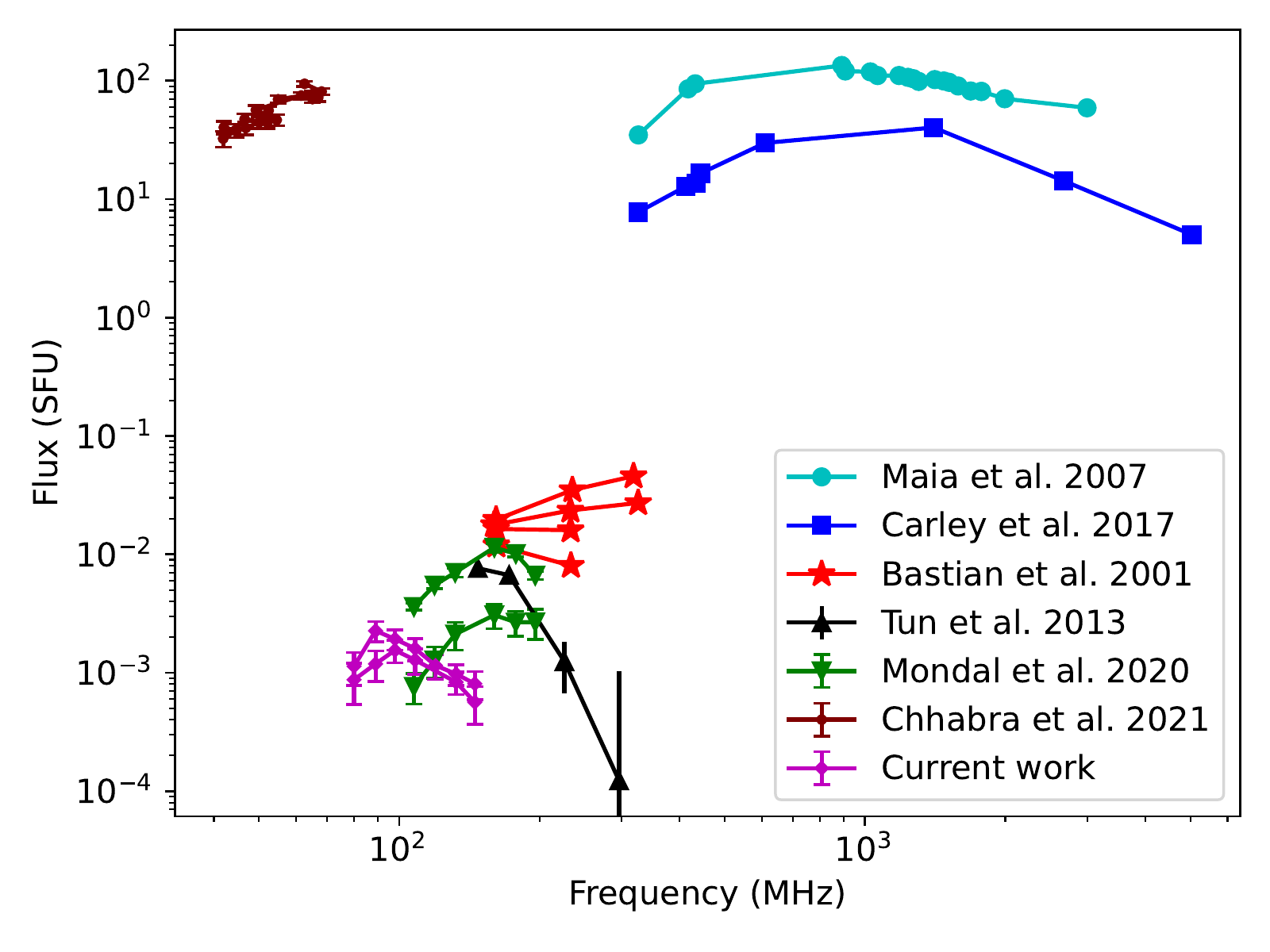}
    \caption{{\bf Comparison of spectra of gyrosynchrotron emission from CME plasma of previous and present works.} Magenta points represent sample spectra from current work, which are fainter compared to flux density observed in previous works.}
    \label{fig:past_works}
\end{figure}
\begin{figure*}[htp]
    \centering
    \includegraphics[trim={1.5cm 0cm 3cm 0cm},clip,scale=0.38]{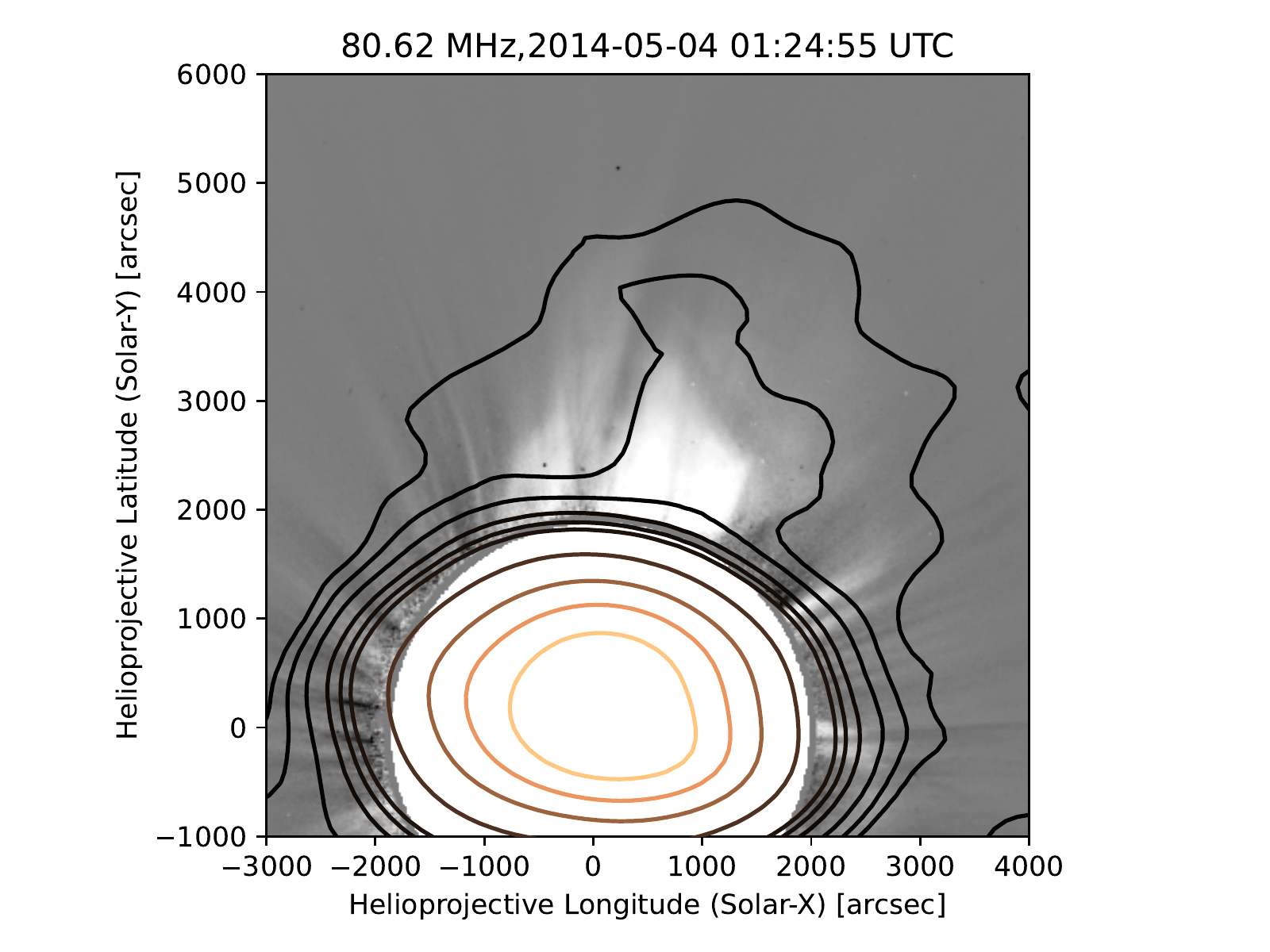}\includegraphics[trim={1.5cm 0cm 3cm 0cm},clip,scale=0.38]{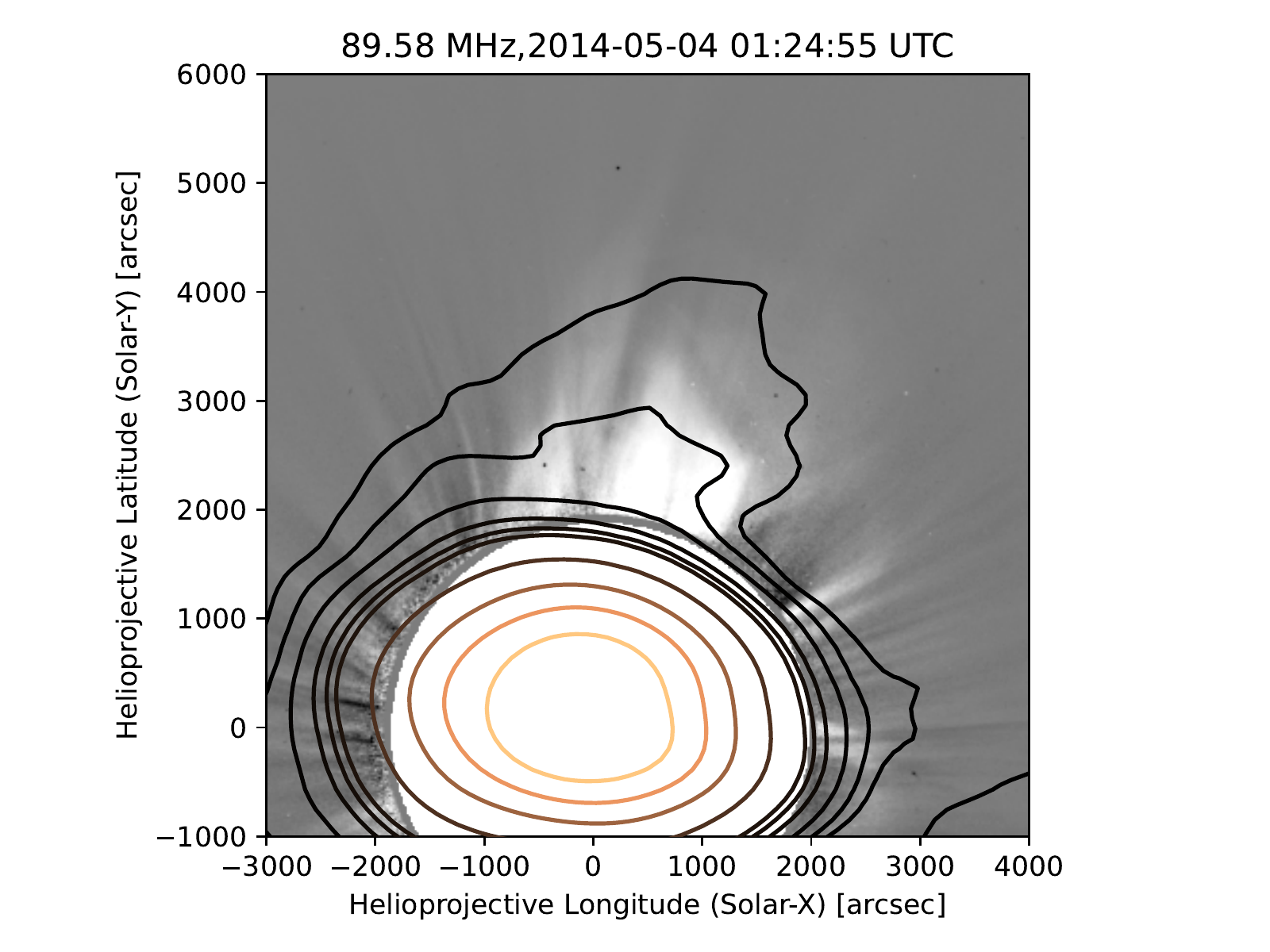}\includegraphics[trim={1.5cm 0cm 3cm 0cm},clip,scale=0.38]{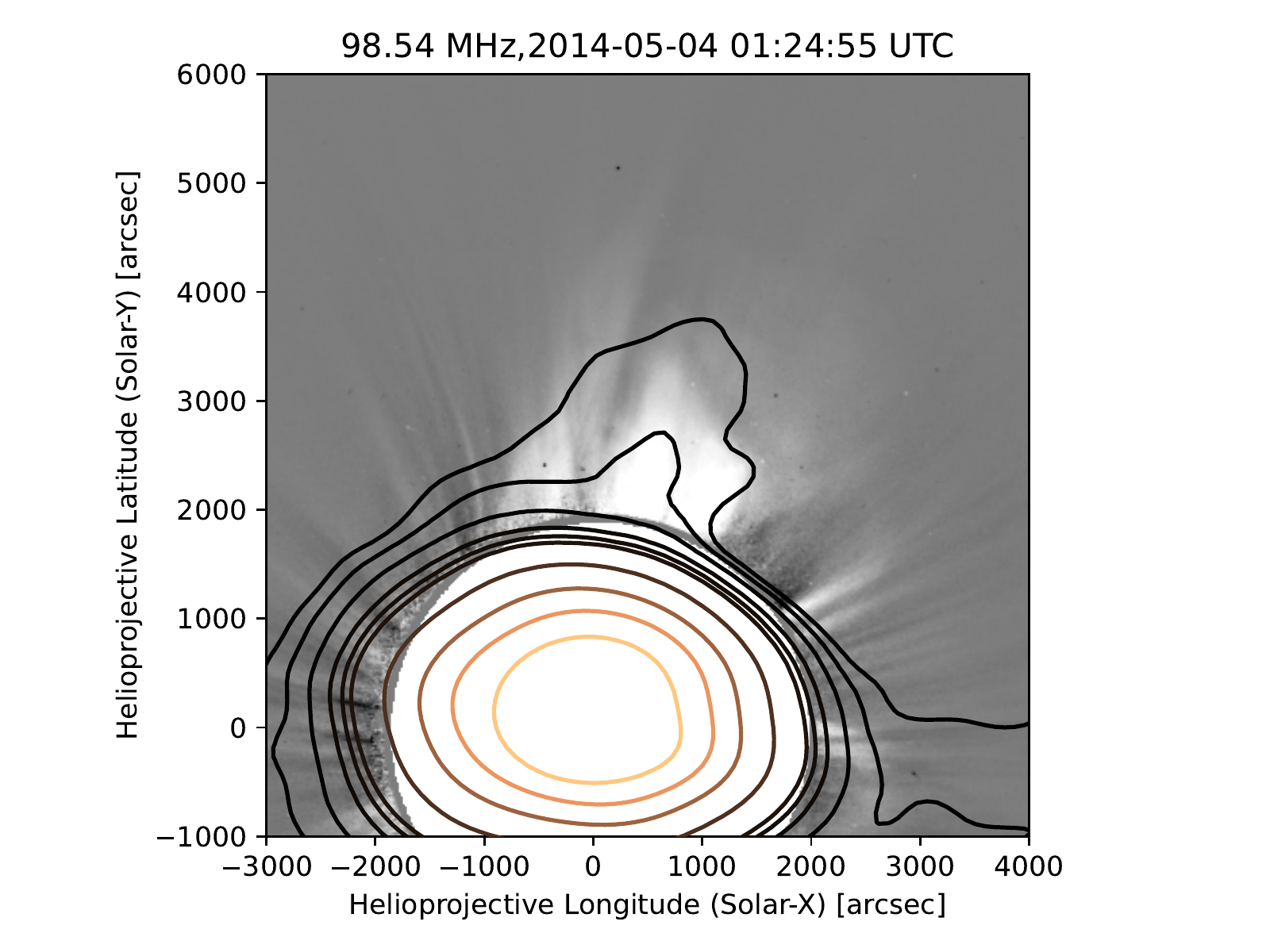}
    \includegraphics[trim={1.5cm 0cm 3cm 0cm},clip,scale=0.38]{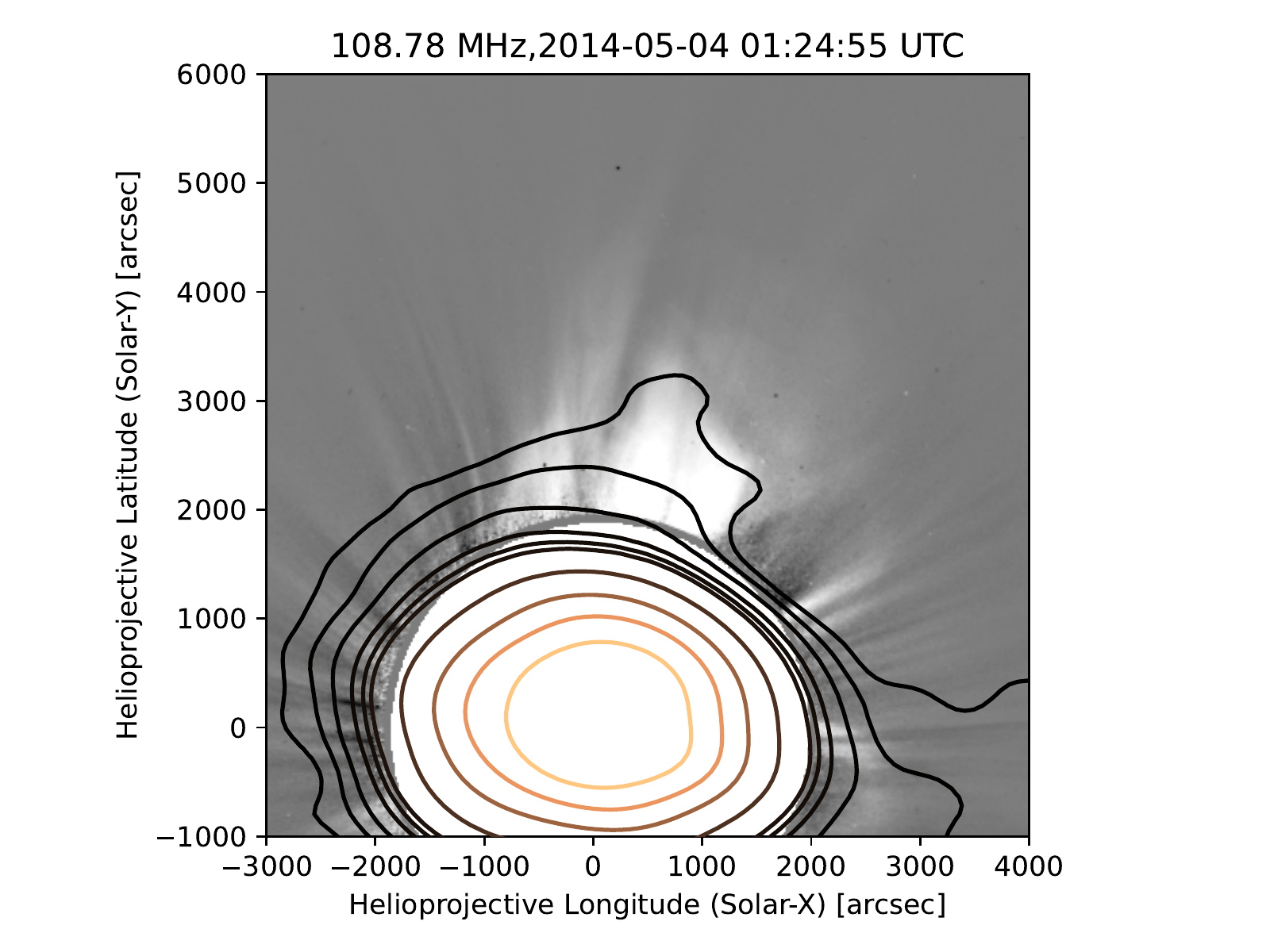}\\
    
    \includegraphics[trim={1.5cm 0cm 3cm 0cm},clip,scale=0.38]{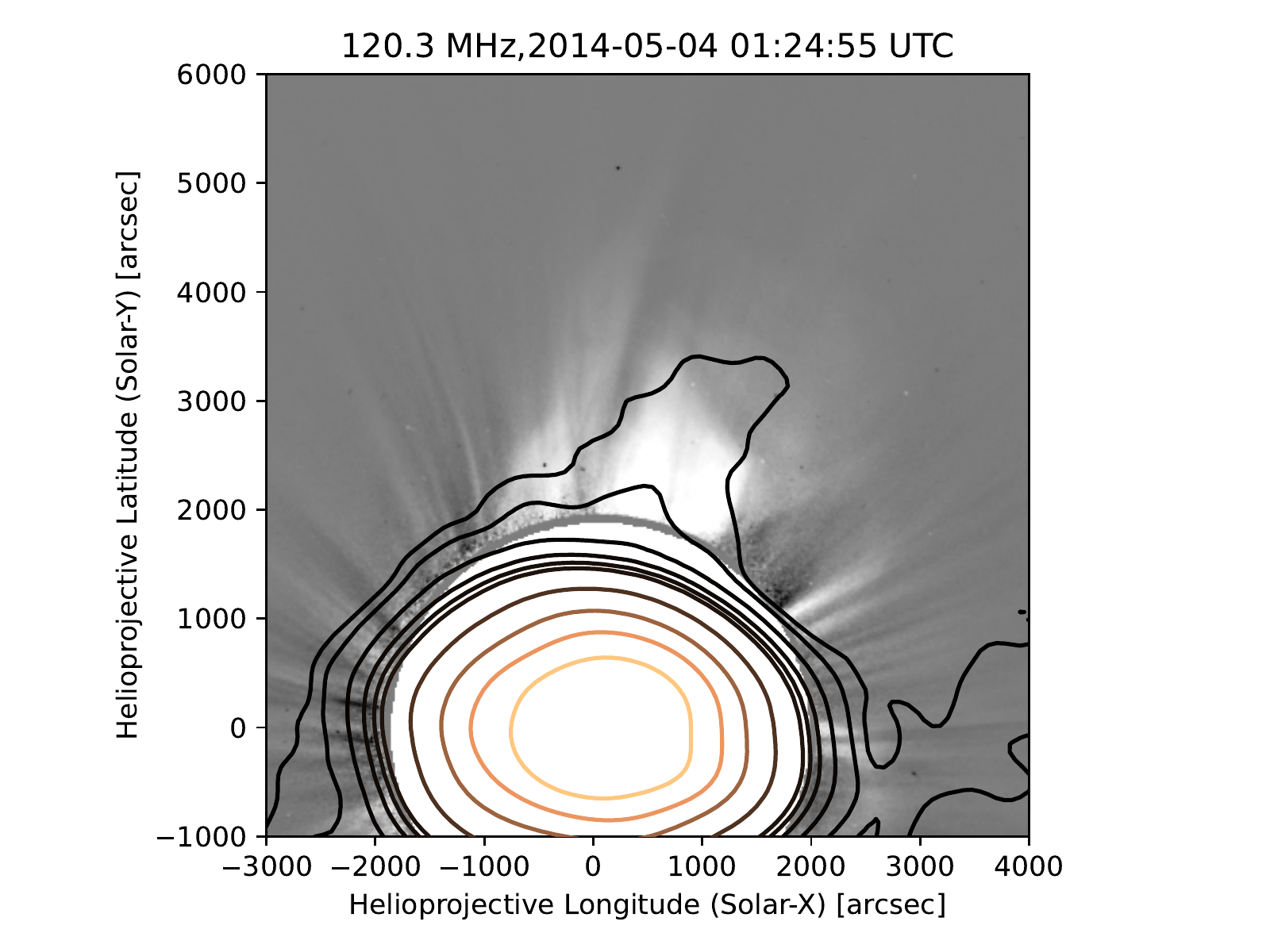}\includegraphics[trim={1.5cm 0cm 3cm 0cm},clip,scale=0.38]{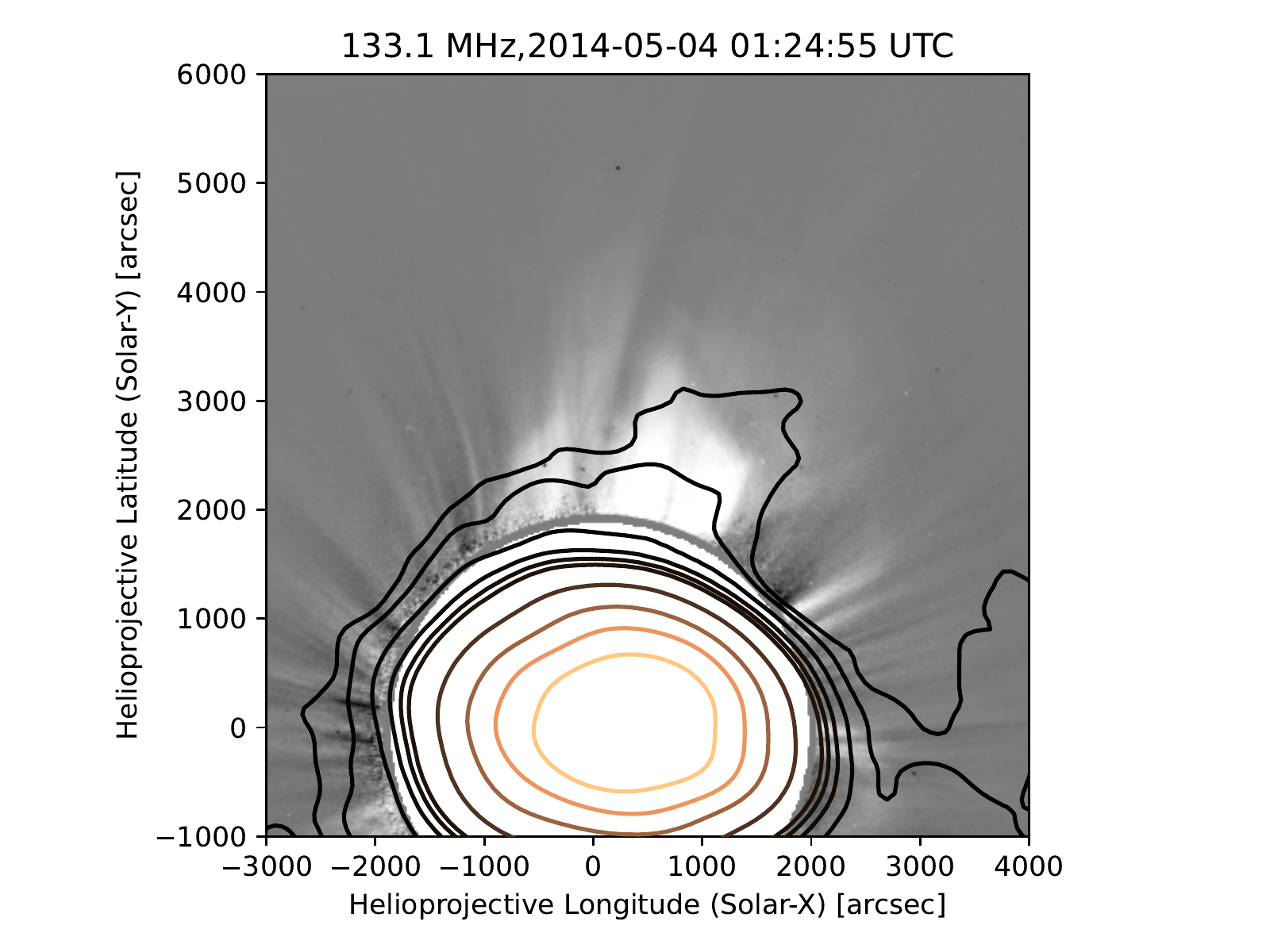}
    \includegraphics[trim={1.5cm 0cm 3cm 0cm},clip,scale=0.38]{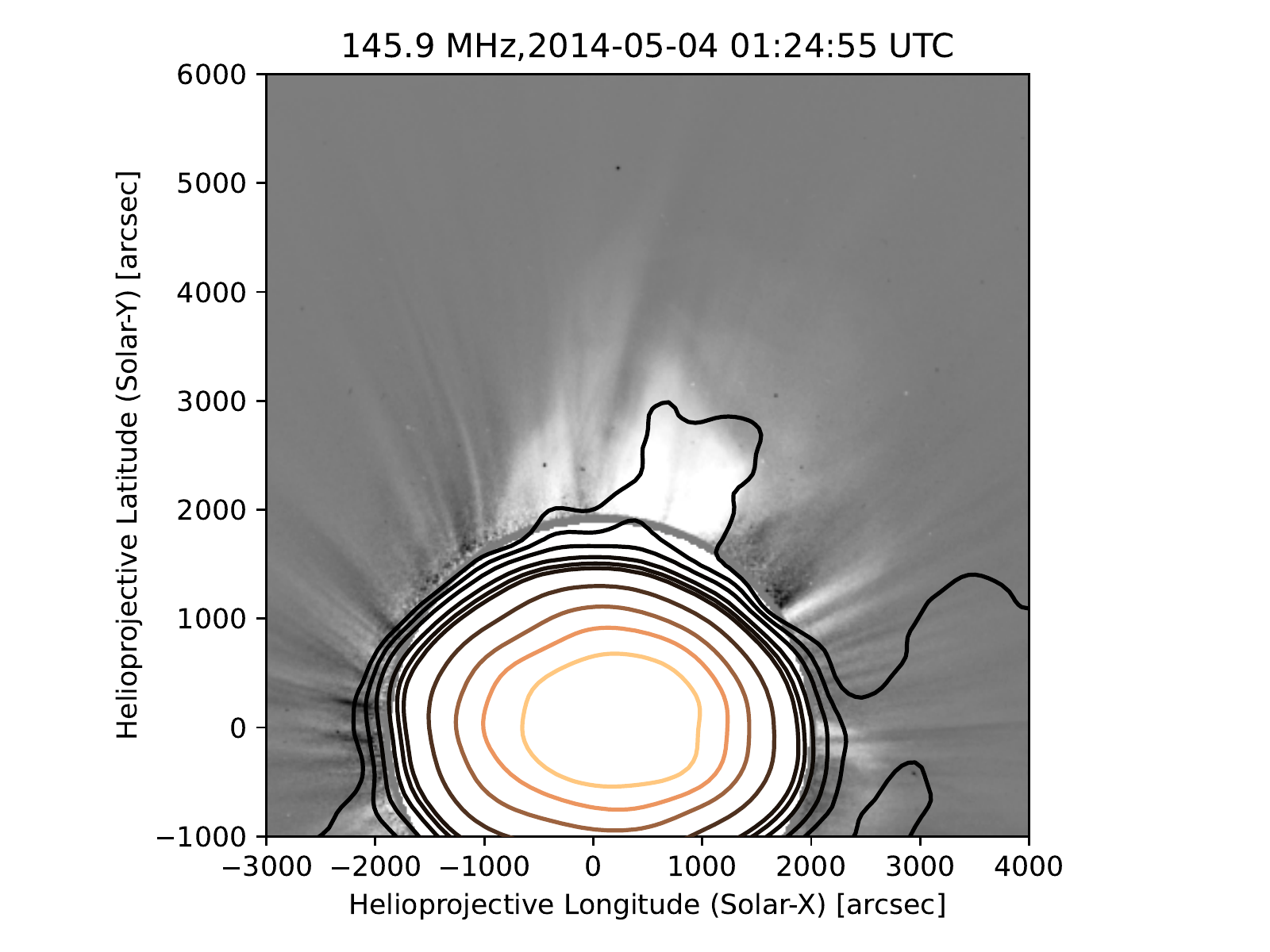}\includegraphics[trim={1.5cm 0cm 3cm 0cm},clip,scale=0.38]{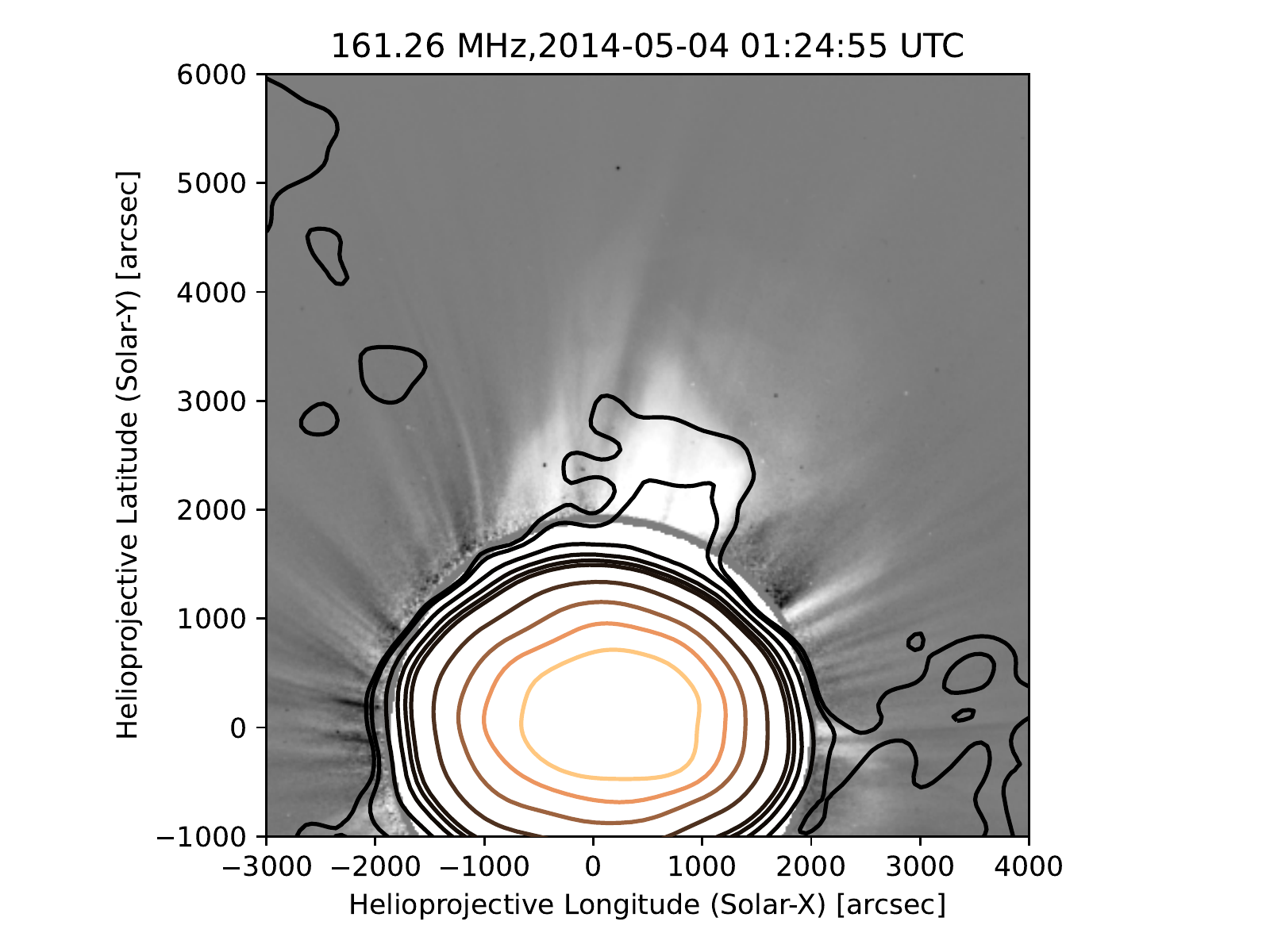}\\
    
    \includegraphics[trim={1.5cm 0cm 3cm 0cm},clip,scale=0.38]{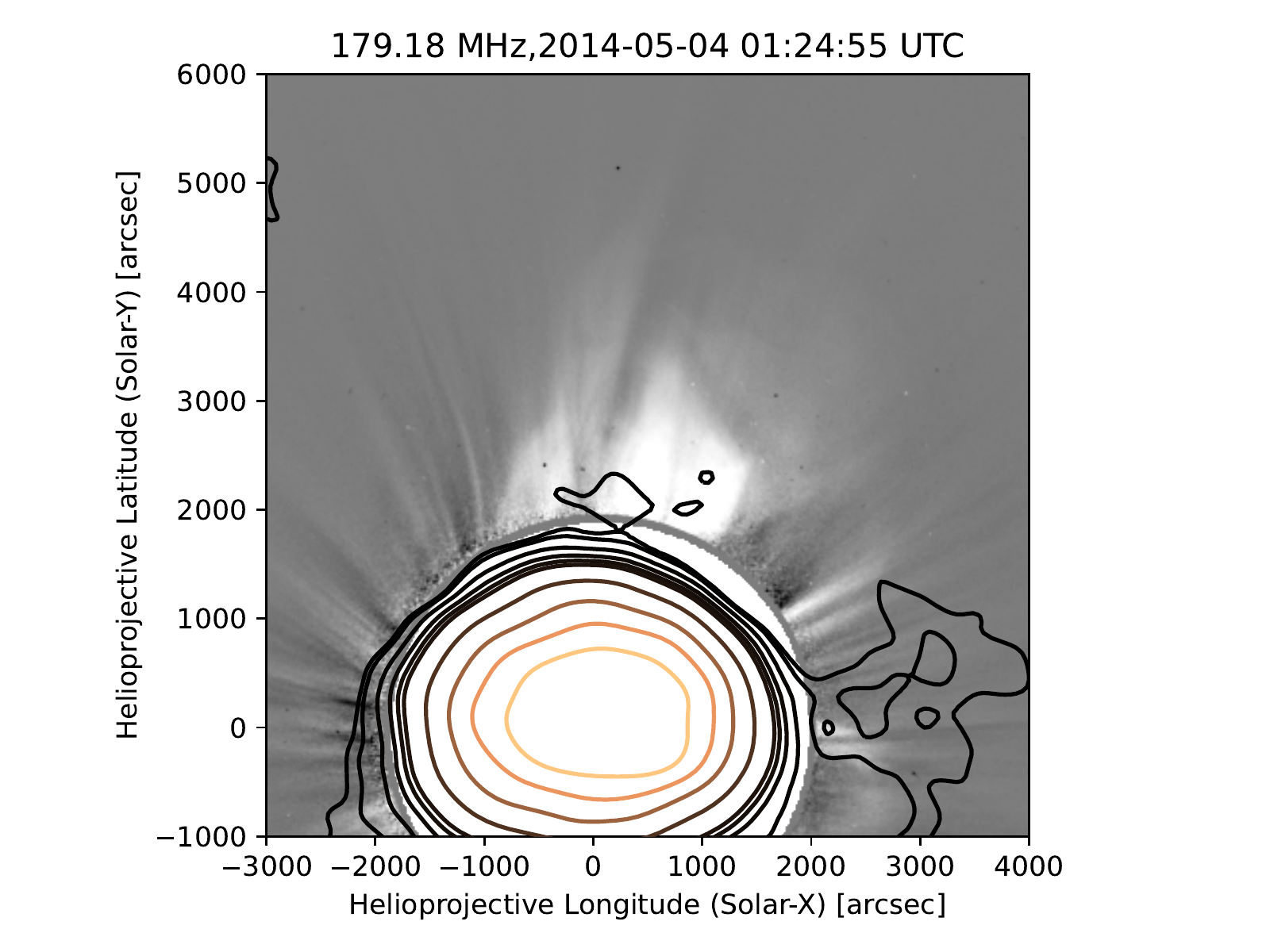}\includegraphics[trim={1.5cm 0cm 3cm 0cm},clip,scale=0.38]{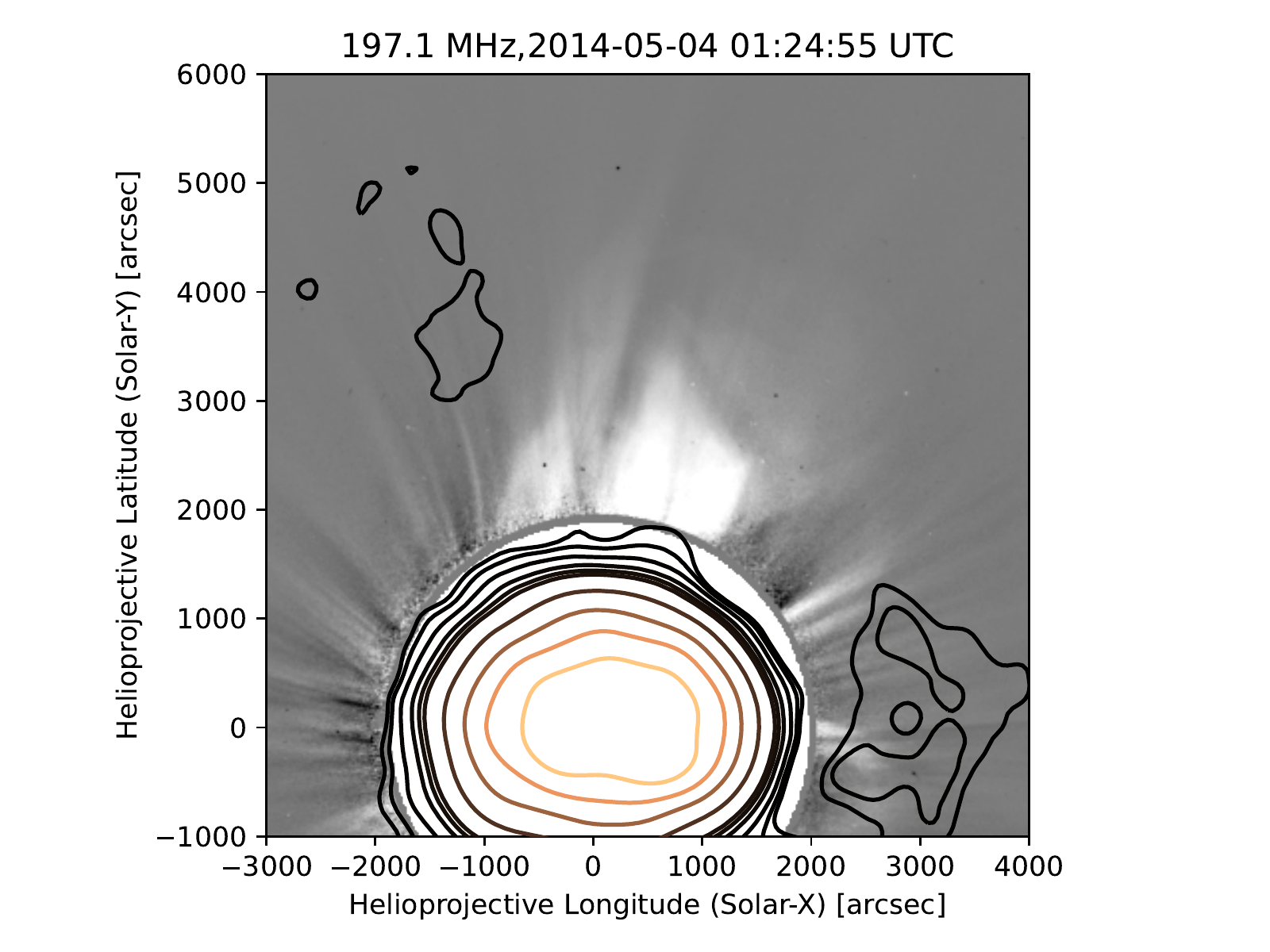}
    \includegraphics[trim={1.5cm 0cm 3cm 0cm},clip,scale=0.38]{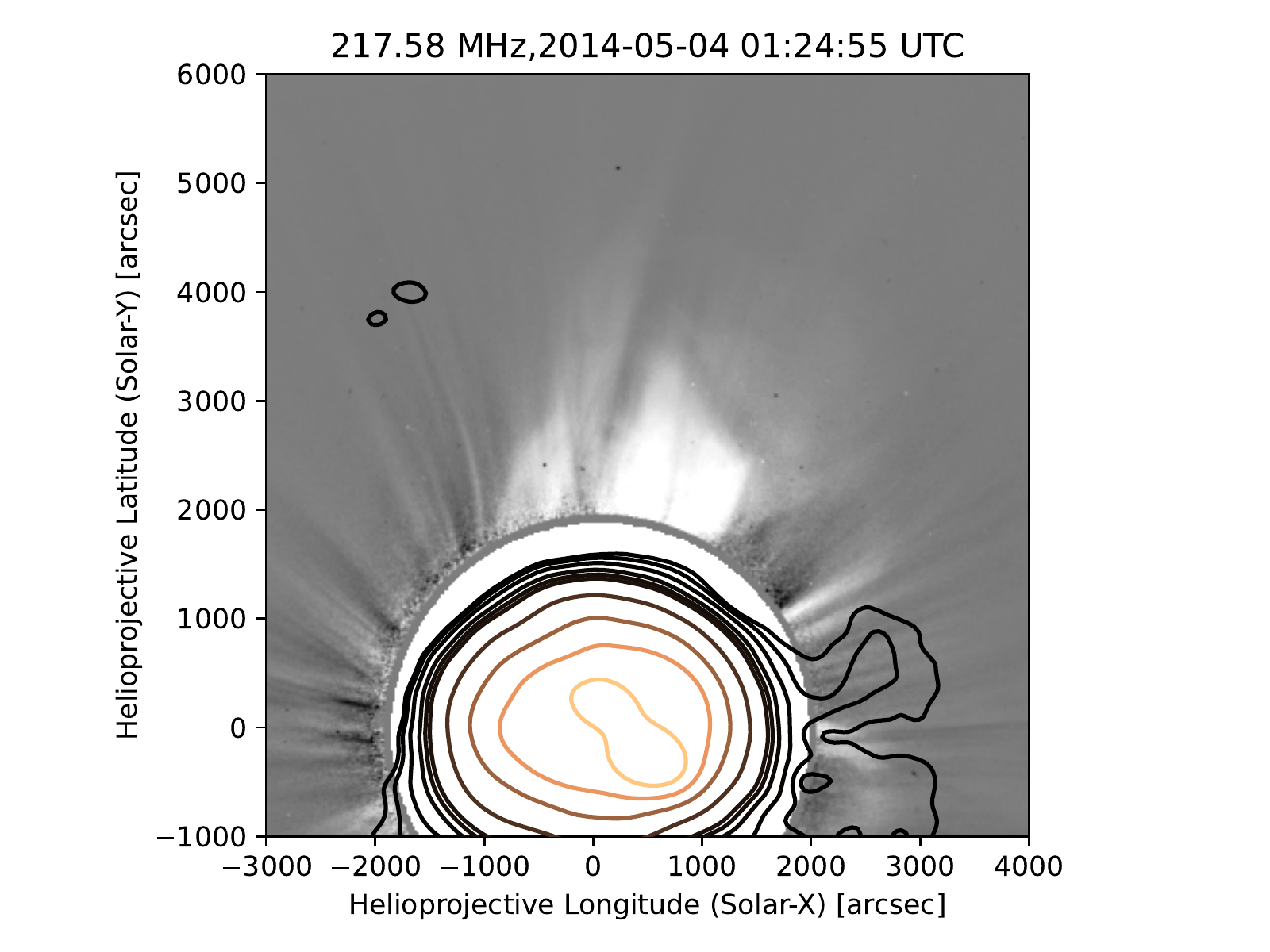}\includegraphics[trim={1.5cm 0cm 3cm 0cm},clip,scale=0.38]{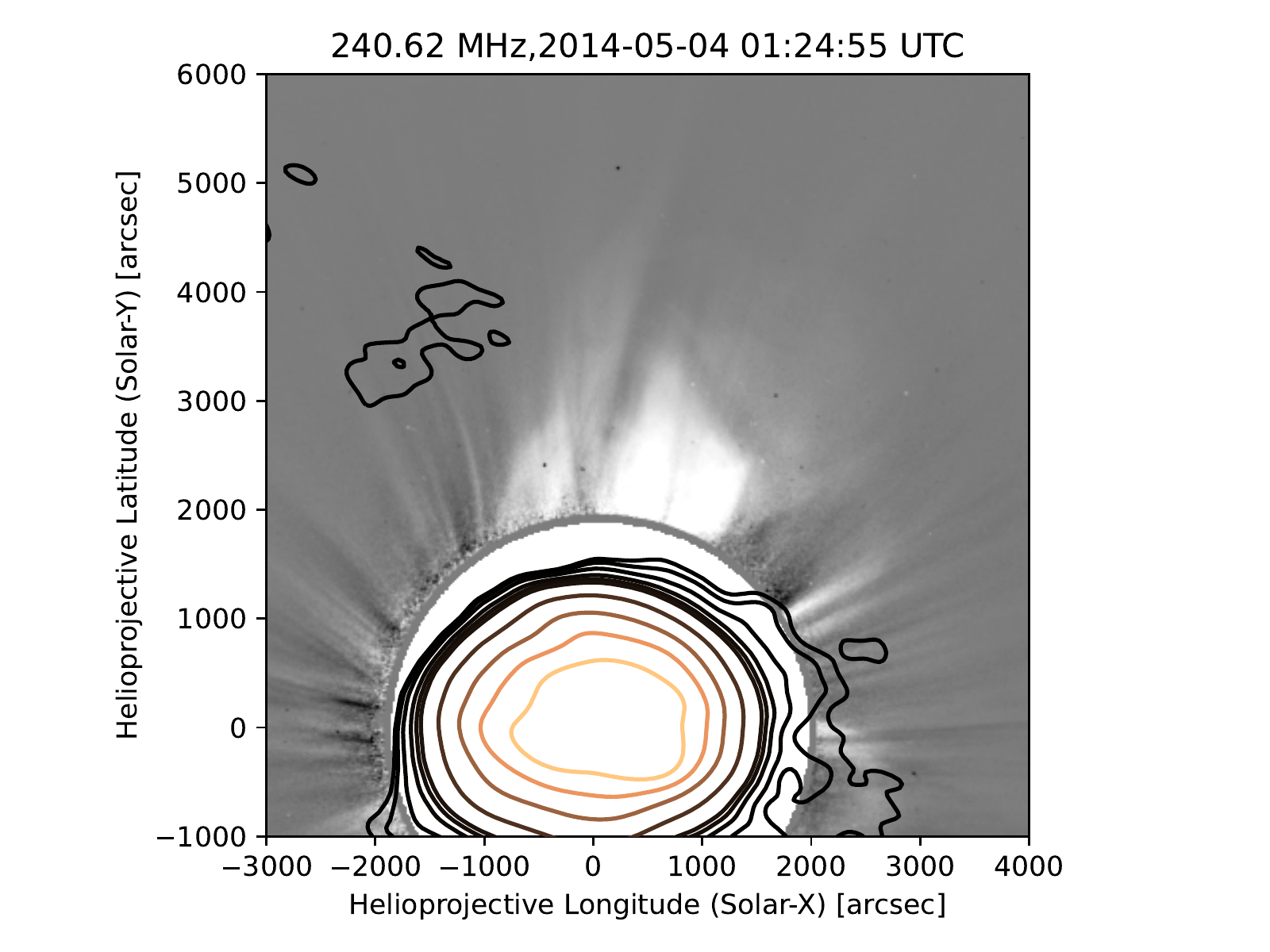}
    \caption{\textbf{Stokes I radio emission from CME-1 at multiple frequency bands of the MWA.} Frequency increases from the top left panel of the image to bottom right panel. Radio emission from CME-1 is detected upto 161 MHz. Contour levels are at 0.5, 1, 2, 4, 6, 8, 20, 40, 60, 80 \% of the peak flux density.}
    \label{fig:c2_c3_comp_freq}
\end{figure*}

\begin{figure*}
    \centering
    \includegraphics[trim={1.2cm 0.3cm 2cm 0cm},clip,scale=0.7]{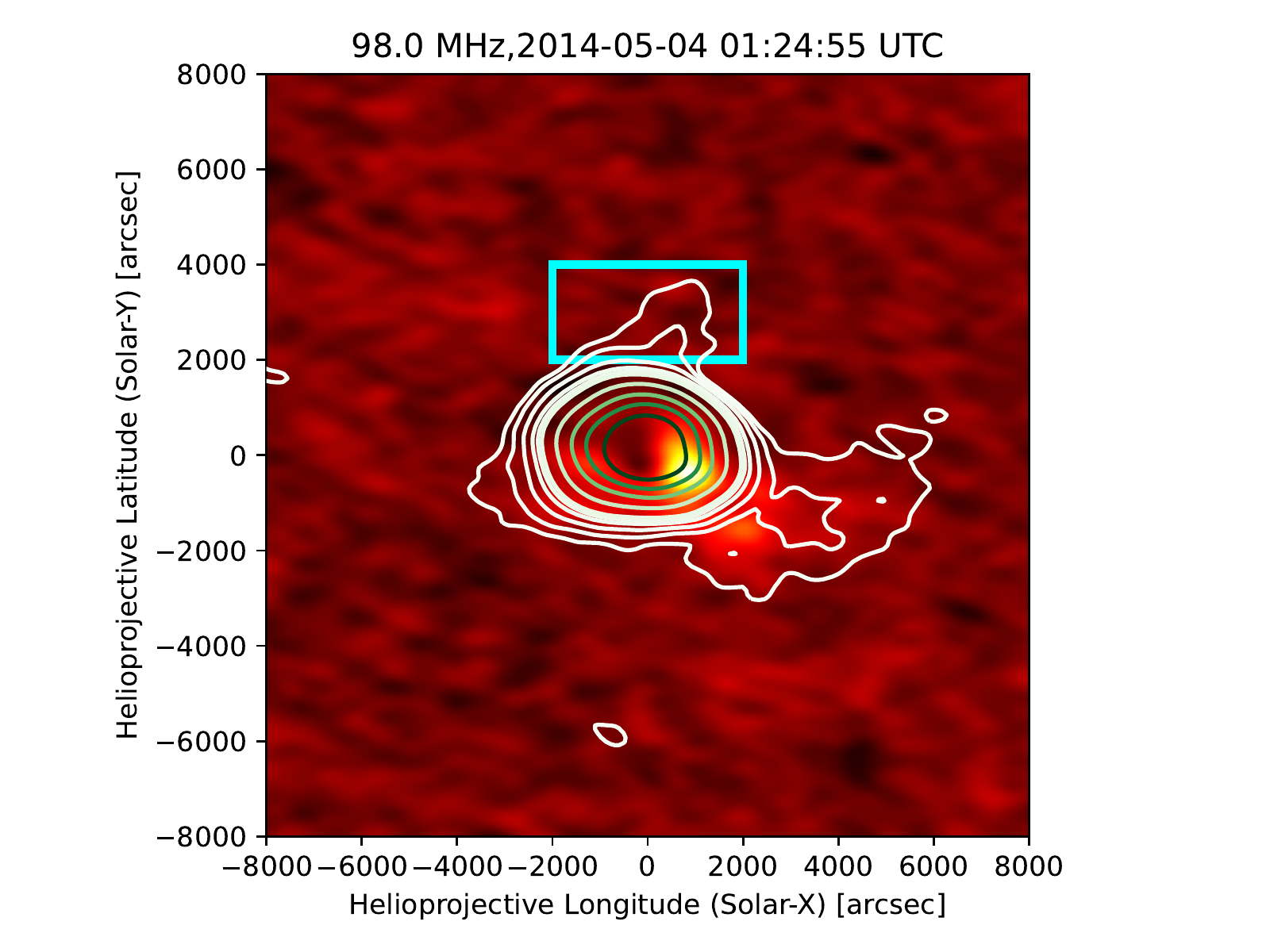}\includegraphics[trim={1.2cm 0.3cm 2cm 0.2cm},clip,scale=0.7]{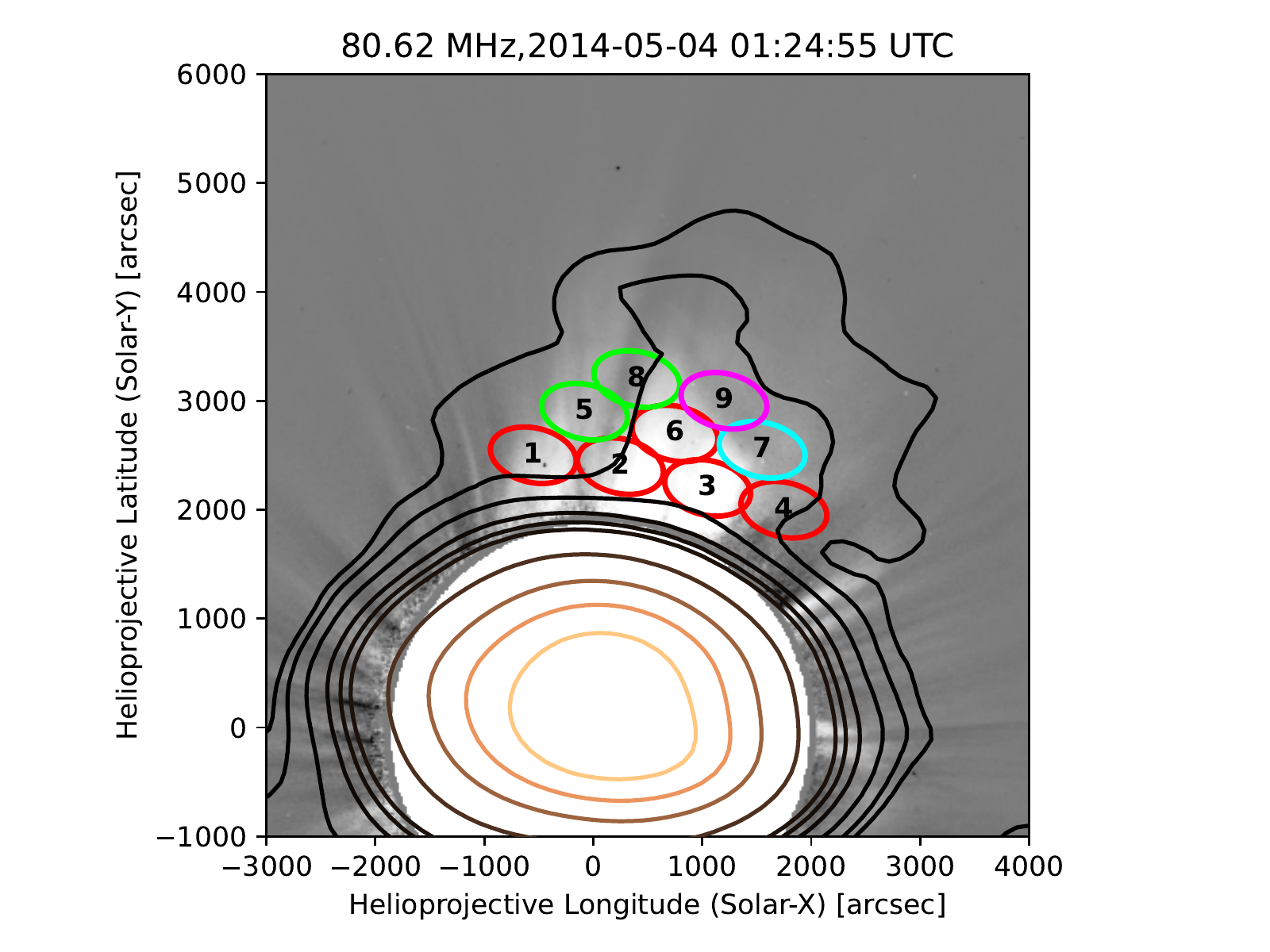}
    \caption{\textbf{Left panel: Circular polarization image for CME-1.} A sample image at 98 MHz is shown. The background image is circular polarization (Stokes V) and Stokes I emission is shown by the contours. Contours at  0.5, 1, 2, 4, 6, 8, 20, 40, 60, 80 \% level of the peak flux density. No Stokes V emission is detected from the CME-1 marked by the cyan box. The background is noise-like and there is no systematic imaging artifacts. \textbf{Right panel: Regions where spectra have been extracted.} Red regions are those where spectrum fitting is done and spectrum fitting is not done for green regions. Spectrum fitting is also done for region 7 marked by cyan keeping some parameters fixed. Region 9 marked by magenta only have a single spectral point.}
    \label{fig:circular_pol}
\end{figure*}

\subsection{Radio Observation and Data Analysis}\label{subsec:radio_data_analysis}
CME-1 was observed at meter-wavelength radio bands using the MWA. On 2014 May 04, the MWA observed the Sun from 00:48 UTC to 07:32 UTC under the project ID  G0002\footnote{\url{http://ws.mwatelescope.org/metadata/find}}. The MWA observations were done in 12 frequency bands, each of width 2.56 MHz, and centered around 80, 89, 98, 108, 120, 132, 145, 161, 179, 196, 217, and 240 $\mathrm{MHz}$. The temporal and spectral resolution of the data were 0.5 $\mathrm{s}$ and 40 $\mathrm{kHz}$, respectively. CMEs are often associated with a variety of active solar emissions -- types II, III and IV radio bursts \citep{Gopalswamy2011_CME_radio,Carley2020}.
The radio dynamic spectrum from the Learmonth Solar Spectrograph, however does not show any evidence of associated radio emission in the 
25--180 MHz band from 00:00 UTC to 03:00 UTC (Figure \ref{fig:learmonth}). 
No signature of coherent radio emission is seen in the more sensitive data from the MWA either. We also searched for any signature of type-II or interplanetary type-II bursts in WAVES radio data \citep{Bougeret1995} onboard the WIND and STEREO-A and B spacecrafts and did not find any in data from all three spacecraft. We also do not find any evidence of white-light shock in either COR-1 image from STEREO-B (Figure \ref{fig:north_cme_eruption}) or LASCO-C2 image (Figure \ref{fig:north_cme}).

We have performed polarization calibration and made full Stokes images from the MWA observation using P-AIRCARS. 
Flux density calibration was done using the technique presented by \citet{Kansabanik2022}, which is implemented in P-AIRCARS.
Integration of 10 s and 2.56 MHz was used for imaging for all 12 frequency bands. We follow the IAU/IEEE convention of Stokes parameters \citep{IAU_1973,Hamaker1996_3}. 

\section{Results}\label{sec:result}
This section presents the detection of spatially resolved GS radio emission from CME plasma using wideband spectro-polarimetric imaging observation from the MWA. 

\subsection{Radio Emission from CME-1}\label{sec:radio_detection}
Figure \ref{fig:north_cme} shows a sample Stokes I radio image at 80.62 MHz overlaid on the closest LASCO C2 and C3 base difference images.
This work focuses on the radio emission from CME-I, marked by the cyan box.
Studies of the other extended radio emissions seen in Fig. \ref{fig:north_cme}, which arise from a different CME (south-west) and a streamer (south-east), are beyond the scope of this work.

\cite{Mondal2020a} (referred to as M20 hereafter) detected spatially resoloved radio emission from CME plasma upto 4.73 $R_\odot$. 
At the time of publication, these detections were at the lowest flux densities and farthest solar distances.
Two sample spectra from the CME-1 are shown in Figure \ref{fig:past_works}.
The flux density of the radio emission from CME-1 is comparable to the weakest flux density detected by M20.
The radio emission is detected out to 5.2 $R_\odot$ (Figure \ref{fig:north_cme}), a bit beyond the maximum detection height reported by M20.

Extended radio emissions are detected at multiple frequencies from the regions co-located with CME-1. The evolution of the radio emission from CME-1 with frequency for a single time slice centered at 01:24:55 UTC is shown in Figure \ref{fig:c2_c3_comp_freq}. Frequency increases from the top left to the bottom right of the figure. Radio emission from the CME-1 is detected upto 161 MHz with more than $5\sigma_I$ significance, where $\sigma_I$ is the Stokes I map rms in a region close to the Sun. 
We also notice the spatial extent of radio emission shrinks with increasing frequency. At the lowest frequency, 80 MHz, the radio emission extends across the entire white-light CME, while at 161 MHz the emission is present only over the a small part of it. We noticed that this is not due to dynamic range limitation.

\subsection{Circularly Polarized Radio Emission from CME-1}\label{subsec:circular_pol}
Most of the previous studies \citep{Bain2014,Carley2017,Mondal2020a} did not include polarization measurements. \citet{bastian2001} observed low degree of circular polarization using NRH, but no quantitative information was reported. 
\citet{Tun2013} reported a high degree of circular polarization but did not quantify the instrumental polarization leakage.
In this work, we present high fidelity full Stokes images made using P-AIRCARS. The background color map shown in the left panel of Figure \ref{fig:circular_pol} is a sample Stokes V image at 98 MHz and the contours represent the Stokes I emission. 

\subsubsection{Estimating Stringent Upper Limits of Stokes V Emission}\label{subsec:upperlimits_estimation}
Any radio polarization measurement has two primary contributions to its uncertainty -- a fundamental limit imposed by the thermal noise of the measurement and the other arising due to imperfections in correcting for instrumental leakage. Robust polarization calibration provided by P-AIRCARS ensures that the errors introduce due to uncorrected instrumental polarization leakage is extremely small (typically less than 0.1\%, as described in \cite{Kansabanik2022_paircarsI}). In addition, there can also be systematic artifacts in the image due to errors incurred during the deconvolution process which radio imaging relies upon \citep{Cornwell1999}. The dense array footprint of the MWA provides an extremely well-behaved point-spread-function (PSF) \citep{Kansabanik_principle_AIRCARS}, which reduces deconvolution errors to a level  below those from other sources \citep{Mondal2019}. This is evident from the Stokes V map (Figure \ref{fig:circular_pol}, left panel), which clearly shows that the background is noise-like and no systematic artifacts are seen in the image. The measured rms in the Stokes V image ($\sigma_\mathrm{V}$) is only about 1.3 times the expected instrumental thermal noise, further attesting to the high quality calibration and imaging. The rms values vary with frequency and are listed in Table \ref{table:stokesV_noise}.
The Stokes V emission from CME-1 is too weak to be detected at any of the observing bands.
Nonetheless, the noise like nature of these images at the location of CME-1 and the low values of instrumental leakage enable us to place robust upper limits on the absolute value of the Stokes V emission \citep[e.g.,][etc.]{Bastian2000,Lynch2017,Lenc2018,Cendes_2022} at each of the frequency bands as discussed further in Section \ref{sec:spectroscopy}.

\begin{table}
\centering
    \renewcommand{\arraystretch}{1.4}
    \begin{tabular}{|p{1.5cm}|p{1.5cm}|p{1.5cm}|p{1.5cm}|}
    \hline
       Frequency (MHz) & $\sigma_\mathrm{V}$ (Jy) & Frequency (MHz) & $\sigma_\mathrm{V}$ (Jy)\\ \hline \hline 
        80 & 4.55 & 145 &  1.52\\
        \hline
        89 & 4.38 & 161 &  1.24\\
        \hline
        98 & 3.75 & 179 &  0.71\\
        \hline
        108 & 3.20 & 197 & 0.45\\
       \hline
       120 & 2.18 & 217 & 1.74\\
       \hline
       132 & 1.82 & 240 & 0.95\\
       \hline
    \end{tabular}
    \caption{\textbf{Measured rms noise from the Stokes V maps at 12 spectral bands.}}
    \label{table:stokesV_noise}
\end{table}

\subsection{Spatially Resolved Spectroscopy}\label{sec:spectroscopy}
Wideband imaging observations allow us to perform spatially resolved spectroscopy of the radio emission from CME-1. We have extracted spectra from regions with size equal to the size of the PSF  at the lowest observing frequency of 80 MHz. These regions are shown in the right panel of Figure \ref{fig:circular_pol} and have been chosen to ensure that the Stokes I emission is seen at 0.5\% level or more in at least two spectral bands.

We have calculated rms noise ($\sigma$) and mean ($\mu$) over a comparatively large region close to the Sun. We have also calculated the deepest negative ($n$) over a region close to the CME, and rms noise ($\alpha$) far away from the Sun. The flux density ($f$) for a region at a given frequency is considered to be a reliable detection, if all of the 
following three criteria are satisfied:
\begin{enumerate}
    \item $f>\mu+5\sigma$
    \item $f>5\alpha$
    \item $f>5|n|$
\end{enumerate}
These stringent selection criteria ensure that we do not include any spectral points which are prone to imaging artifacts. 

The uncertainty of the Stokes I flux density, $\sigma_\mathrm{I}$, is estimated as,
\begin{equation}
    \sigma_\mathrm{I}=\mathrm{max}(\mu,\sigma).
\end{equation} 
The uncertainty of Stokes V is also estimated in a similar fashion. For Stokes V image, $\mu_\mathrm{V}$ is close to zero, and $\alpha$ is comparable to $\sigma$. Hence, we only consider the rms noise calculated from the Stokes V image close to the Sun as the uncertainty, $\sigma_\mathrm{V}$.  As we do not have any Stokes V detection and Stokes V can not be more than Stokes I, we use $V_u=\mathrm{min(5\sigma_\mathrm{V},I)}$ as the upper limit on absolute value of Stokes V for each of the frequency bands. 

Spectra are fitted for the red regions which have detections more than five spectral bands (Figure \ref{fig:circular_pol}, right panel). For these regions we fit for five GS model parameters as discussed in Section \ref{sec:spectrum_modeling}. Region 7 marked by cyan in the same figure has a clear peak in the spectrum, but is detected only at four spectral bands. 
Hence spectral fitting for region 7 is performed holding some additional GS model parameters constant. For the regions marked in green, the magnetic field strength is estimated as discussed in Section \ref{subsec:magnetic_green}. Emission from region 9 is detected at two spectral bands, but one of them falls short of meeting all of the selection criteria. 

\subsection{Emission Mechanism}\label{subsec:emission_mechanism}
\begin{figure}
   \centering
    \includegraphics[trim={0.2cm 0.5cm 0cm 0cm},clip,scale=0.54]{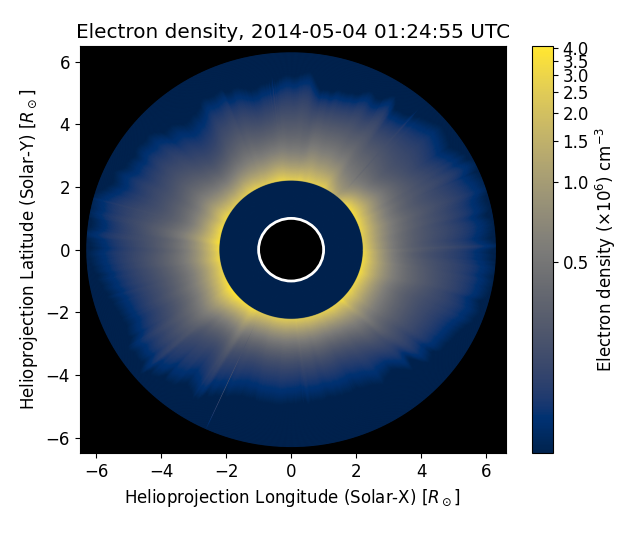}
    \caption{\textbf{Total coronal electron density at 01:24:55 UTC obtained from LASCO-C2 white-light image.} Electron density is estimated using the inversion method developed by \cite{Hayes2001}.}
    \label{fig:electron_density}
\end{figure}

Possible mechanisms for explaining radio emissions from CMEs are -- plasma emission, free-free emission and GS emission. All of these mechanism have dependence on the local plasma density. We have estimated the coronal electron density from the LASCO-C2 white light coronagraph images using the inversion method developed by \cite{Hayes2001}. Coronal electron density map is shown in Figure \ref{fig:electron_density}.

Average electron density over the region of CME-1 is about $10^6\ \mathrm{cm^{-3}}$ and leads to a corresponding plasma frequency of about $8.5\ \mathrm{MHz}$. However, the radio emission from CME-1 is detected at more than an order of magnitude higher frequency.
The observing frequency being much larger than the local plasma frequency convincingly rules out plasma emissions as the possible mechanism. 

The next possibility we examine is free-free emission. Considering the coronal plasma temperature $T_\mathrm{e}\approx10^6\ \mathrm{K}$, and neglecting magnetic fields, free-free optical depth is given by \citep{Gary1994},
\begin{equation}
    \tau_\nu\approx0.2\ \frac{\int n_e^2\ dl}{\nu^2\ T_e^{\frac{3}{2}}}.
    \label{eq:free_free_optical_depth}
\end{equation}
Assuming uniform density along the part of the LoS intersecting the CME, $\int n_e^2dl=n_e^2L$, where, $L$ is the LoS depth. Assuming, $L=1\ R_\odot$, $\tau_\nu$ becomes unity at 3.7 MHz. Since the frequency of observation is more than 10 times higher than this value, optically thick free-free emission is also ruled out.
For optically thin free-free emission, the brightness temperature  is proportional to $\nu^{-2}$ (or a flat flux density spectrum). 
As the observed spectra show well defined peaks, optically thin free-free emission can also be ruled out.
The only likely emission mechanism remaining is the GS emission.

\section{Gyrosynchrotron Emission : Parameter Sensitivity}\label{sec:spectrum_sensitivity}
Mildly relativistic electrons gyrating in magnetic field emit GS emission. GS emission mechanism is well understood theoretically \citep{Melrose1968,Ramaty1969}. 
However, using the exact expressions is computationally very expensive.
Over last decade or so fast gyrosynchrotron codes have been developed \citep{Fleishman_2010,Kuznetsov_2021}.
These codes are versatile and can produce GS spectra for any given distribution of energy and pitch angles of non-thermal electrons.
\cite{Fleishman_2010} quantified the differences between the spectra using exact and approximate expressions and \cite{Kuznetsov_2021} quantified the effects of pitch angle distributions. 
Building on this and benefiting from the significantly reduced computation time, here we explore the phase space of GS model parameters.

The electron distribution can be described by the expression,
\begin{equation}
    f(E,\mu)=u(E)\ g(\mu),
\end{equation}
where $u(E)$ is electron energy distribution function, $g(\mu)$ is the electron pitch angle distribution function, $E$ is the energy of the electron, $\mu=cos\alpha$, $\alpha$ being the electron pitch angle. 
The normalization conditions for $u(E)$ and $g(\mu)$ are:
\begin{equation}
    \int_{E_\mathrm{min}}^{E_\mathrm{max}} u(E)\ dE=\frac{n_e}{2\pi};\\
    \int_{-1}^{+1} g(\mu)\ d\mu=1
\end{equation}
where, $n_e$ is the non-thermal electron density.
For simplicity, we assume an isotropic distribution of pitch angle.
The simplest form of $u(E)$ is a single power-law. 
\begin{equation}
    u(E) = NE^{-\delta},\ \mathrm{for}\ E_\mathrm{min}<E<E_\mathrm{max},
\end{equation}
where $E_\mathrm{min}$ is the minimum and $E_\mathrm{max}$ is the maximum energy cutoff and $N$ is a normalization constant. 
We also assume homogeneous distribution along the relevant part of the LoS.

\begin{figure*}
    \centering
    \includegraphics[trim={0.5cm 0.5cm 0.5cm 0.5cm},clip,scale=0.29]{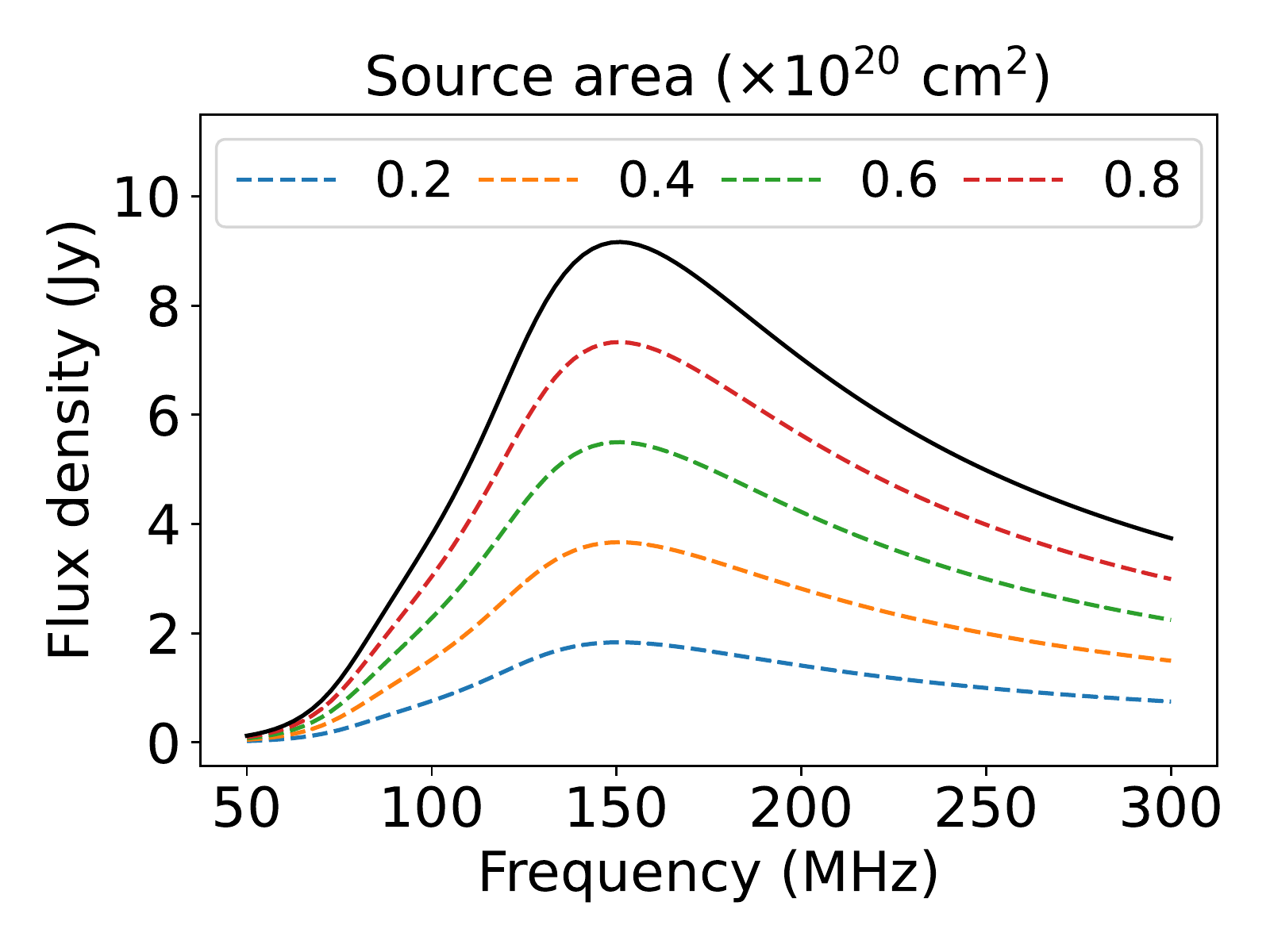}\includegraphics[trim={0.5cm 0.5cm 0.5cm 0.5cm},clip,scale=0.29]{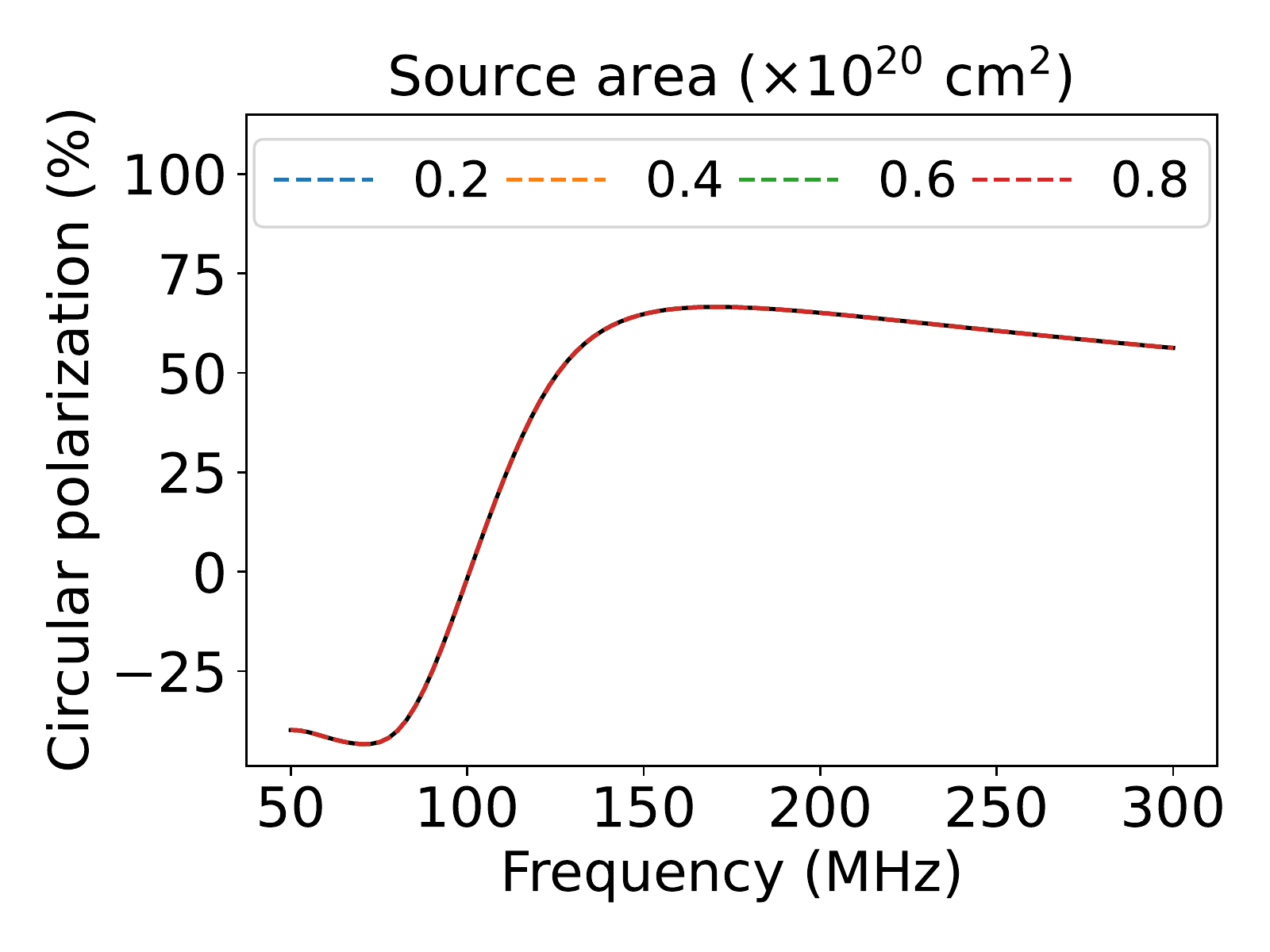}\includegraphics[trim={0.5cm 0.5cm 0.5cm 0.5cm},clip,scale=0.29]{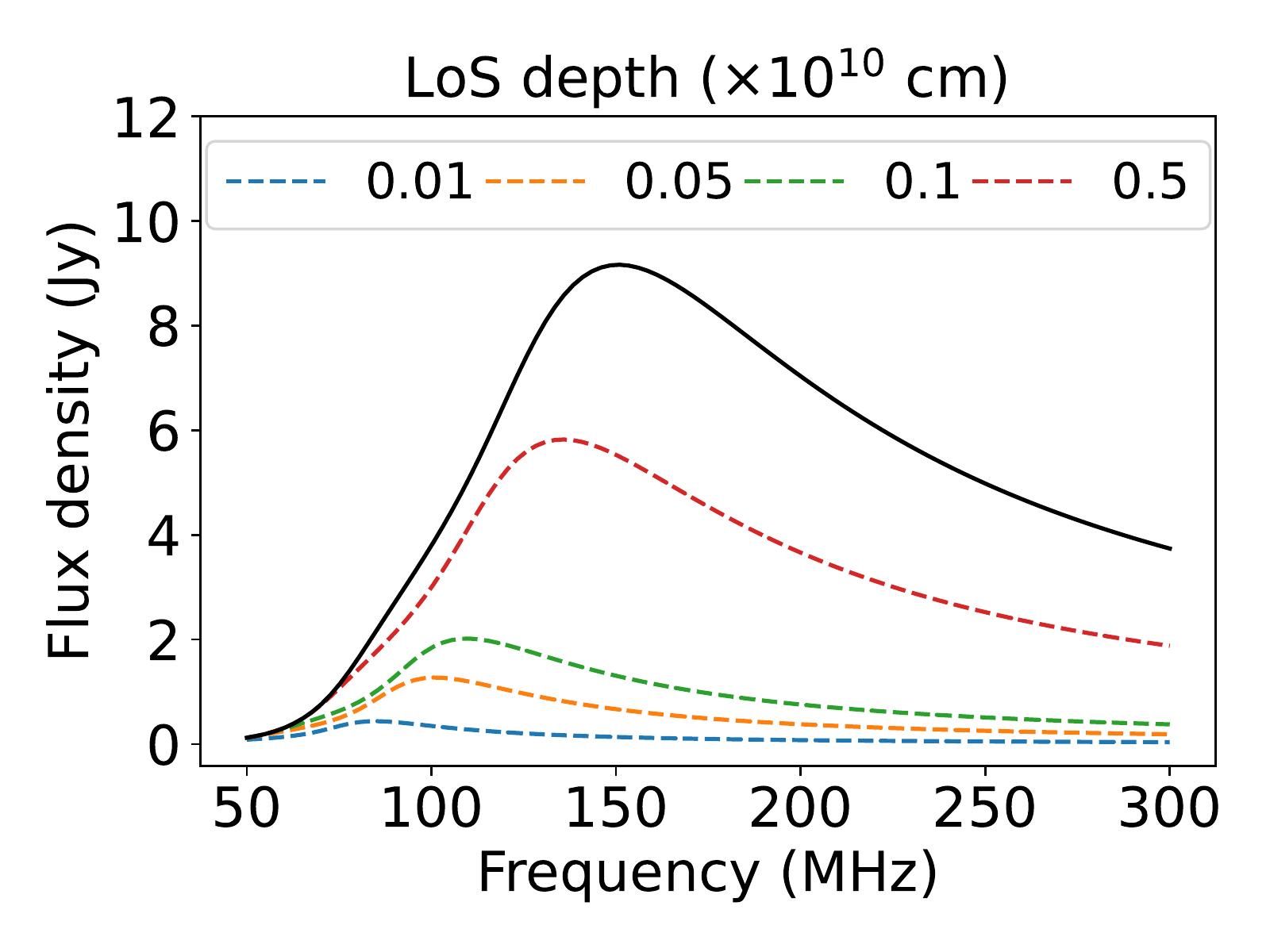}\includegraphics[trim={0.5cm 0.5cm 0.5cm 0.5cm},clip,scale=0.29]{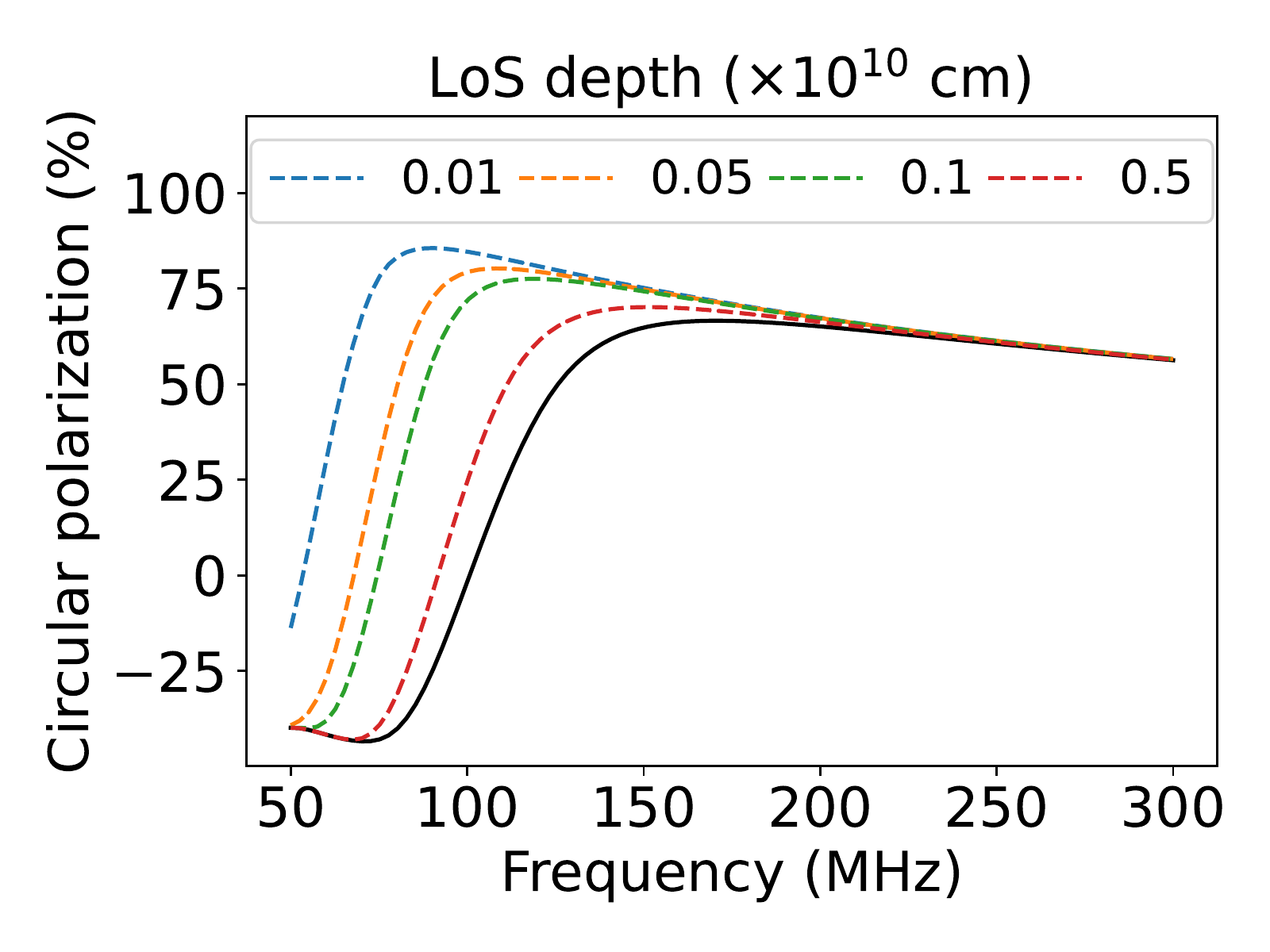}\\
    
    \includegraphics[trim={0.5cm 0.5cm 0.5cm 0.5cm},clip,scale=0.29]{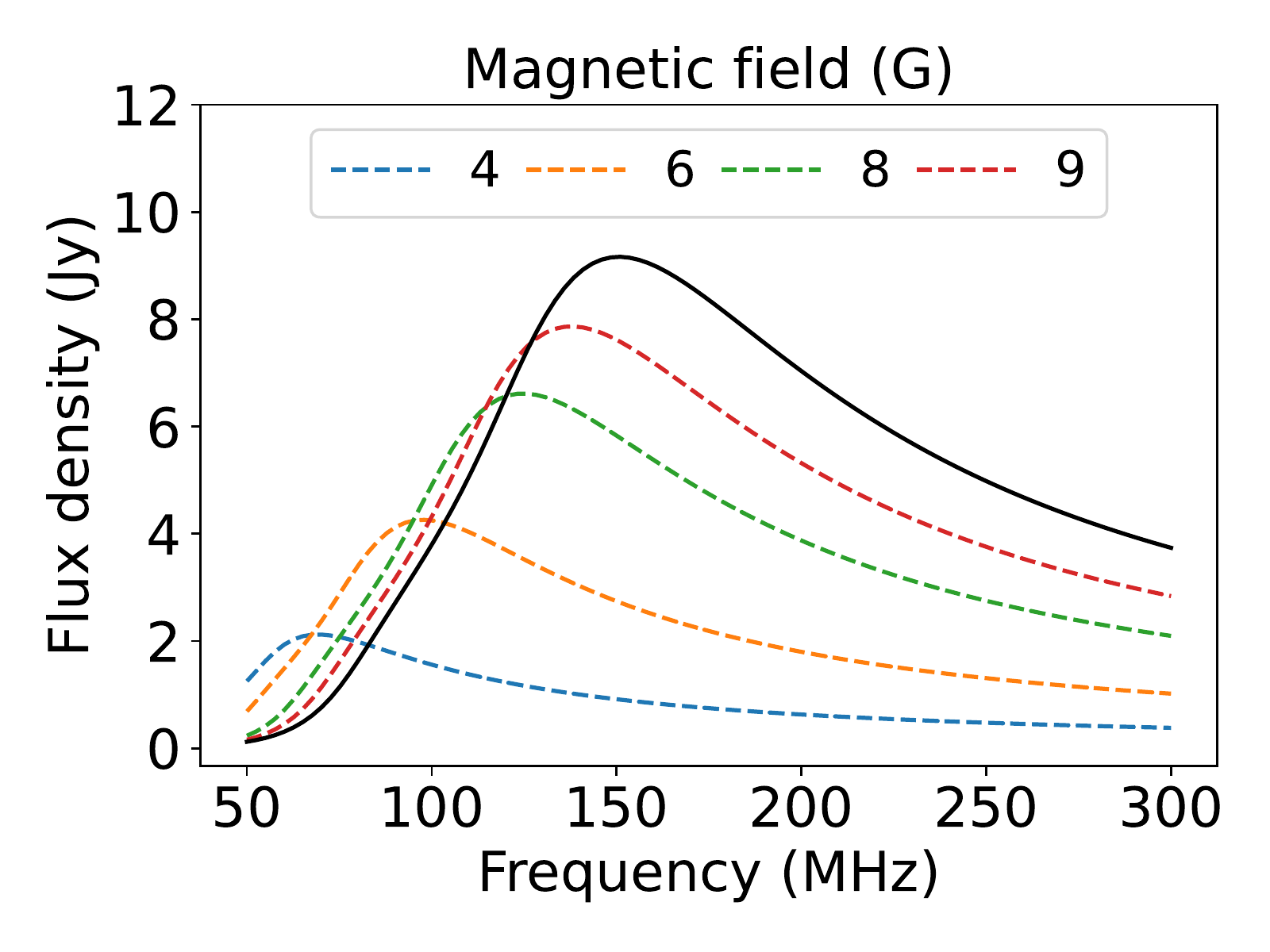}\includegraphics[trim={0.5cm 0.5cm 0.5cm 0.5cm},clip,scale=0.29]{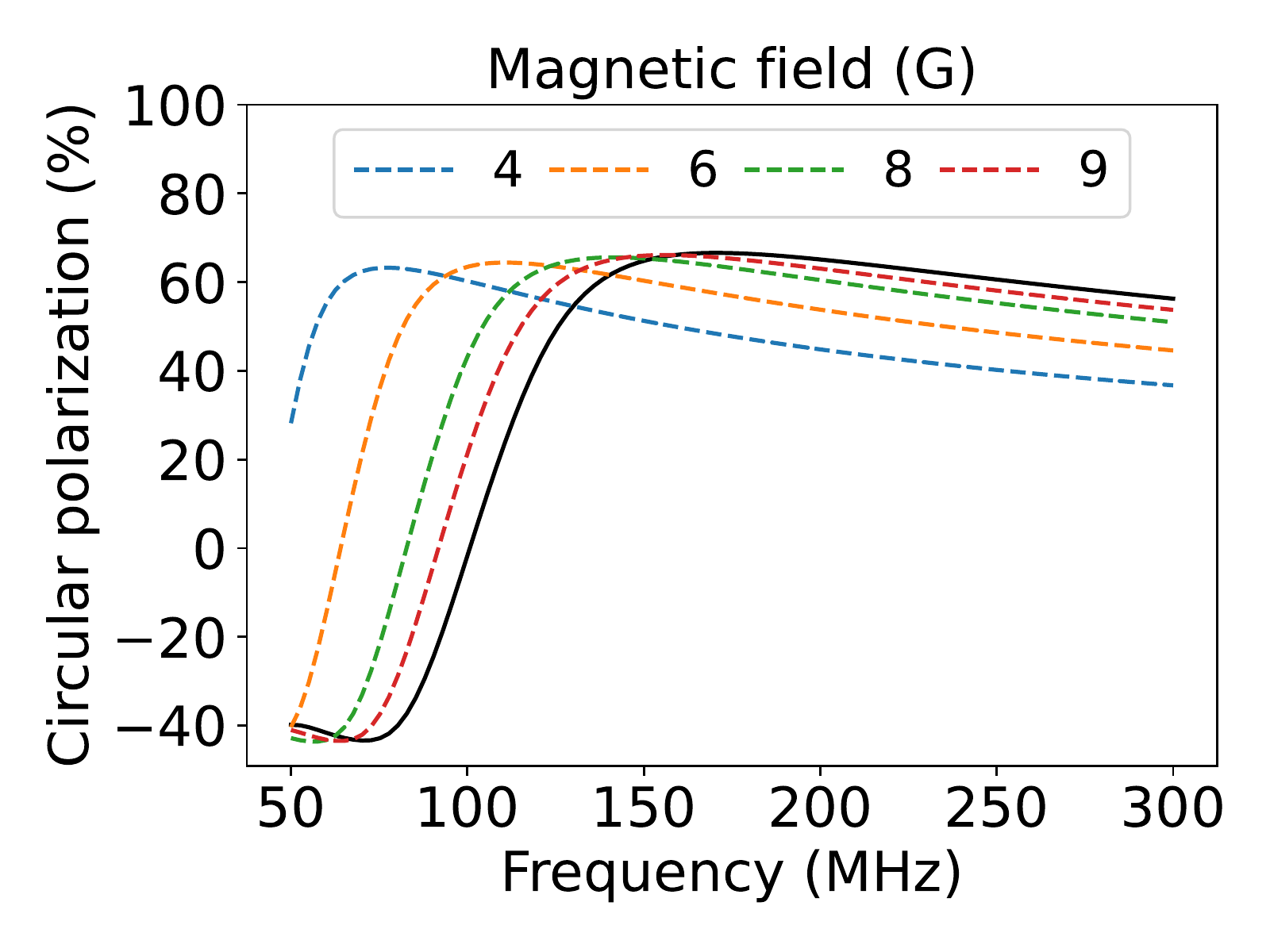}\includegraphics[trim={0.5cm 0.5cm 0.5cm 0.5cm},clip,scale=0.29]{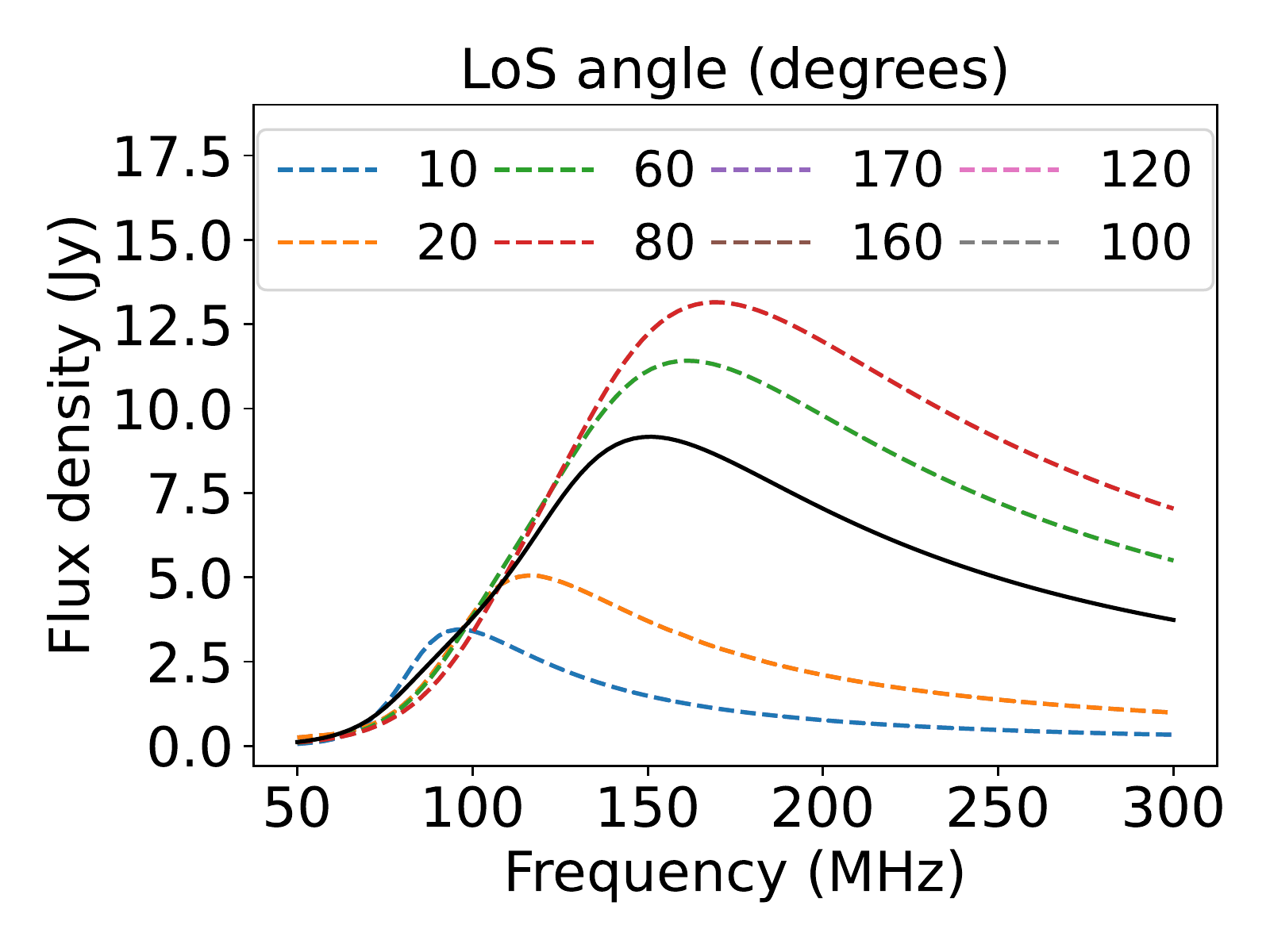}\includegraphics[trim={0.5cm 0.5cm 0.5cm 0.5cm},clip,scale=0.29]{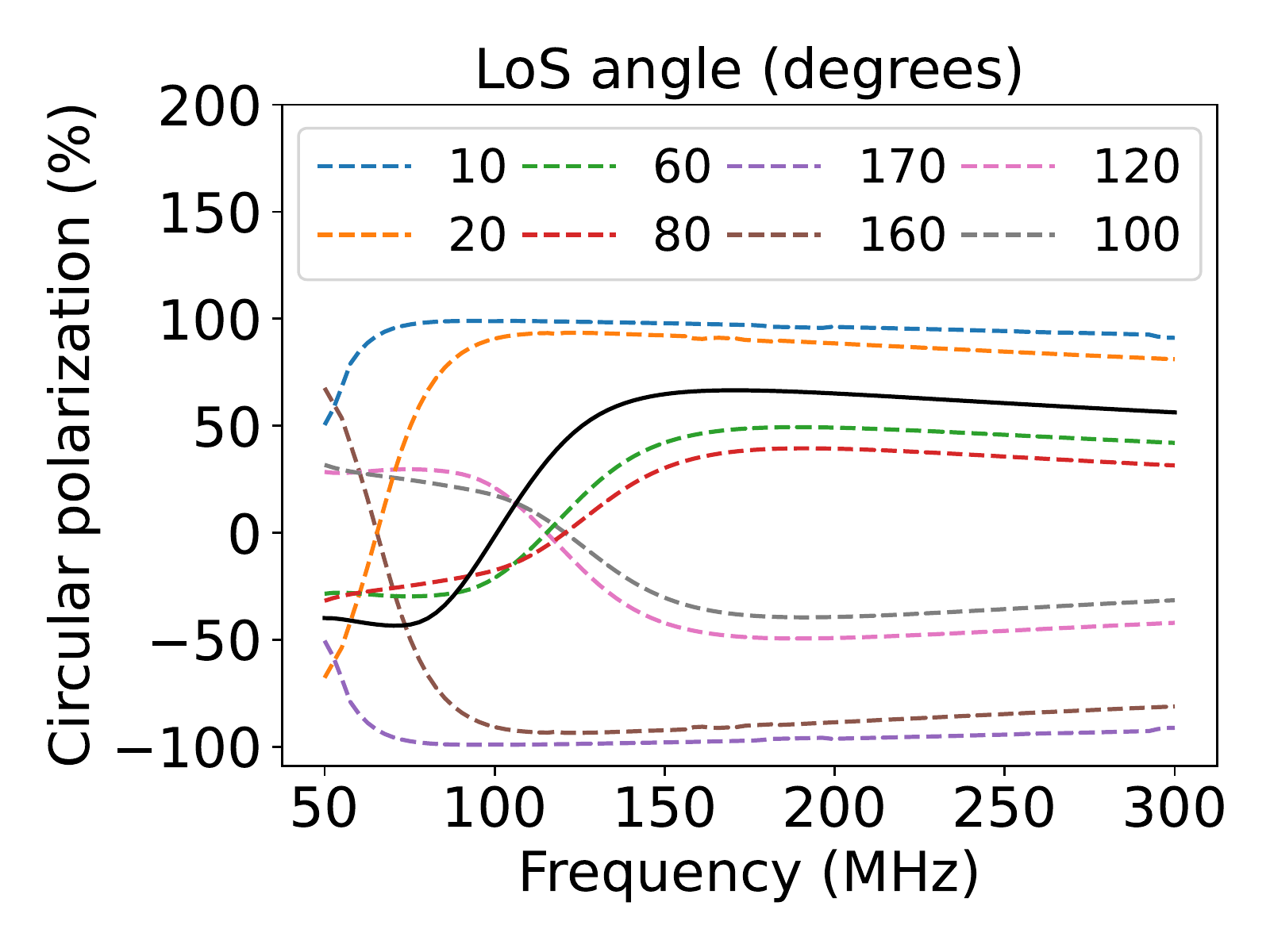}\\
    
   \includegraphics[trim={0.5cm 0.5cm 0.5cm 0.5cm},clip,scale=0.29]{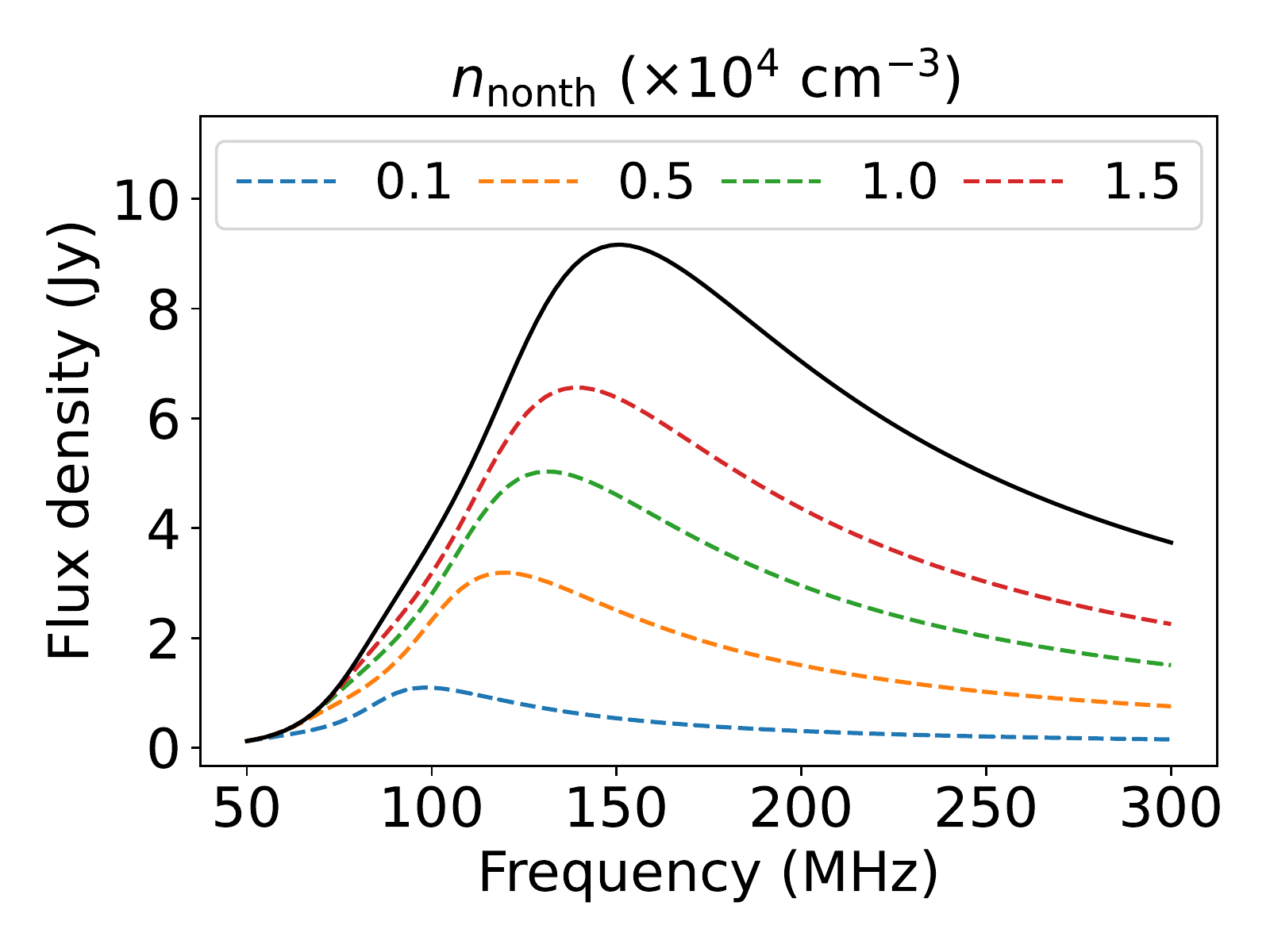}\includegraphics[trim={0.5cm 0.5cm 0.5cm 0.5cm},clip,scale=0.29]{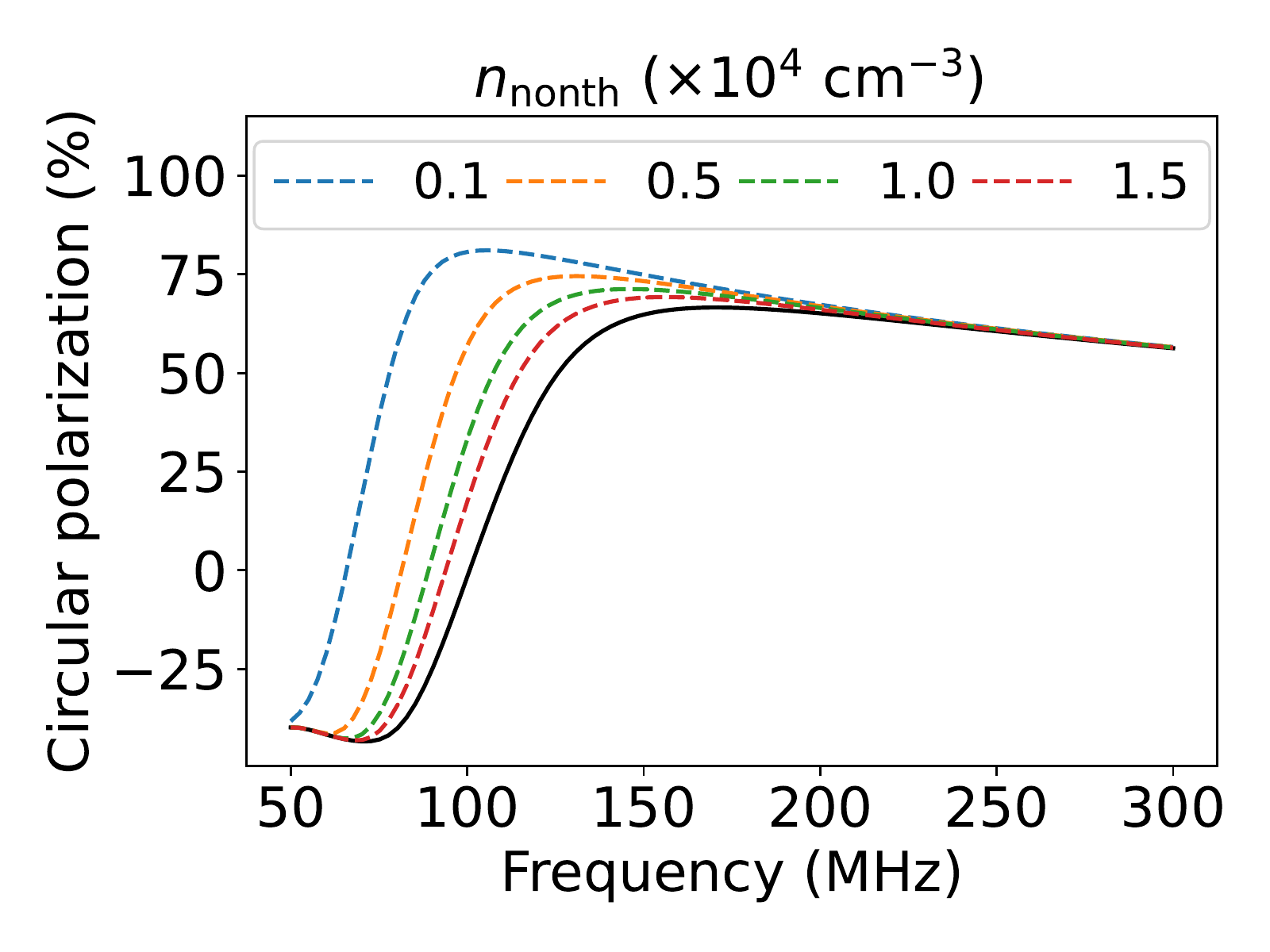}\includegraphics[trim={0.5cm 0.5cm 0.5cm 0.5cm},clip,scale=0.29]{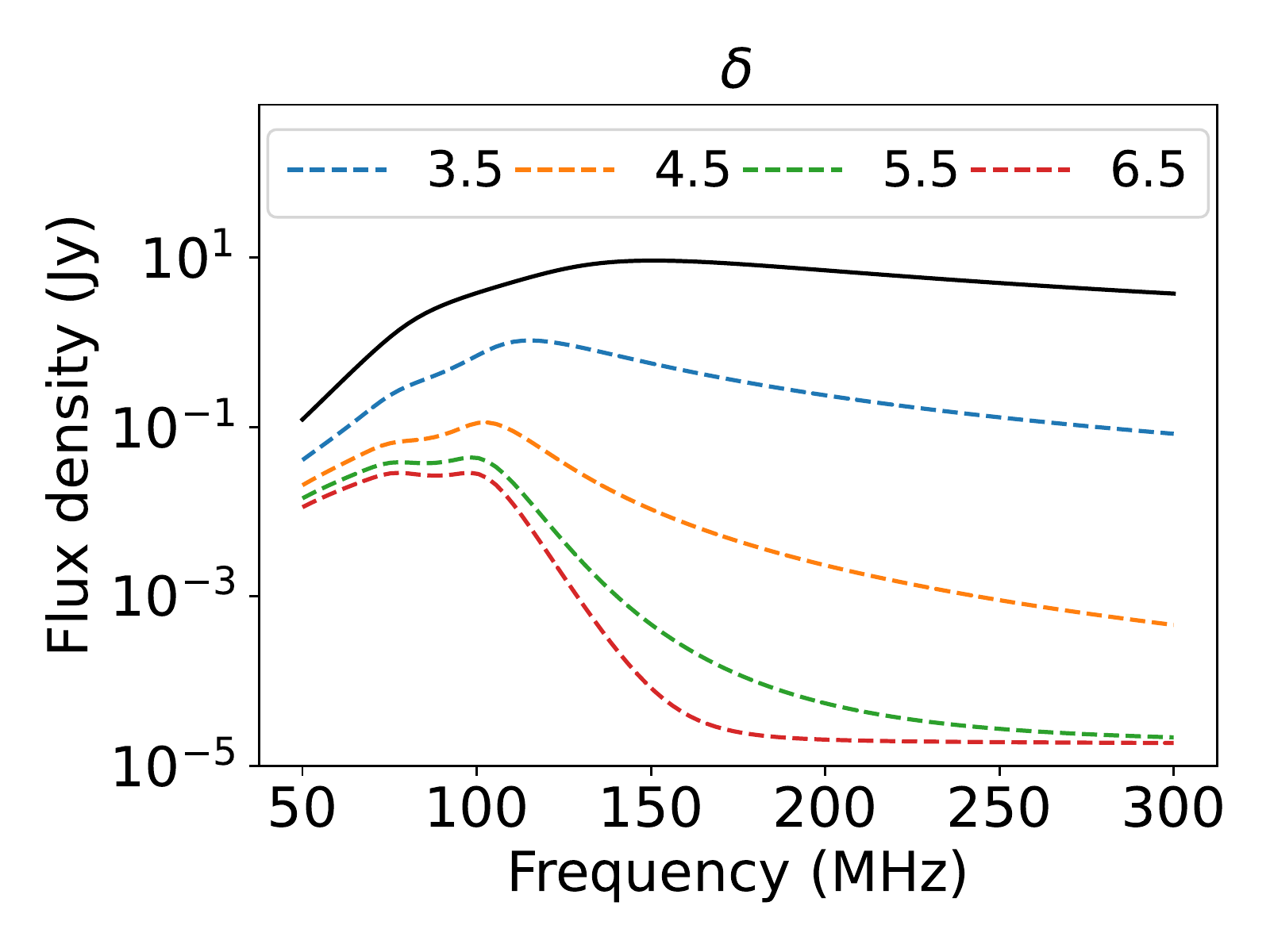}\includegraphics[trim={0.5cm 0.5cm 0.5cm 0.5cm},clip,scale=0.29]{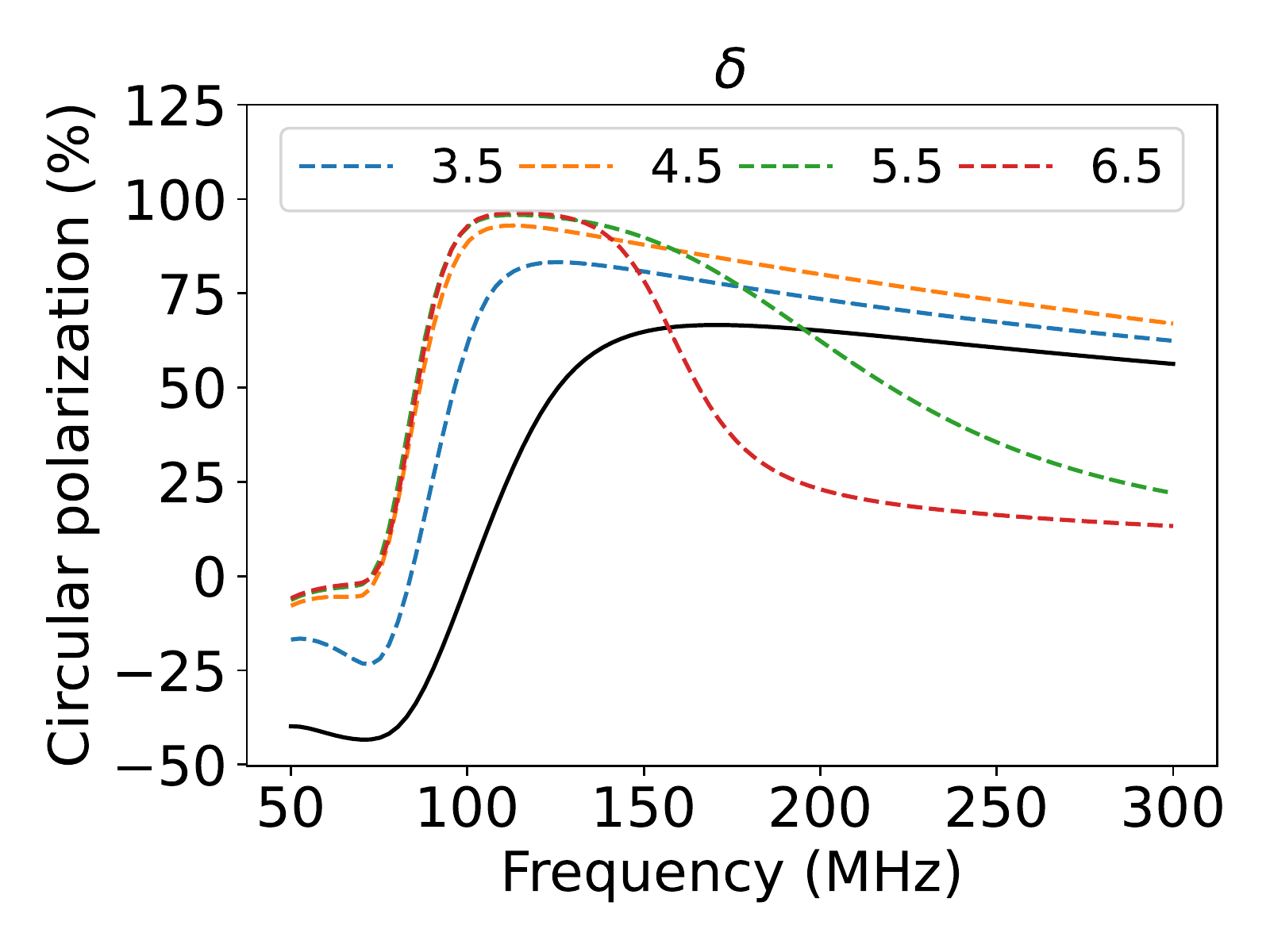}\\
    
   \includegraphics[trim={0.5cm 0.5cm 0.5cm 0.65cm},clip,scale=0.29]{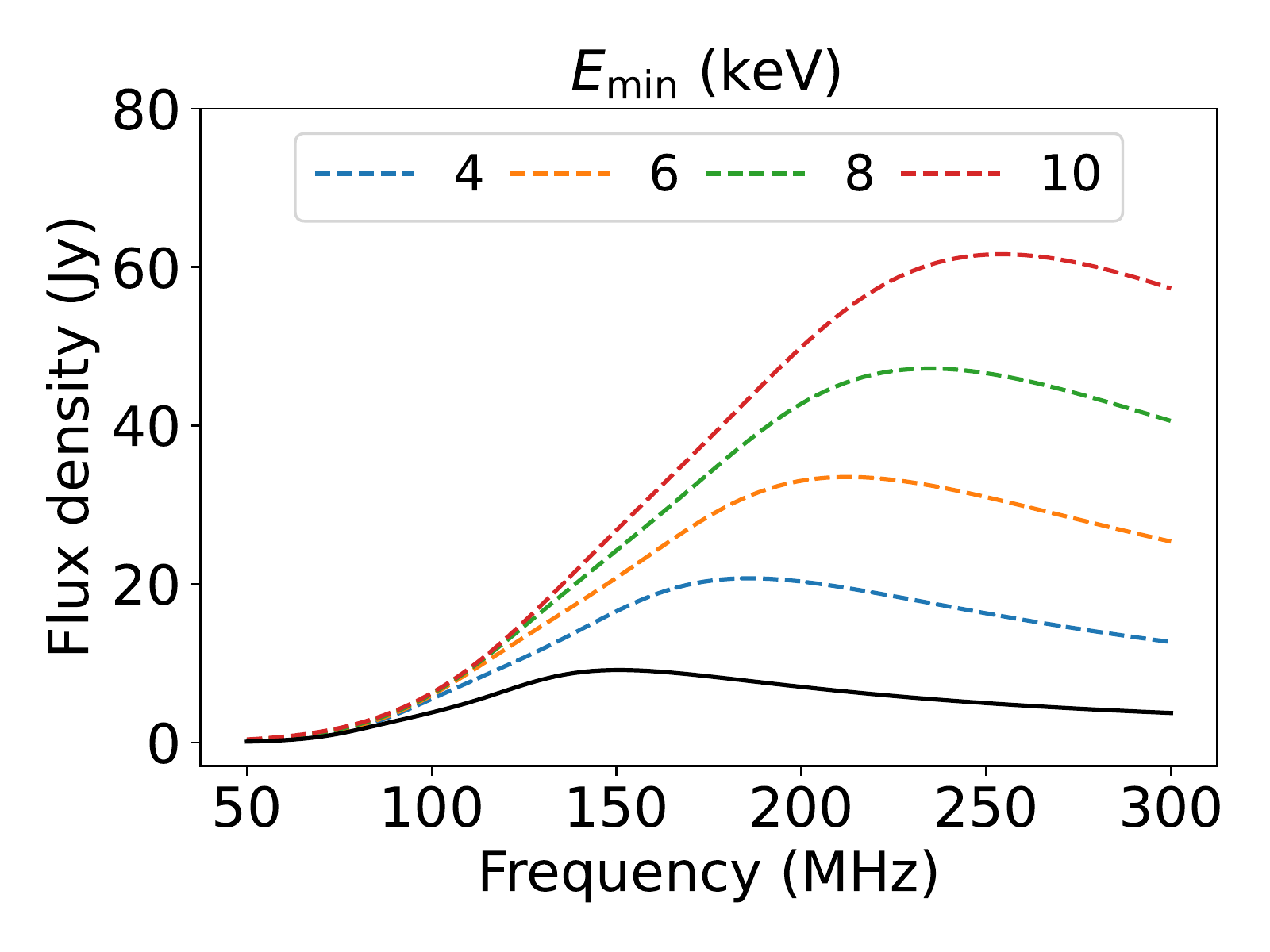}\includegraphics[trim={0.5cm 0.5cm 0.5cm 0.6cm},clip,scale=0.29]{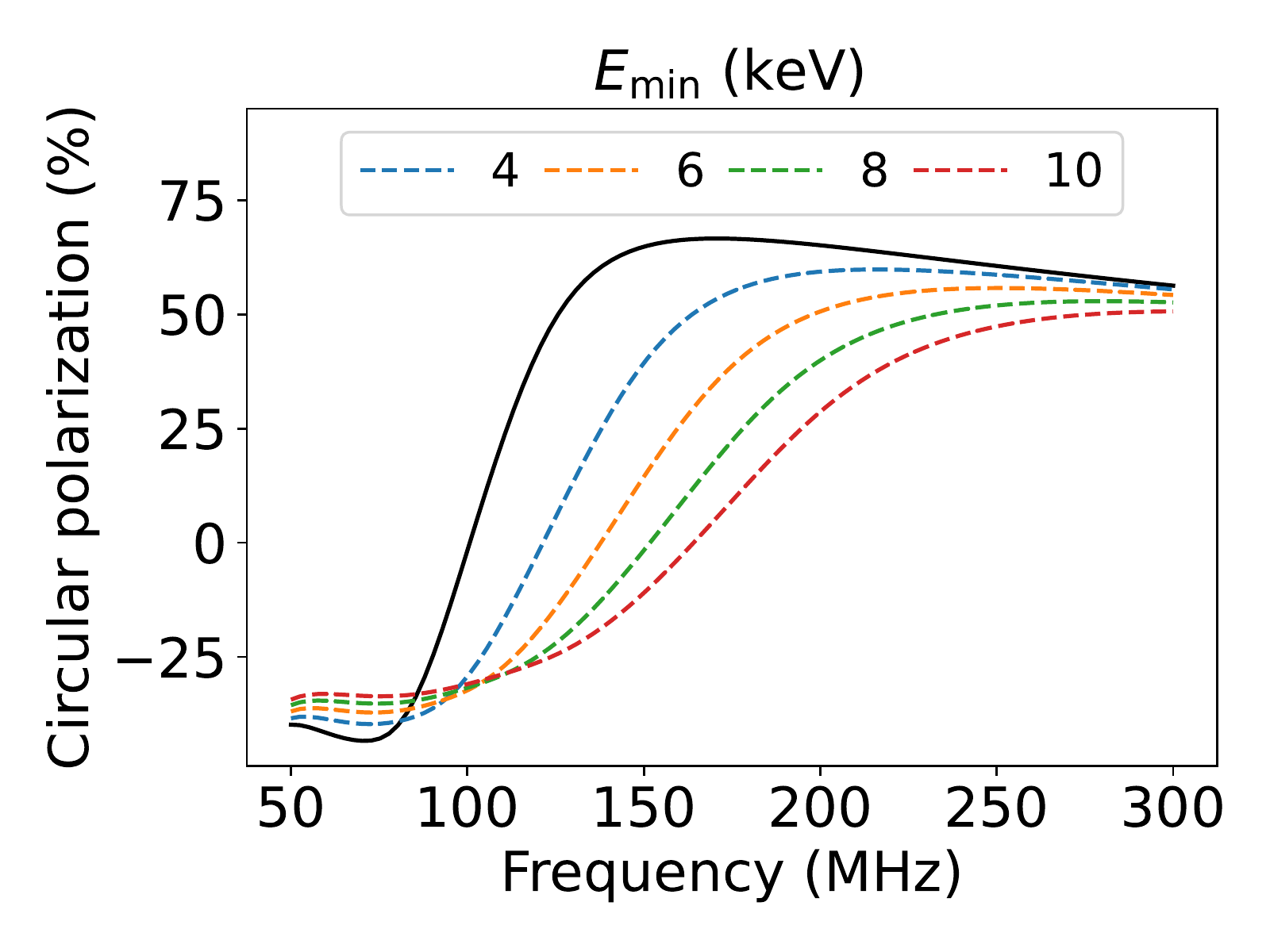} \includegraphics[trim={0.5cm 0.5cm 0.5cm 0.6cm},clip,scale=0.285]{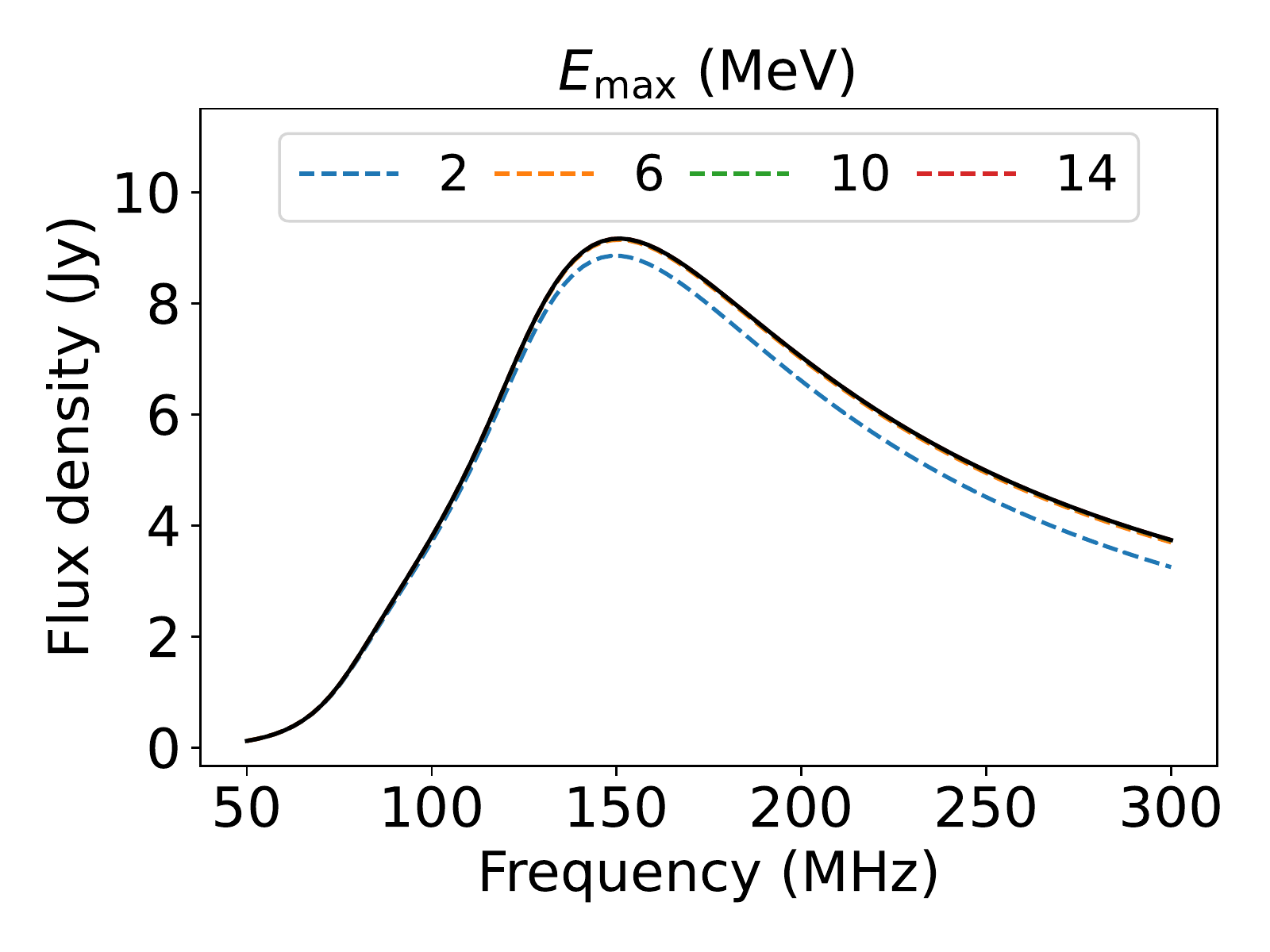}\includegraphics[trim={0.5cm 0.5cm 0.5cm 0.6cm},clip,scale=0.285]{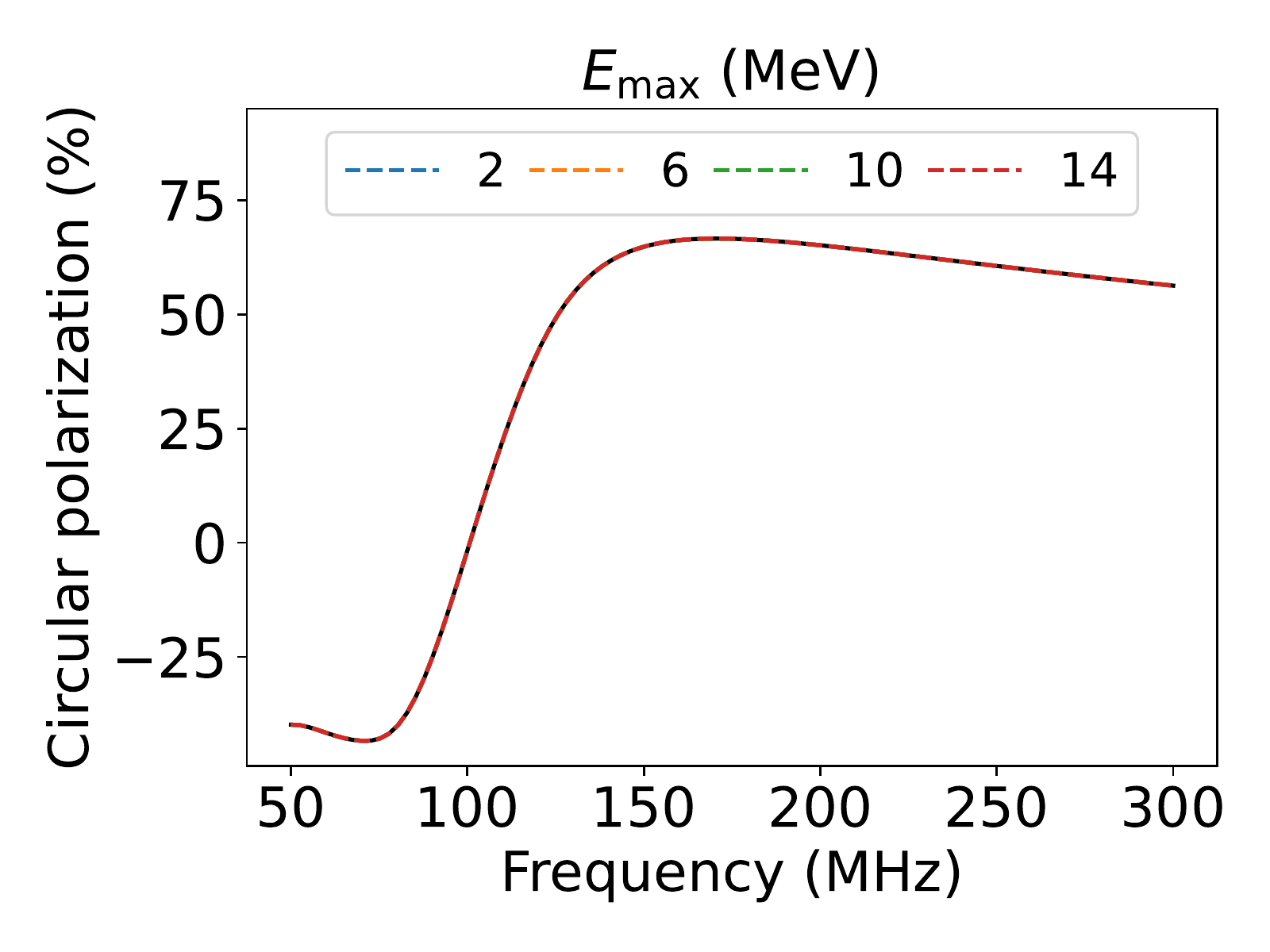}\\
   
    \includegraphics[trim={0.5cm 0.5cm 0.5cm 0.5cm},clip,scale=0.29]{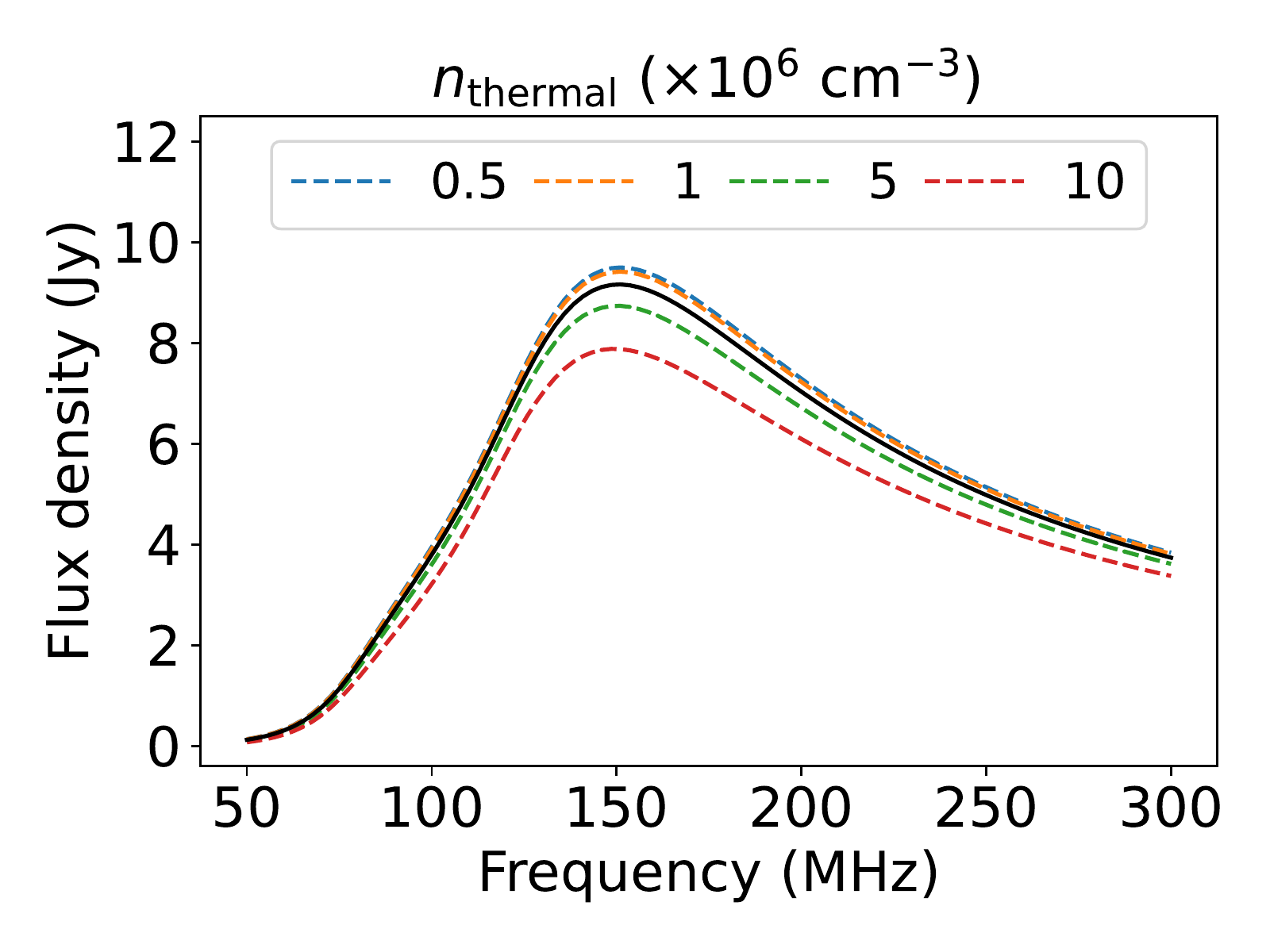}\includegraphics[trim={0.5cm 0.5cm 0.5cm 0.5cm},clip,scale=0.29]{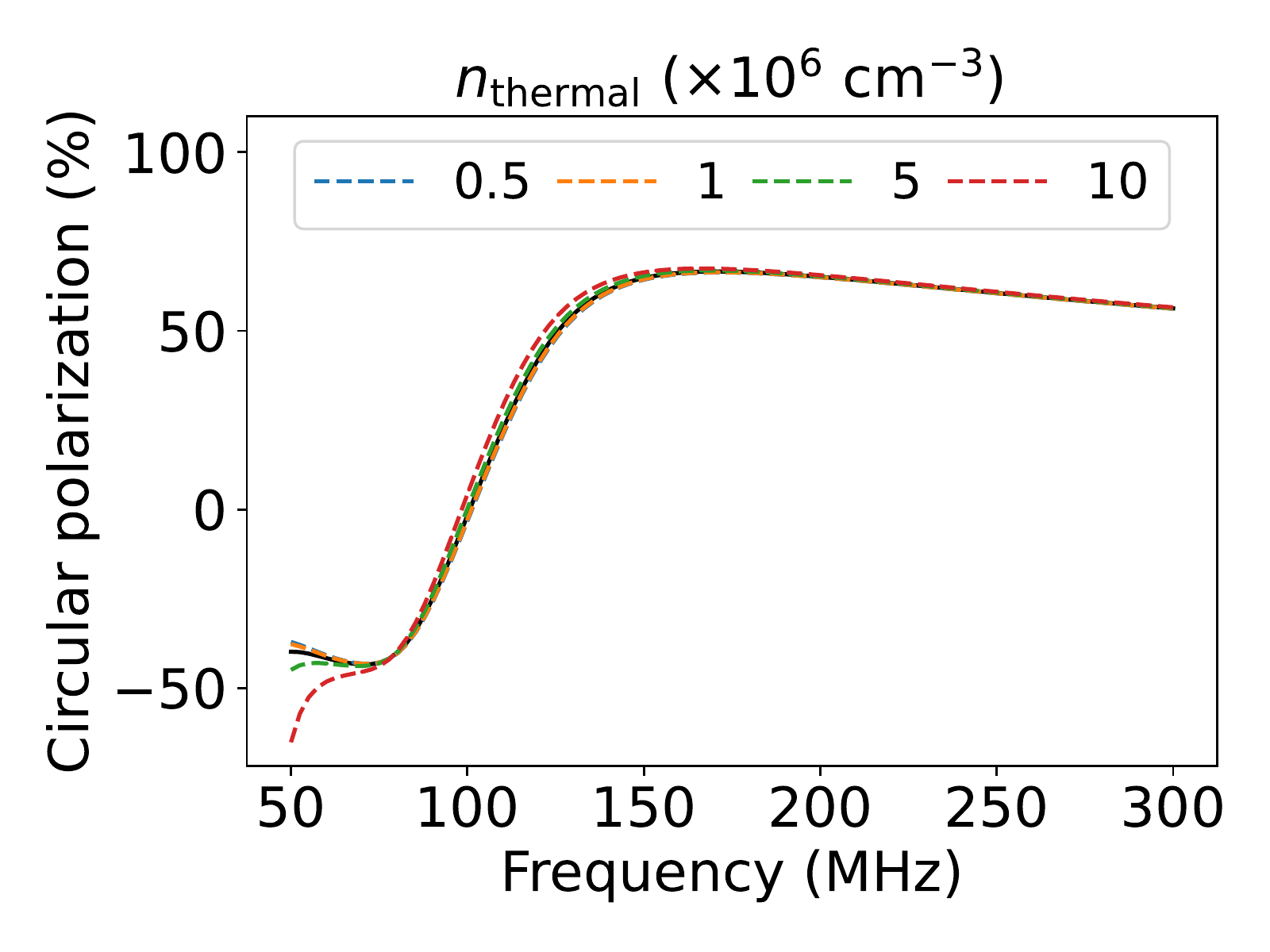}\includegraphics[trim={0.5cm 0.5cm 0.5cm 0.5cm},clip,scale=0.29]{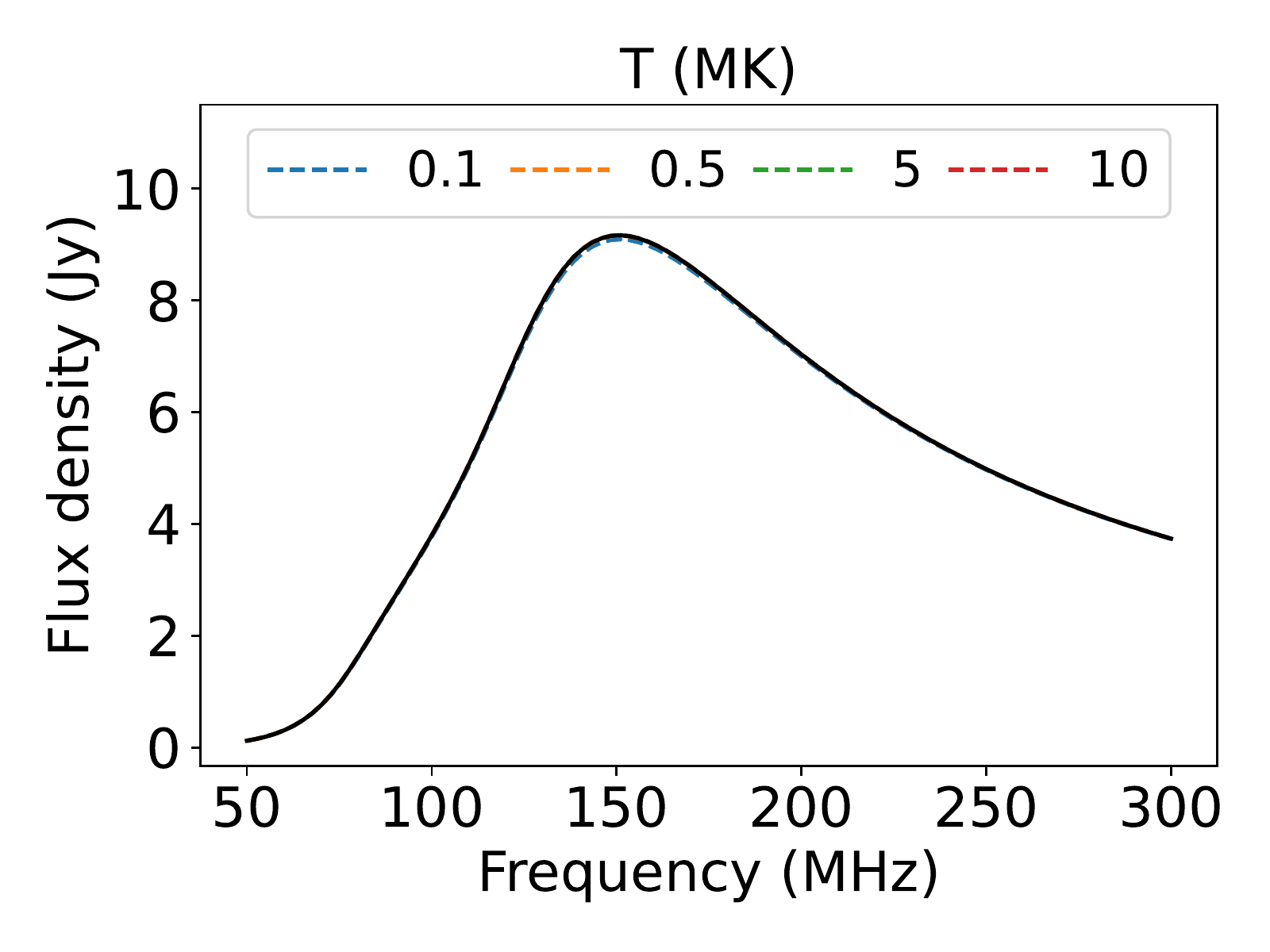}\includegraphics[trim={0.5cm 0.5cm 0.5cm 0.5cm},clip,scale=0.29]{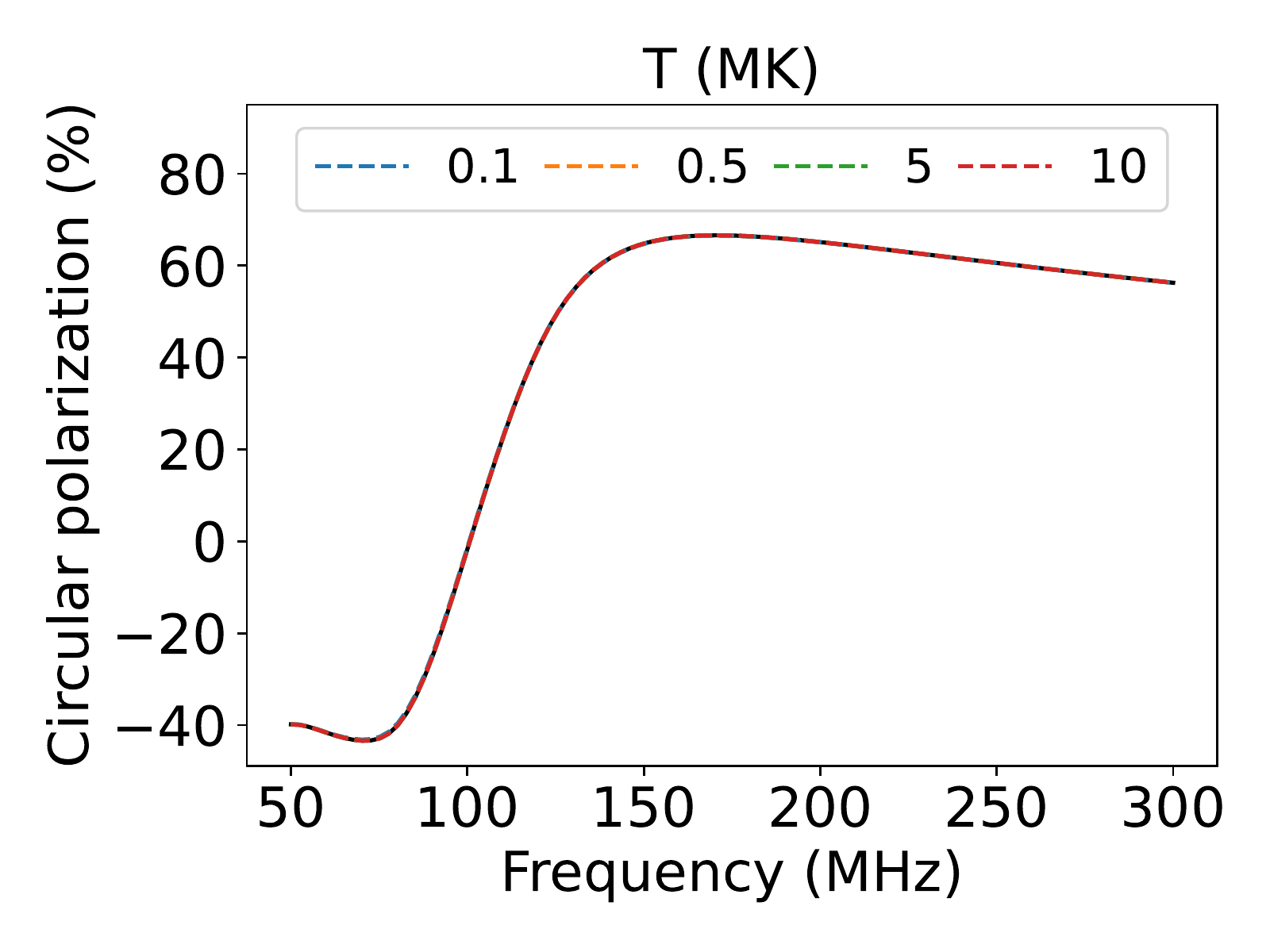}
    \caption{\textbf{Sensitivity of the Stokes I and fractional Stokes V 
    spectra on various GS model parameters.} First and third columns show the Stokes I spectra, and second and fourth columns show spectra for Stokes V fraction for different model parameters. The black solid line in different panels represents the GS spectrum for the reference parameters.}
    \label{fig:param_sensitivity}
\end{figure*}

Even for this simplistic case, the GS model requires ten independent parameters -- magnetic field strength ($B$), angle between the line-of-sight (LoS) and the magnetic field ($\theta$), area of emission ($A$), LoS depth through the GS emitting medium ($L$), temperature ($T$), thermal electron density ($n_{thermal}$), non-thermal electron density ($n_{nonth}$), power-law index of non-thermal electron distribution ($\delta$),  $E_\mathrm{min}$, and $E_\mathrm{max}$. 
Varying each of these parameters leads to its own specific change in the GS spectra, and the effects on the Stokes I and V spectra can be different.
Given the limited number of spectral measurements usually available, it is not feasible to simultaneously constrain all of these model parameters.
In addition, there are intrinsic degeneracies in the GS model, which limit the ability to independently constrain the parameter values.
This has lead the earlier studies to try to constrain some of these parameters using independent measurements (e.g. estimating the thermal electron density from coronagraph observations) and assume reasonable values for some others (e.g. LoS depth, non-thermal electron density, etc.). 

In order to quantitatively explore the impact of variation of each of these parameters independently we carry out a systematic exploration of the GS model parameters, where one parameter is varied over a reasonable range while all others are held constant. 
The ranges of the parameters are motivated by their values explored and estimated in earlier studies of the GS emission from CME plasma at the meter-wavelengths \citep{bastian2001,Tun2013,Mondal2020a}.
Within this range, we make a fiducial choice of a certain value of each of the parameters as the reference value to make comparison convenient. The chosen reference values are -- i) $B=10\ \mathrm{G}$, ii) $\theta=45^{\circ}$, iii) $a = 10^{20}\ \mathrm{cm^{2}}$, iv) $T= 10^6\ \mathrm{K}$, v) $n_\mathrm{thermal}= 2.5\times10^6\ \mathrm{cm^{-3}}$, vi) $n_\mathrm{nonth}= 2.5\times10^4\ \mathrm{cm^{-3}}$, vii) $\delta= 2.8$, viii) $L= 10^{10}\ \mathrm{cm}$, ix) $E_\mathrm{min}=2\ \mathrm{keV}$ and x) $E_\mathrm{max}= 15\ \mathrm{MeV}$. 

A limited exploration of the phase space of GS parameters has been carried out by earlier studies in context of flare observations at microwave regime \citep{Bastian_2007,Zhou_2006,Wu_2019}.
To the best of our knowledge, in the context of the CME plasma, no such explorations have been done.
This section presents an exhaustive exploration of the impact of variations in the physical parameters of GS model on Stokes I and Stokes V spectra.

\begin{figure*}
\centering
    \includegraphics[trim={1.3cm 2cm 7cm 1cm},clip,scale=0.85]{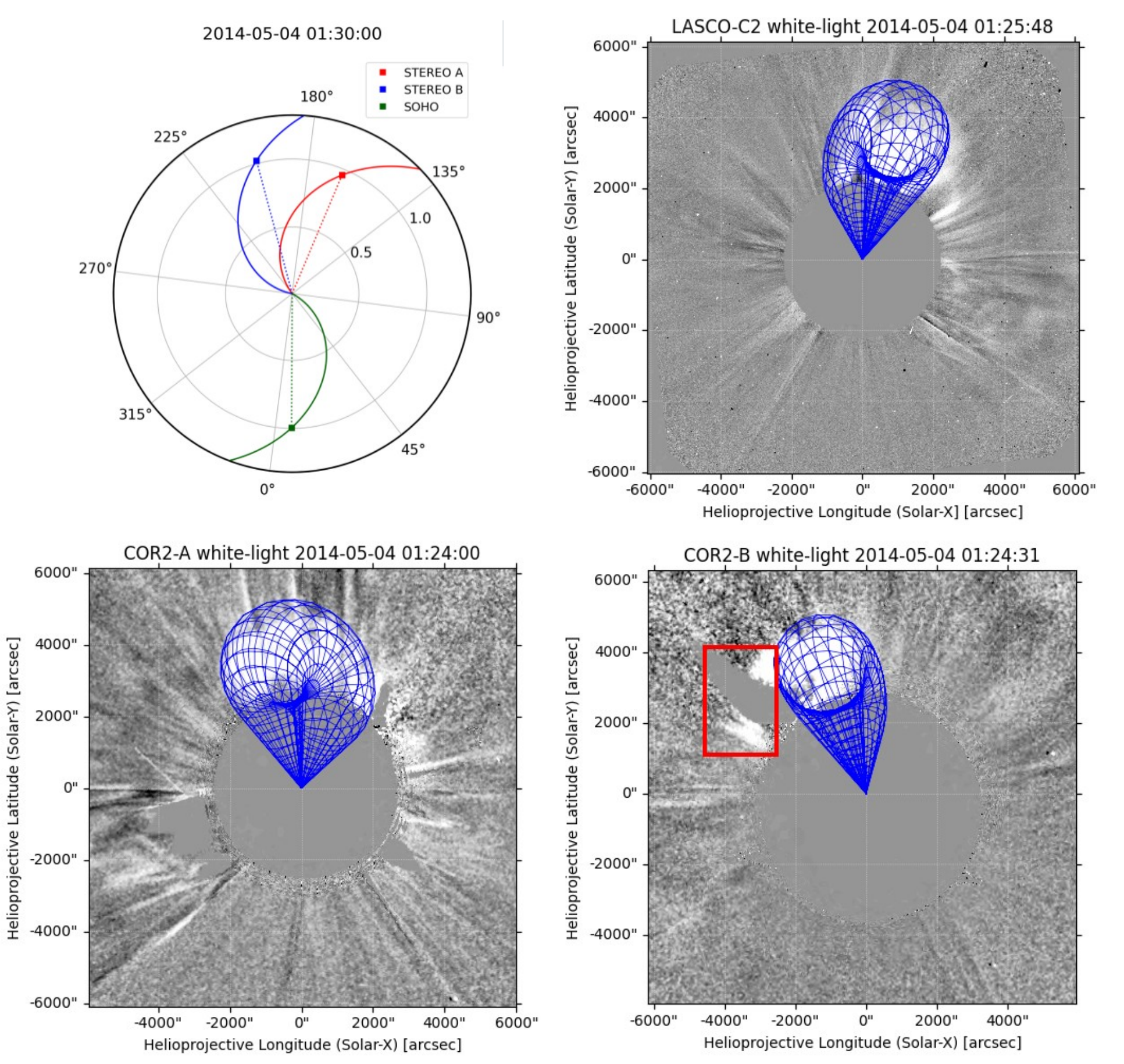}\\
     \caption{\textbf{Three dimensional reconstruction of the CME-1 using Graduated Cylindrical Shell (GCS) model using three vantage point observations. Top left panel: } Position of SOHO, STEREO-A and STEREO-B spacecraft. STEREO-A and STEREO-B were behind the Sun on 2014 May 04. Positions are marked by squares, and the curved lines represent the Parker spiral connected to the each spacecraft. \textbf{Top right panel:} It shows the GCS model of the CME-1 at about 01:25 UTC using the LASCO-C2. \textbf{Bottom panel: }GCS model on COR-2 coronagraph images onboard STEREO-A and STEREO-B spacecraft. A streamer was present at the region marked by red box in the STEREO-B, and hence not considered in the GCS modeling. In LASCO and STEREO-A images the streamer was not bright.}
    \label{fig:gcs}
\end{figure*}

\subsection{Sensitivity of Stokes I Spectra to GS Model Parameters}
\label{subsec:stokesI_sensitivity}
The sensitivity of Stokes I spectra on different GS model parameters are shown in the first and third column of Figure \ref{fig:param_sensitivity}. It is evident from this figure that Stokes I spectra are not sensitive to two of the GS model parameters -- $E_\mathrm{max}$  and $T$. Peak flux density can vary by multiple orders of magnitude as a function of $\delta$, with the peak frequency decreasing with increasing $\delta$. Peak flux density increases with the increase in $A,\ B,\ \theta,\ n_\mathrm{nonth}$ and $E_\mathrm{min}$. On the other hand, peak flux density decreases with the increase in $L$ and $n_\mathrm{thermal}$. Peak frequency is independent of $A$ and $n_\mathrm{thermal}$, while it increases with the increase in $B,\ \theta,\ L,\ n_\mathrm{nonth}$ and $E_\mathrm{min}$. The nature of these variations in the Stokes I spectra imply that there exist degeneracies between values of $B,\ \theta,\  L$, $n_\mathrm{nonth}$ and $E_\mathrm{min}$, in the GS model parameters.  

\subsection{Sensitivity of Stokes V Spectra to GS Model Parameters}
\label{subsec:stokesV_sensitivity}
Sensitivities of Stokes V spectra on different GS model parameters are shown in the second and fourth columns of Figure \ref{fig:param_sensitivity}. Some of the GS parameters -- $A$, $n_\mathrm{thermal}$, $E_\mathrm{max}$ and $T$, do not have any noticeable effect on the Stokes V spectra. $B$, $\theta$ and $\delta$ show significant impacts on both the optically thin and thick parts of the Stokes V spectra. $\delta$ has strong impacts on determining the spectral shape of the Stokes V spectra. The polarization fraction in optically thin part increases with the increase in $B$, while it decreases with the increase in $\theta$. $L,\ E_\mathrm{min}$ and $n_\mathrm{nonth}$ impact only on the optically thick part of the Stokes V spectra and fractional polarization decreases with the increase in each of these parameters. 

\subsection{Resolving the Degeneracy of GS Model Parameters using Stokes V Spectra}\label{subsec:stokes_V_break_degeneracy}
Different natures of impacts of $B$ and $\theta$ on the optically thin part of the Stokes V spectra breaks the degeneracy between them observed in Stokes I spectra. The sign of the circular polarization depends on the whether $\theta$ value is less than or greater than 90$^{\circ}$ degrees. For both $\theta$ and $180^{\circ}-\theta$, the Stokes I spectra are similar, but the Stokes V spectra are inverted.
$L,\ E_\mathrm{min}$ and $n_\mathrm{nonth}$ show similar effects of both Stokes I and V spectra (Figure \ref{fig:param_sensitivity}). But the availability of multi-vantage point observations allowed us to provide a strong upper limit on the $L$, while no such direct observational constraints are available for $E_\mathrm{min}$ and $n_\mathrm{nonth}$. Hence, geometrical constraints of $L$ allowed us to break the degeneracy. However, the degeneracy between $n_\mathrm{nonth}$ and $E_\mathrm{min}$ can not be resolved even when using both the Stokes I and Stokes V spectra.

\begin{figure}
\centering
    \includegraphics[trim={5cm 3cm 5cm 0.8cm},clip,scale=0.42]{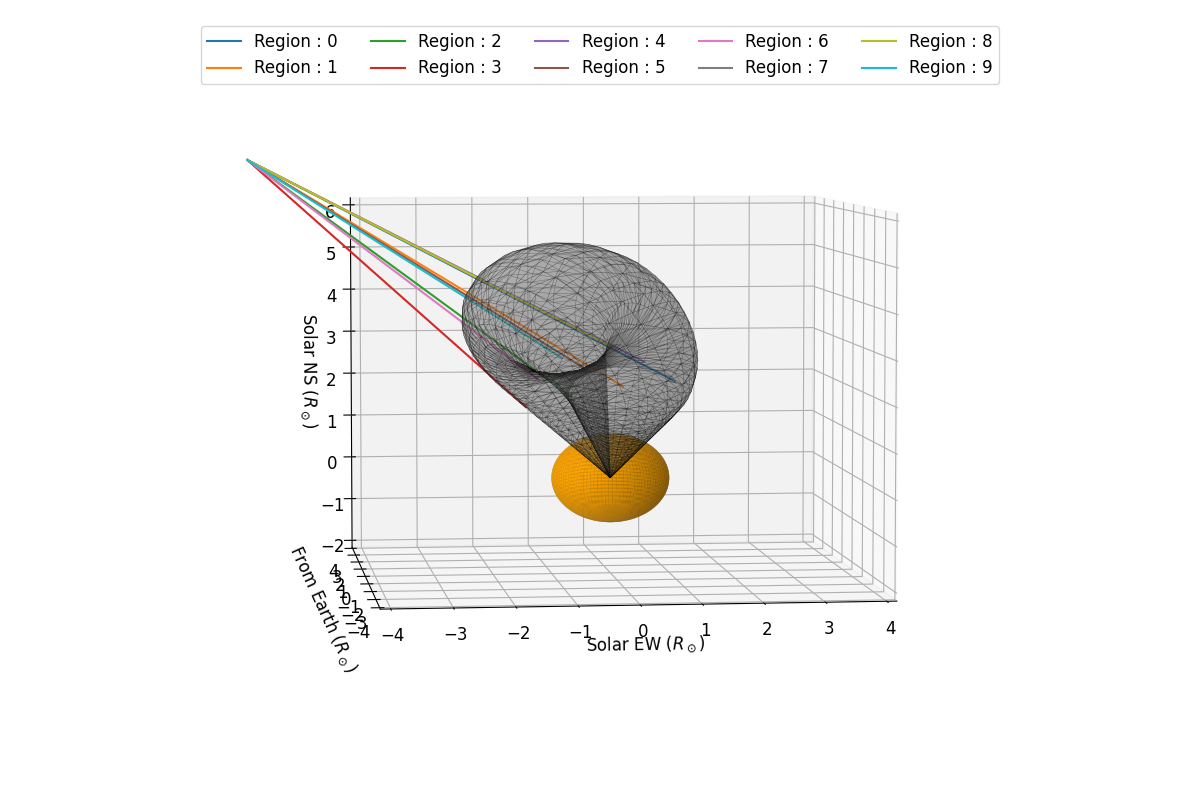}
    \caption{\textbf{Ray-tracing through the Graduated Cylindrical Shell (GCS) model flux rope for different PSF-sized regions.} Different rays originated from the Earth at 214 $R_\odot$ is traced through the GCS flux rope to find out the geometrical LoS depth of a certain PSF-sized region. Orange sphere represent the Sun, and GCS flux rope is shown by grey mesh. Rays are shown by colored lines.}
    \label{fig:gcs_ray}
\end{figure}

\begin{figure*}[!htp]
    \centering
     \includegraphics[trim={0.3cm 0.5cm 0.0cm 0.3cm},clip,scale=0.37]{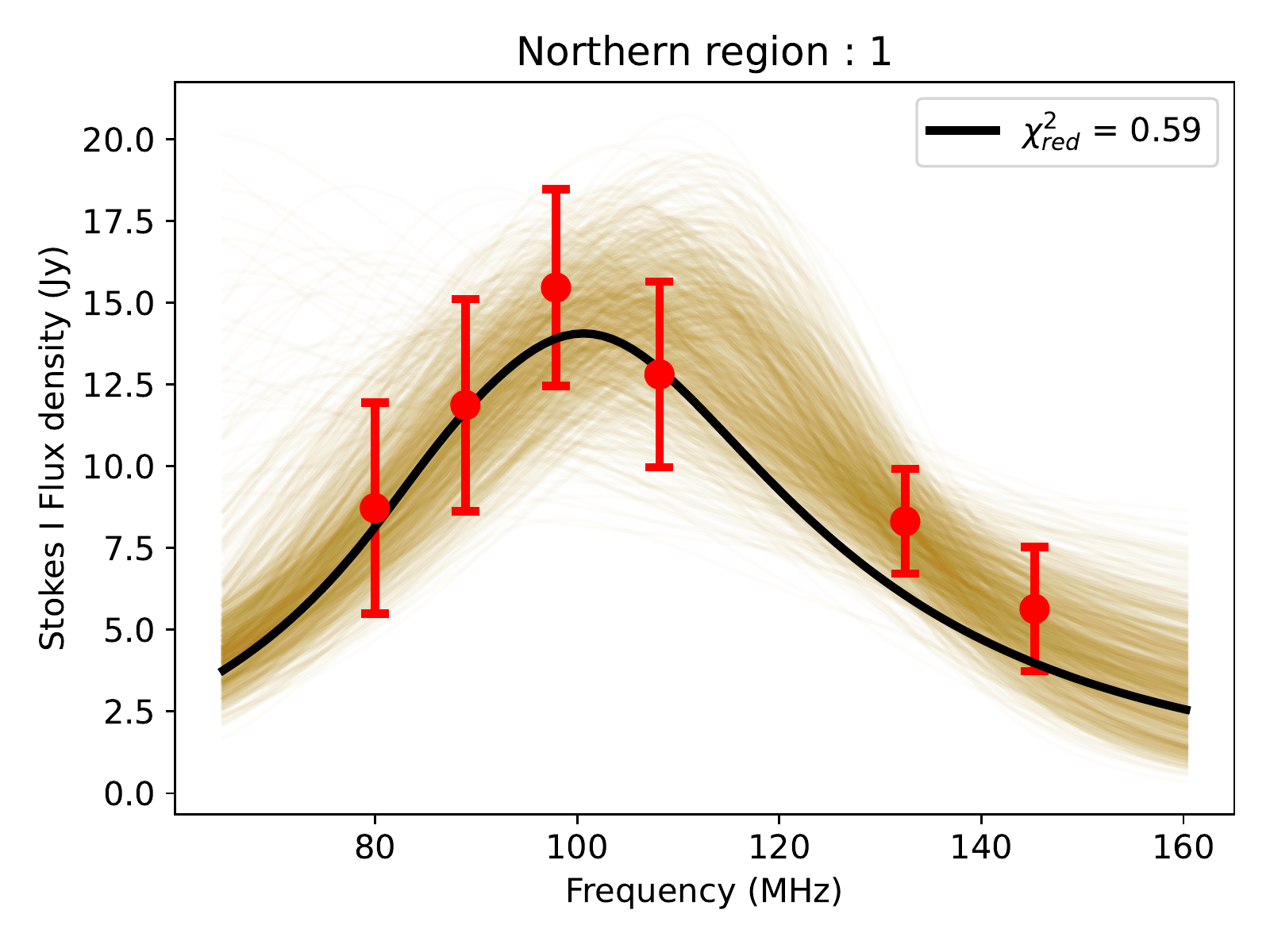}\includegraphics[trim={0.3cm 0.5cm 0.0cm 0.3cm},clip,scale=0.37]{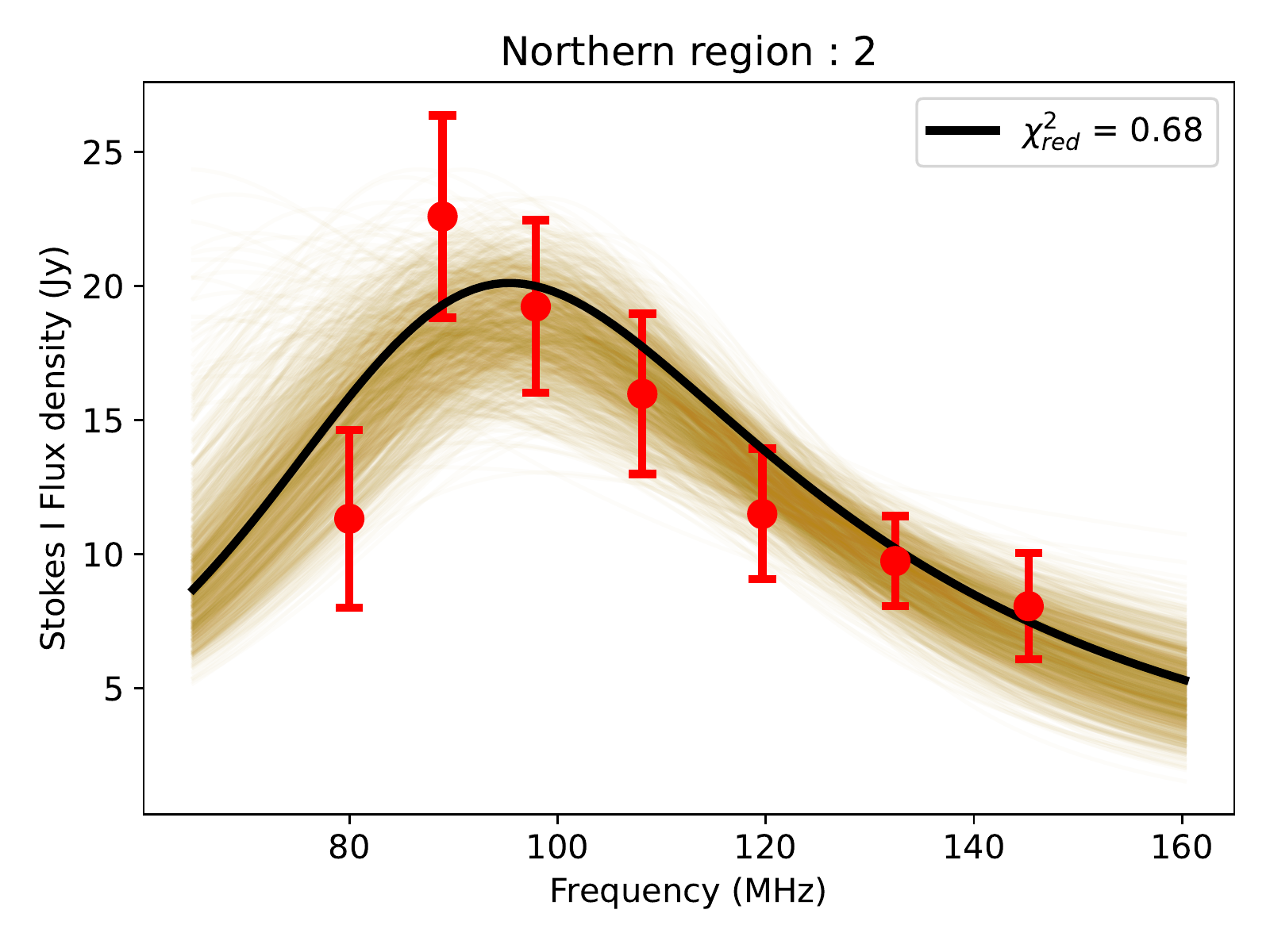}\includegraphics[trim={0.3cm 0.5cm 0.0cm 0.3cm},clip,scale=0.37]{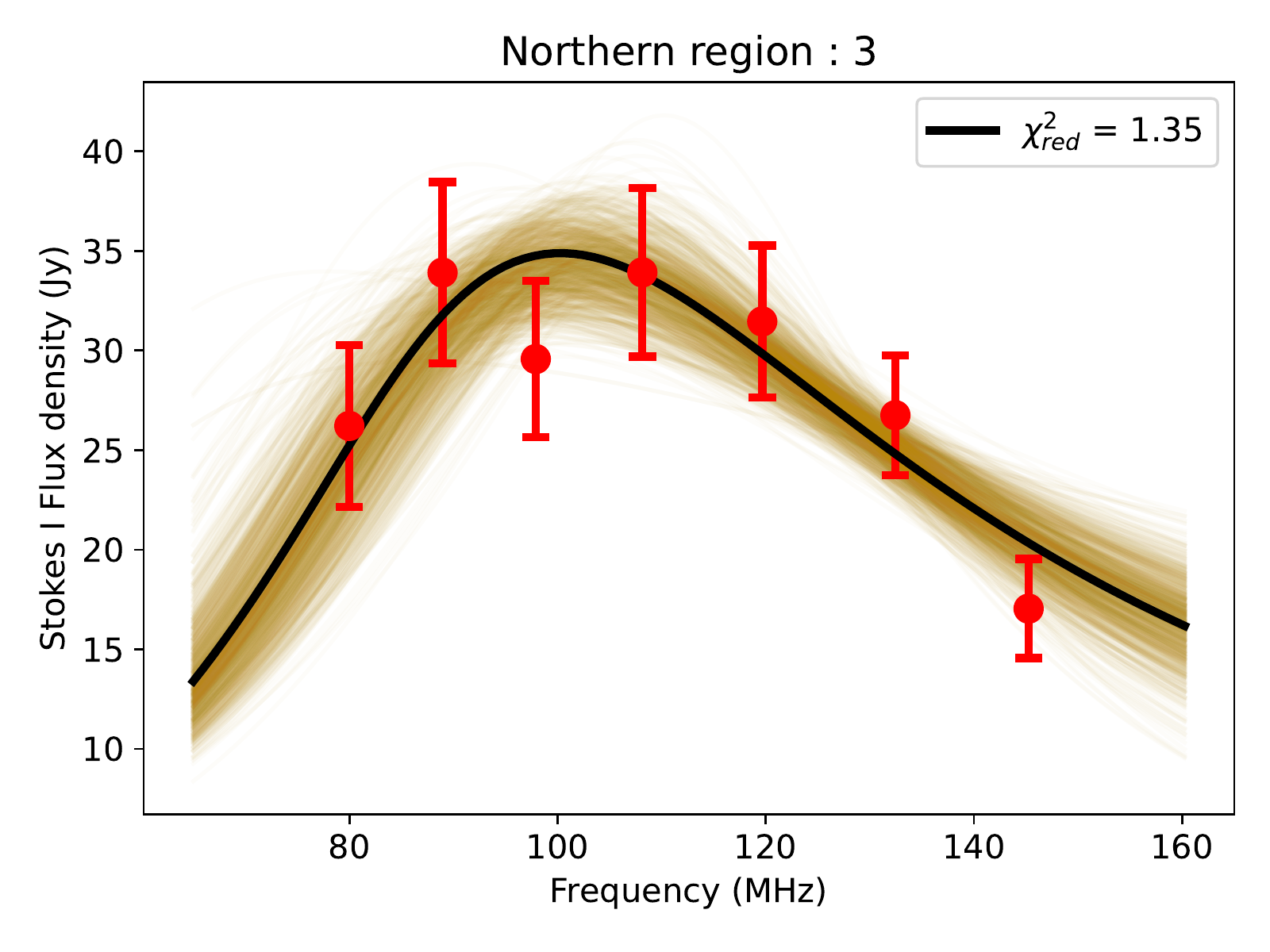}
     \\
    \includegraphics[trim={0.3cm 0.5cm 0.0cm 0.3cm},clip,scale=0.37]{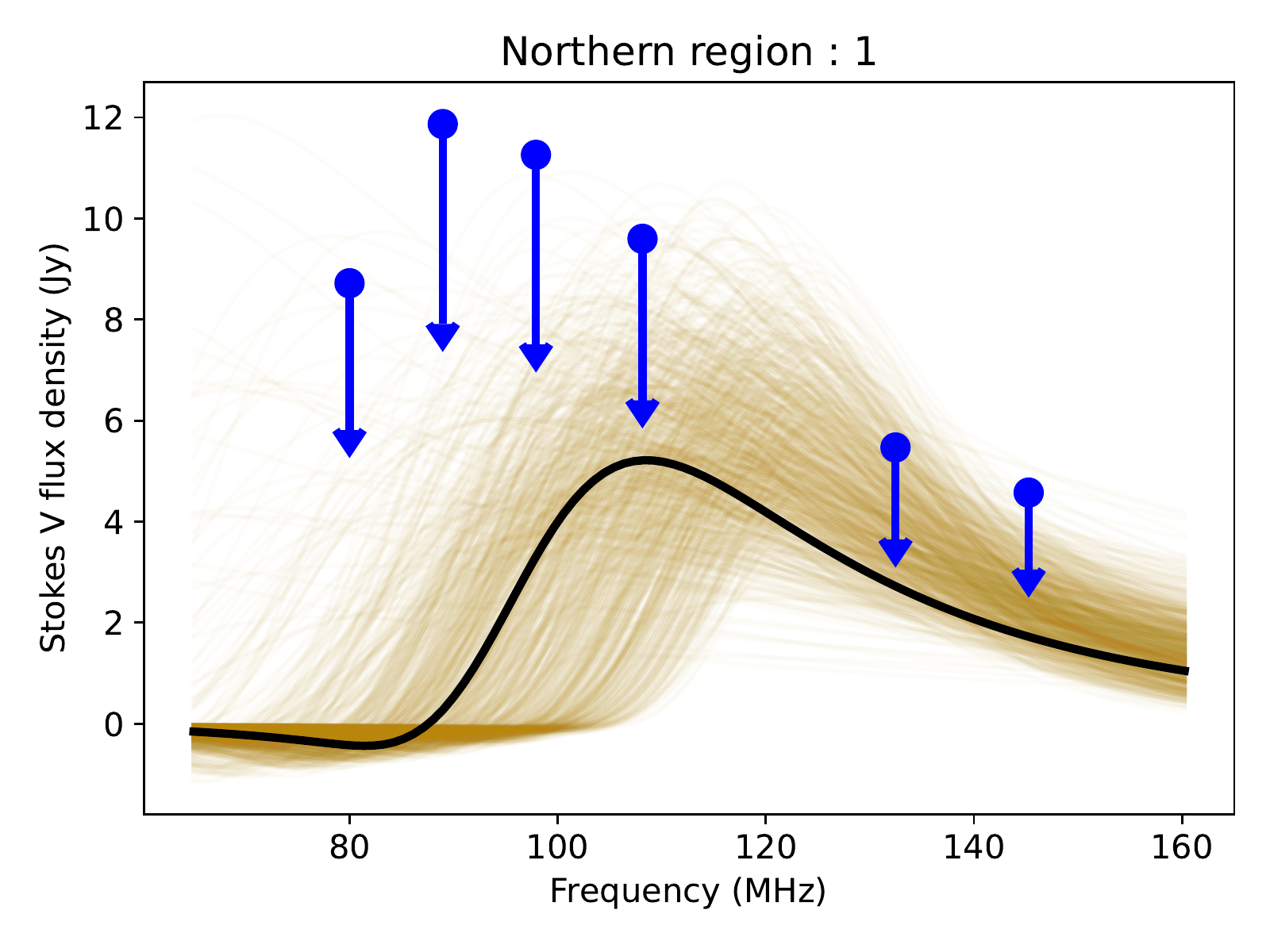}\includegraphics[trim={0.3cm 0.5cm 0.0cm 0.3cm},clip,scale=0.37]{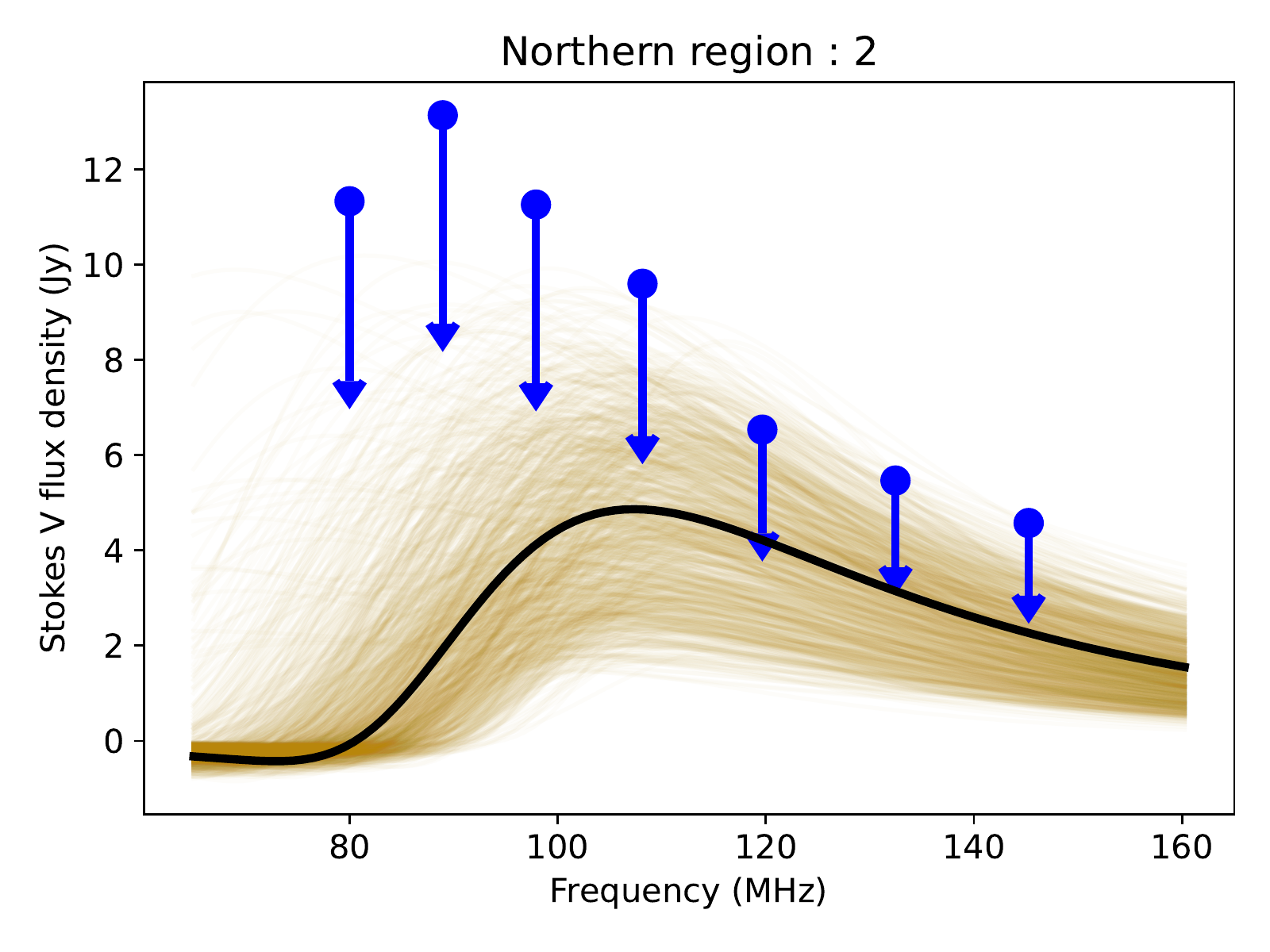}\includegraphics[trim={0.3cm 0.5cm 0.0cm 0.3cm},clip,scale=0.37]{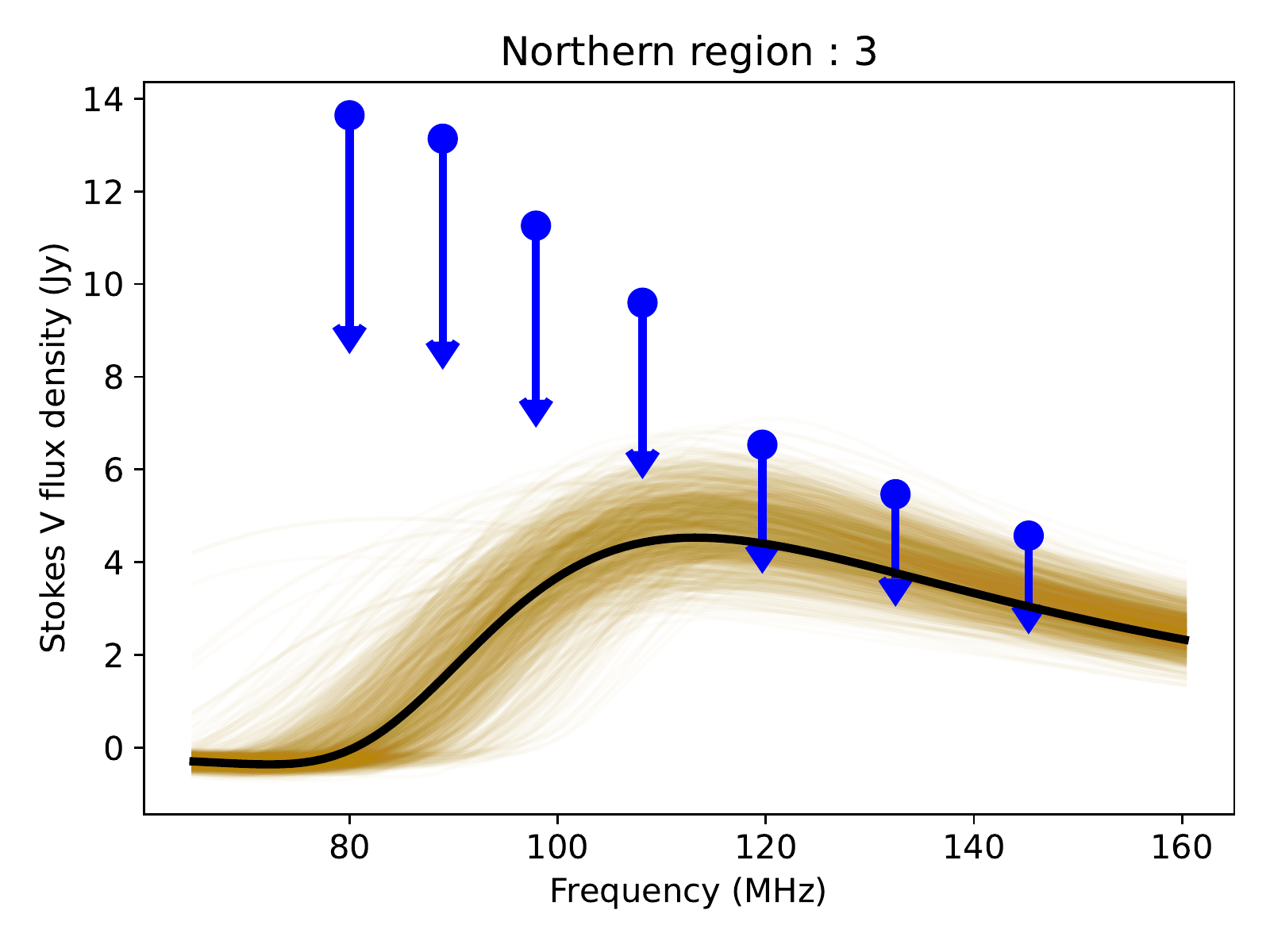}\\
    
   \includegraphics[trim={0.3cm 0.5cm 0.0cm 0.3cm},clip,scale=0.37]{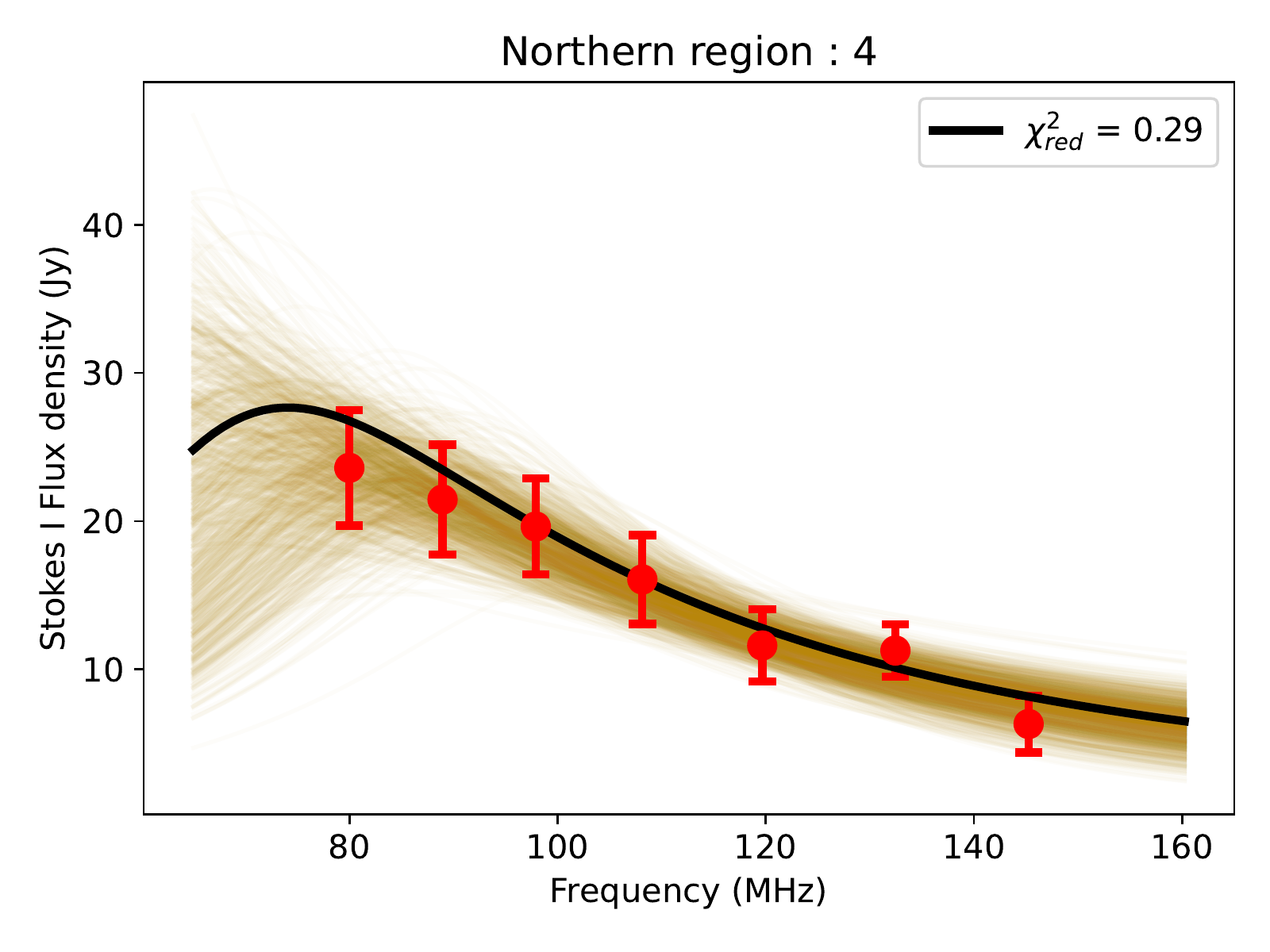}\includegraphics[trim={0.3cm 0.5cm 0.0cm 0.3cm},clip,scale=0.37]{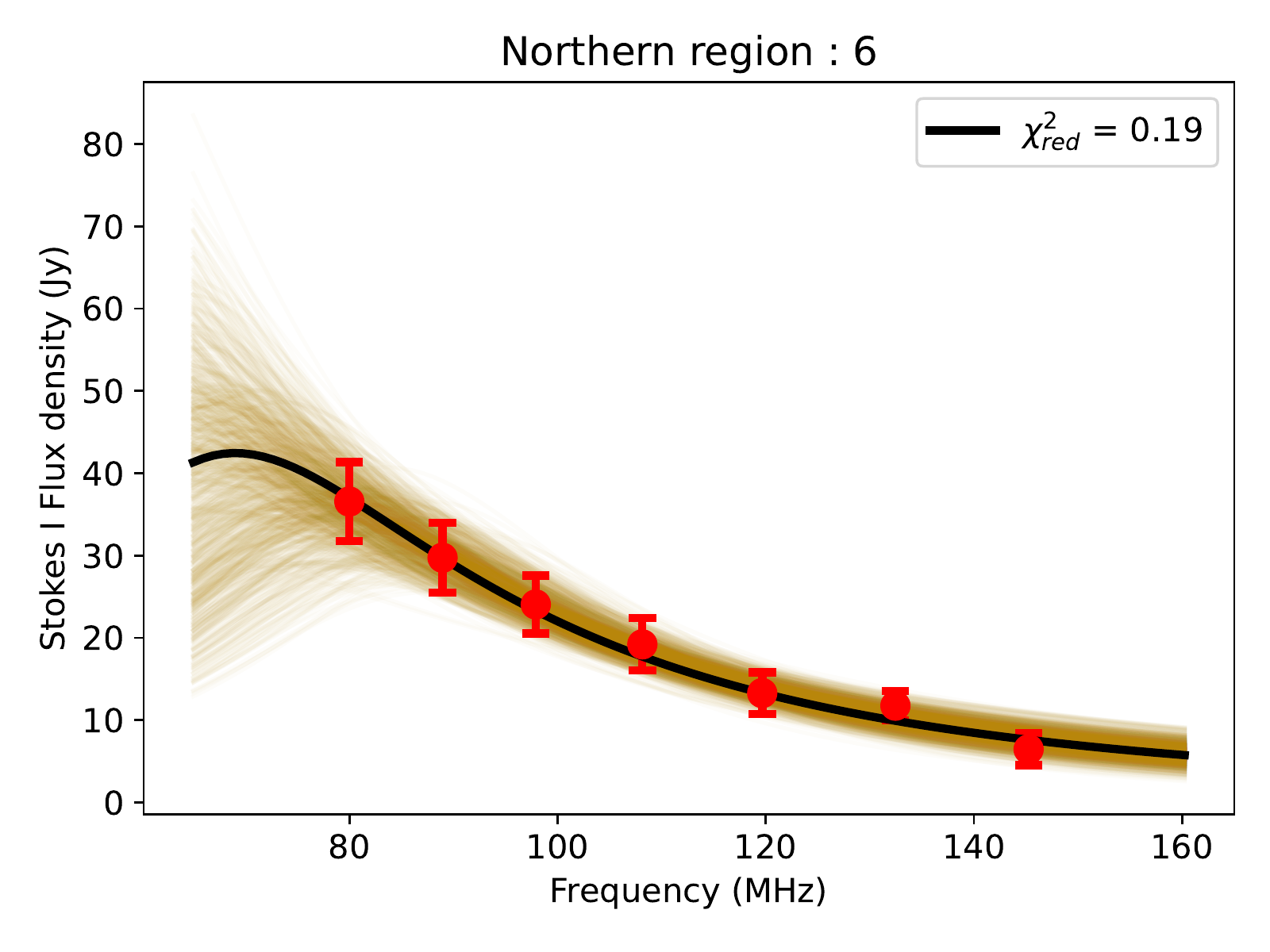}\includegraphics[trim={0.3cm 0.5cm 0.0cm 0.3cm},clip,scale=0.37]{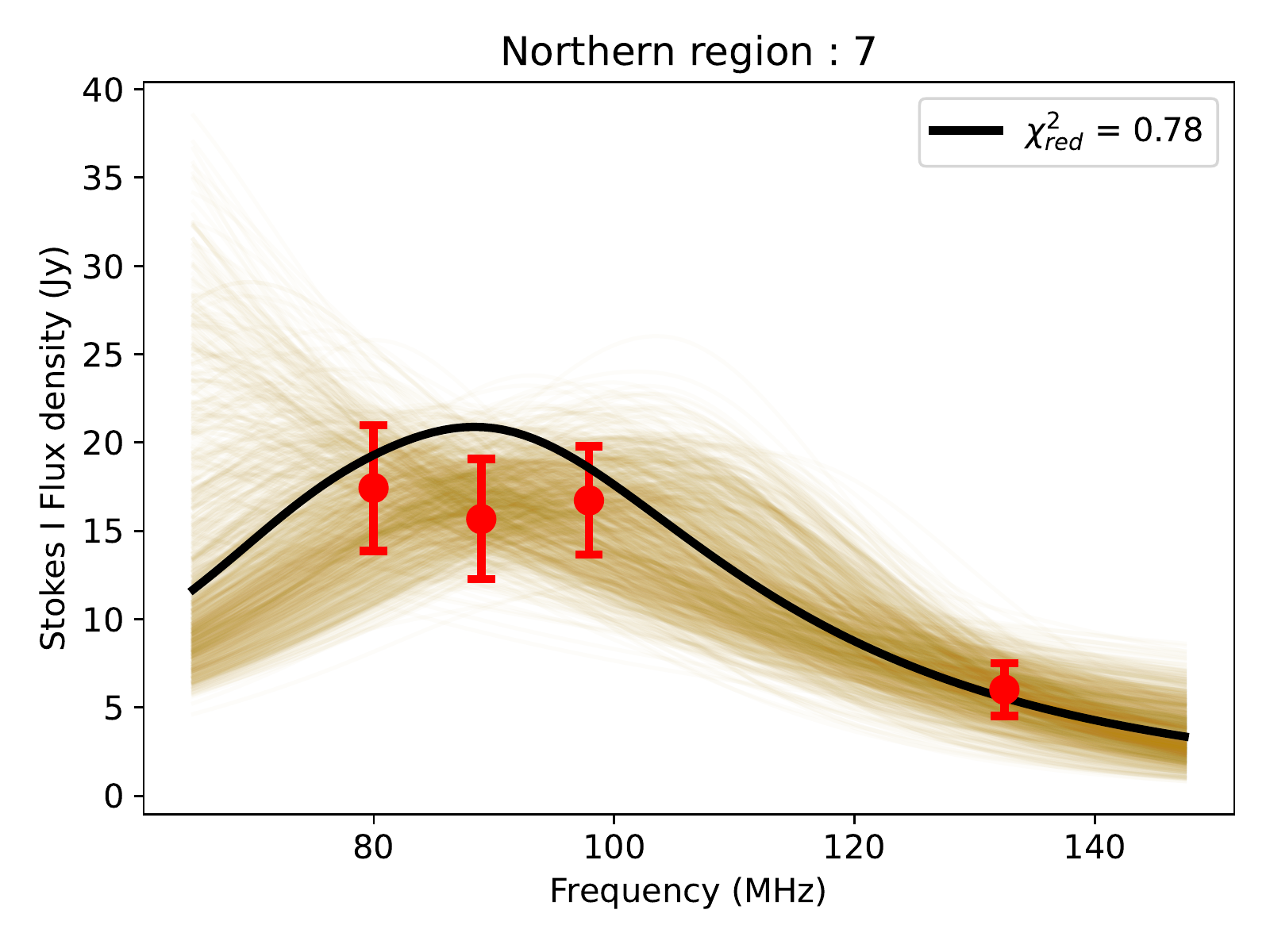}
     \\
    \includegraphics[trim={0.3cm 0.5cm 0.0cm 0.3cm},clip,scale=0.37]{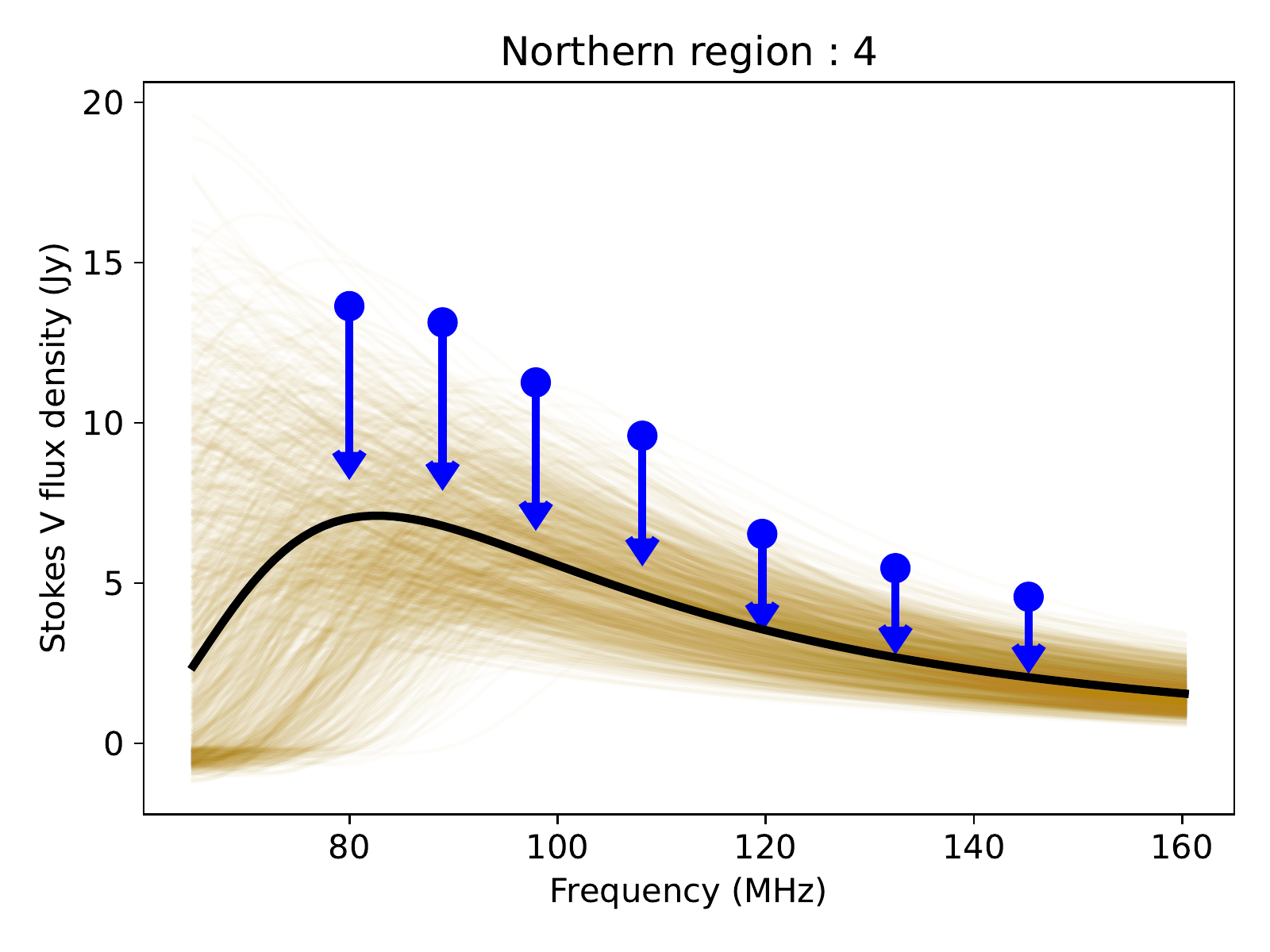}\includegraphics[trim={0.3cm 0.5cm 0.0cm 0.3cm},clip,scale=0.37]{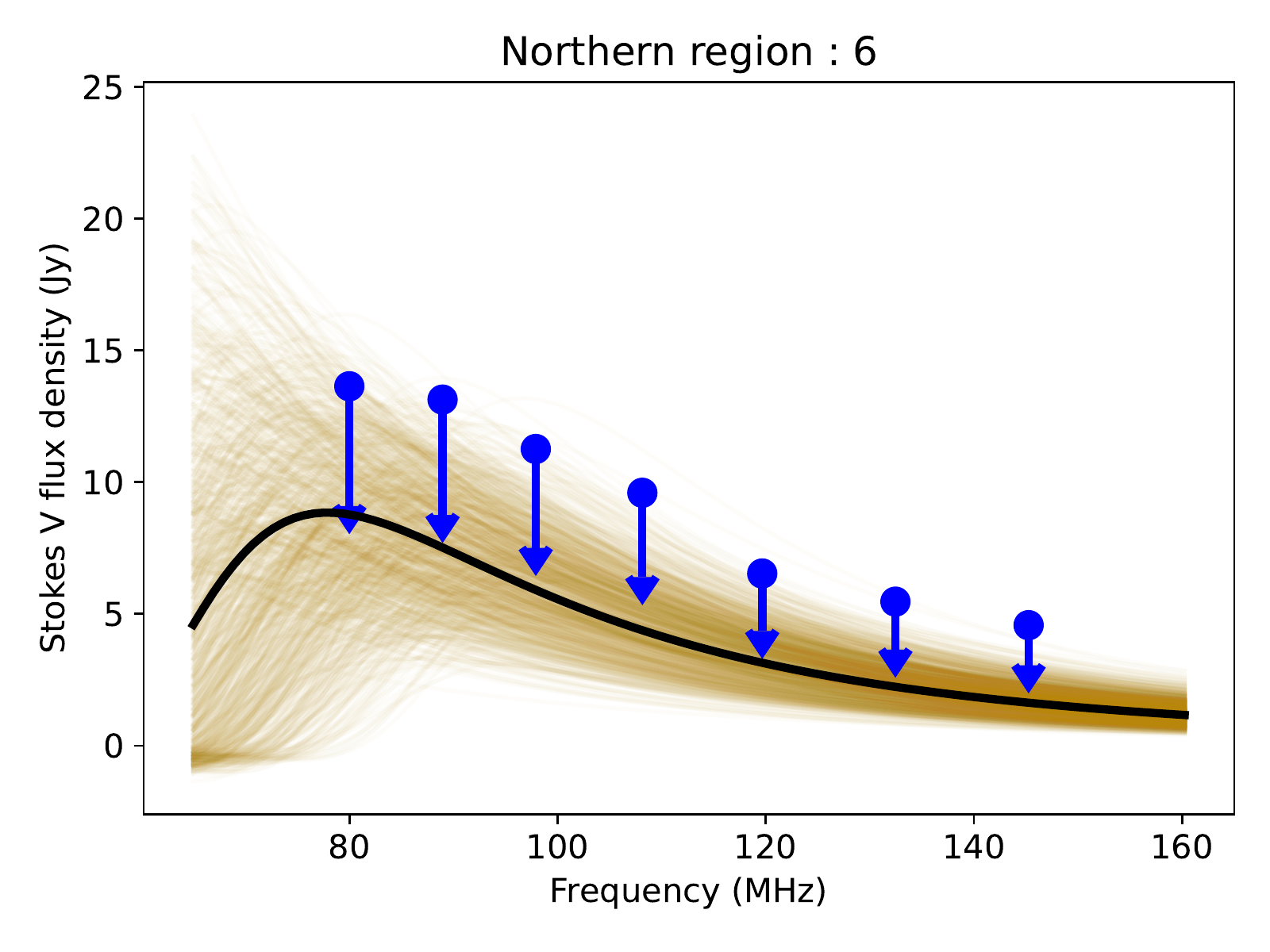}\includegraphics[trim={0.3cm 0.5cm 0.0cm 0.3cm},clip,scale=0.37]{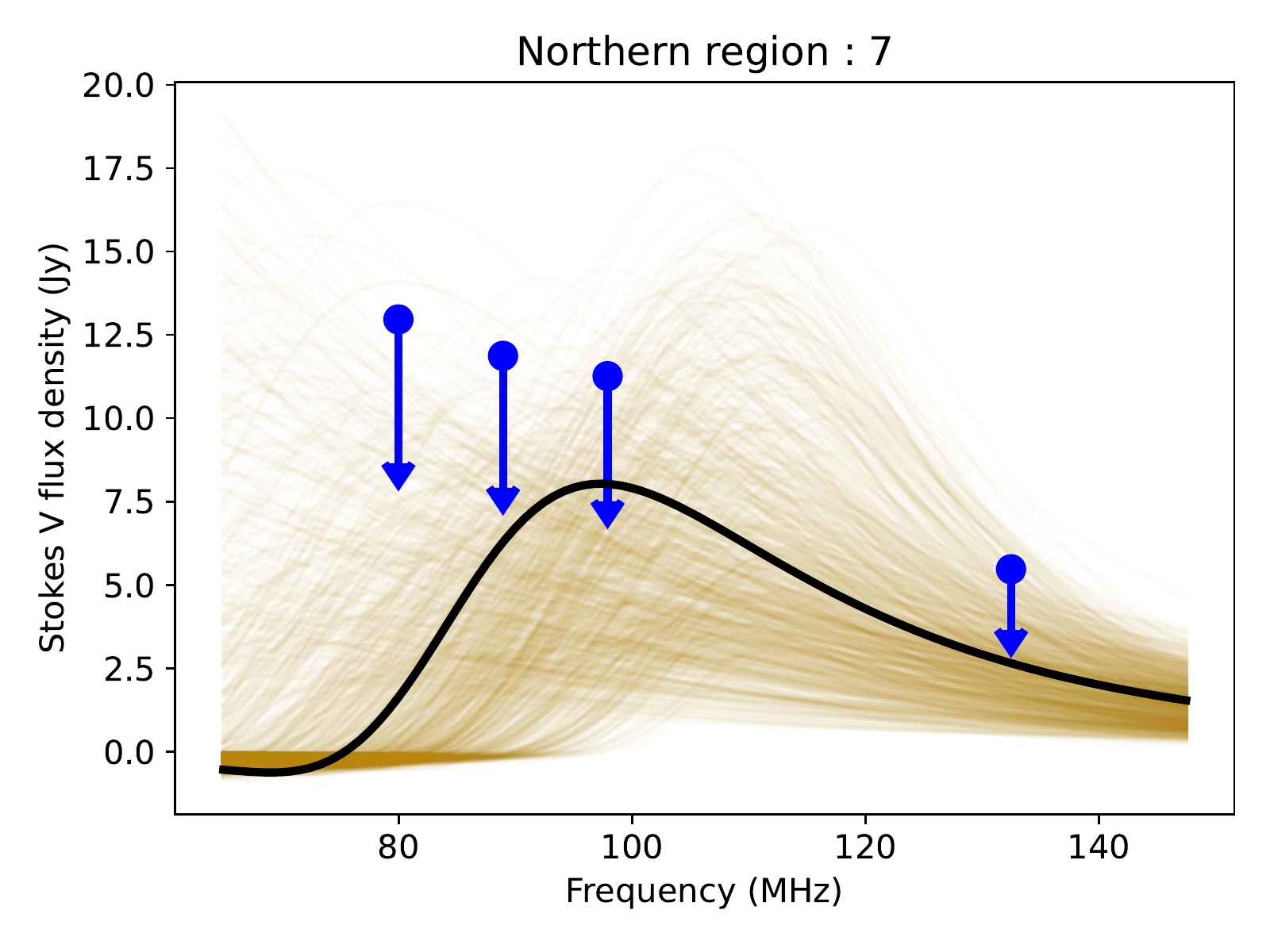}\\
    \caption{\textbf{Observed and fitted spectra of red and cyan regions of CME-1. First and third rows : } Stokes I spectra are shown. Red points represent the observed flux densities. \textbf{Second and fourth rows :} Stokes V spectra are shown. Blue points represent the upper limits at each of the frequencies. The black lines represent the Stokes I and V GS spectra corresponding to GS parameters reported in Table \ref{table:north_params}. Light yellow lines show the GS spectra for 1000 realizations chosen randomly from the posterior distributions of the GS model parameters. Sample posterior distributions for regions 1 and 4 are shown in Figure \ref{fig:corner_northern} and \ref{fig:corner_northern_4}, respectively.}
    \label{fig:spectra}
\end{figure*}

\section{Constraining Model Parameters using Upper/Lower Limits}\label{subsec:upperlimits_methods}
Most often when fitting a model to the data one makes use of well-measured quantities, each with their corresponding measurement uncertainties, and follows the well known $\chi^2$ minimization process \citep{Wolberg2006}.
However, there are often situations, especially when measuring weak signals, when the quantity being measured lies beyond the detection threshold of the measurement process, but the process is able to place a firm upper/lower limit on the quantity of interest.
It seems intuitively reasonable that by constraining the parameters to lie only in the part of the phase space consistent with the limit, the use of such limits should be able to further restrict the allowed parameter space for the model parameters.
Though the use of inequality constraints is not common when using $\chi^2$ minimization approaches, well established techniques for solving such problems exist \citep[see,][for a review]{Borwein2006} and their software implementations are also available in commonly used {\em python} libraries like {\em scipy}\footnote{\href{https://docs.scipy.org/doc/scipy/reference/generated/scipy.optimize.minimize.html}{Link to scipy optimization}} \citep{Scipy2020}. 

In addition to the $\chi^2$ based approaches, there also exist other well established mathematical frameworks for incorporating the constraints from availability of limits. 
A detailed description is available in \citet{Andreon2015_bayesian} along with several examples of applications in physics and astrophysics.
Standalone upper limits and combination of measurements and upper limits have often been used with considerable success to constrain physical systems across diverse areas of astrophysics \citep[e.g.,][etc.]{Aditya2015,Kanekar_2015,Kanekar2016,Montmessin2021,Brasseur2022} and cosmology \citep[e.g.,][etc.]{Planck_Col_2016,Ghara2020,Greig2021a,Bevins2022,Maity2022} including solar physics \citep[e.g.,][etc.]{Leer1979,Benz1996,Klein2003}. 

\subsection{Mathematical Framework}\label{subsec:upperlimtit_mathframe}
A few different mathematical approaches can be used to constrain model parameters using limits \citep{Andreon2015_bayesian}. 
The particular framework suitable for our needs is described in detail by \cite{Ghara2020} and examples of its applications are available in \cite{Greig2021,Maity2022,Maity2022a}. 
To place the following analysis in context, this framework is briefly described below.
This framework is based on the Bayes theorem \citep{Puga2015,Andreon2015_bayes_thereom}. 
Bayes theorem states that
\begin{equation}
    \mathcal{P}(\lambda|\mathcal{D})=\frac{\mathcal{L}(\mathcal{D}|\lambda)\ \pi(\lambda)}{\mathcal{P}(\mathcal{D})},
    \label{eq:bayes_theorem}
\end{equation}
where $\mathcal{D}$s are the data points and $\lambda$s the set of free parameters of the model. 
In this Bayesian framework, the objective is to compute the posterior distribution, $\mathcal{P}(\lambda|\mathcal{D})$, which is the probability of having the set of model parameters $\lambda$ given the data $\mathcal{D}$. $\pi(\lambda)$ is the prior distribution of the model parameters, $\mathcal{P(D)}$ is called the evidence which is the probability distribution of generating observed values given a set of model parameters. 
Evidence is not relevant from a parameter finding perspective in general. 
The standard practise is to set it to unity, implying that a given choice of model parameters leads to a unique set of observed values \citep{brooks2011handbook}. 
$\mathcal{L}(\mathcal{D}|\lambda)$ is the likelihood function which gives the conditional probability distribution of data given the distribution of the model parameters, $\pi(\lambda)$. 
In absence of prior knowledge of the model parameters, the standard practise is to use a uniform distribution \citep[e.g.,][etc.]{Kashyap1998,Middleton2015,Li_2019,Ghara2020,Maity2022} over a physically meaningful range of the model parameters.

For a well-measured quantity, i.e. when the measurement is above the noise threshold, 
the likelihood function is defined as
\begin{equation}
\begin{split}
      \mathcal{L}_\mathrm{1}(\mathcal{D}|\lambda)&=\mathrm{exp}\left(-\frac{1}{2}\sum_{i=1}^n \left[\frac{\mathcal{D}_\mathrm{i}-m_\mathrm{i}(\lambda)}{\sigma_\mathrm{i}}\right]^2\right)\\
      &=\prod_{i=1}^n \mathrm{exp}\left(-\frac{1}{2}\left[\frac{\mathcal{D}_\mathrm{i}-m_\mathrm{i}(\lambda)}{\sigma_\mathrm{i}}\right]^2\right)
\end{split}
\label{eq:likelihood_1}
\end{equation}
where, $n$ is the total number of data points, $\mathcal{D}_\mathrm{i}$, $m_\mathrm{i}(\lambda)$, and $\sigma_\mathrm{i}$ are the observed values, models values and uncertainty on the measurements respectively. 
For the case of upper limits, the likelihood function is defined as follows \citep{Ghara2020,Greig2021,Maity2022},
\begin{equation}
\begin{split}
      \mathcal{L}_\mathrm{2}(\mathcal{D}|\lambda)&=\prod_{i=1}^n \frac{1}{2}\left[1-\mathrm{erf}\left(\frac{\mathcal{D}_\mathrm{i}-m_\mathrm{i}(\lambda)}{\sqrt{2}\sigma_\mathrm{i}}\right)\right],
\end{split}
\label{eq:likelihood_2}
\end{equation}
where $erf$ refers to the error function.
When a mix of detections and upper limits are available, one can define the joint likelihood function as
\begin{equation}
    \mathcal{L}(\mathcal{D}|\lambda)=\mathcal{L}_\mathrm{1}(\mathcal{D}|\lambda)\ \mathcal{L}_\mathrm{2}(\mathcal{D}|\lambda),
    \label{eq:join_likelihood}
\end{equation}
which allows one to use the constraints from the detections as well as the upper limits. 
Using this joint likelihood function in the Monte Carlo Markov Chain \citep[MCMC,][]{brooks2011handbook} analysis allows one to use all available information to and better infer the model parameters.

Unlike $\chi^2$ minimization, MCMC analysis does not yield a unique set of values for model parameters. Instead, it provides the probability distribution of the parameter values, denoted by the {\em posterior distribution} in MCMC analysis. This allows us to fully understand the degeneracies in the parameter space and thus enable a more robust understanding of the underlying physical system.
The {\em true} value of the parameter is close to the value with the highest probability. Thus MCMC based approaches overcome one of the inherent limitations of a $\chi^2$ based approach -- the possibility of converging to one of the many local minimas in the $\chi^2$ space, especially when dealing with a large number of free parameters.

\section{Spectrum modeling}\label{sec:spectrum_modeling}
This section describes our approach to modeling the observed spectra using a GS model to estimate CME plasma parameters.
As demonstrated in Section \ref{sec:spectrum_sensitivity}, in the physically motivated range of parameters explored here, the model GS spectra are quite insensitive to variations in $T$ and $E_\mathrm{max}$. 
Hence, $T$ and $E_\mathrm{max}$ are kept fixed at 1 MK and 15 MeV respectively.
Thermal electron density is estimated independently from inversion of the white light images. The value used at any given radial distance is the average over the entire azimuthal range.
$n_\mathrm{thermal}$ is kept fixed at this value during GS model fitting. 
Among the other seven parameters, we fit $B$, $\theta$, $A$, $\delta$ and $E_\mathrm{min}$, while setting $n_\mathrm{nonth}$ to 1\% of the $n_\mathrm{thermal}$, similar to what has been assumed in previous works \citep{Carley2017,Mondal2020a}. 
For spectra from some regions (regions 2 and 3 in the right panel of Figure \ref{fig:circular_pol}), we explicitly fit for $L$, and for the other regions we fix $L$ to a pre-defined value as detailed in Section \ref{subsec:estimate_gcs}.

\subsection{Estimation of Geometrical Parameters}\label{subsec:estimate_gcs}
A key reason for choosing this CME for a detailed study was that it has coronagraph observations from multiple vantage points, SOHO, STEREO-A and STEREO-B, which enable us to build a well constrained three-dimensional model.
The locations of these spacecraft are shown in the top left panel of Figure \ref{fig:gcs} created using Solar-MACH \citep{solar_mach}\footnote{\url{https://solar-mach.github.io/}}.
We perform a three-dimensional reconstruction of the CME using the Graduated Cylindrical Shell model \citep[GCS;][]{Thernisien_2006,Thernisien_2011} using its {\em python} implementation \citep{gcs_python}. A good visual fit is obtained following the method described by \cite{Thernisien2009}. 
The GCS model arrived at is shown by blue mesh in Figure \ref{fig:gcs}, where different panels show superposition on LASCO-C2 and COR-2 images from STEREO-A and STEREO-B. The best visual fit GCS model parameters are:
\begin{enumerate}
    \item Front height ($h_{front}$) : 5.8 $R_\odot$
    \item Half-angle ($\alpha$) : 21$^{\circ}$
    \item Carrington Longitude ($\Phi$) : 153$^{\circ}$ 
    \item Heliospheric Latitude ($\Theta$) : 65$^{\circ}$ 
    \item Aspect Ratio ($\kappa$) : 0.34
    \item Tilt Angle ($\gamma$) : -32$^{\circ}$
\end{enumerate}

At 80 MHz, the radio emission is detected upto the leading edge observed in LASCO-C2 white-light image (Figure \ref{fig:north_cme}). The projected distance of the radio emission in the sky plane is 5.2 $R_\odot$. 
The corresponding three-dimensional distance computed from the GCS model based on the multi-vantage point observations puts this at $\sim5.8\ R_\odot$. This indicates that the CME-1 lies only about 27$^{\circ}$ out of the plane of the sky. Hence the geometric LoS makes a large angle with the direction of CME propagation.

For LoS originating from the Earth, we ray-traced through the the GCS model and computed the geometrical path length through the CME ($L_\mathrm{geo}$) for each PSF-sized region using {\em python}-based ray-tracing code {\em trimesh} \citep{trimesh}.
The ray paths for different regions are shown in Figure \ref{fig:gcs_ray}. Estimated $L_\mathrm{geo}$ for each of these regions are listed in Table \ref{table:los_depth}. 

Determining the best fit GCS model is not the result of a formal optimization procedure. 
In addition to the limitations imposed by the sensitivity of measurements, it is prone to errors for reasons ranging from human subjectiveness to relative locations of the vantage points.
To quantify these errors, \cite{Thernisien2009,VERBEKE2022} examined a large number of synthetic CMEs of different kinds observed using different numbers and configurations of spacecraft. 
They found that adding observations from a third or more vantage points do not reduce the errors on the model parameters significantly.

As there is no analytic relationship between GCS model parameters and $L_\mathrm{geo}$, usual error propagation cannot be used to estimate the uncertainty on $L_\mathrm{geo}$ ($\sigma(L_\mathrm{geo})$). 
To overcome this limitation, we generated 10,000 realizations of GCS model parameters from independent Gaussian distributions for each of the parameters.
The mean of these distributions was set to the fitted values and the standard deviation to the uncertainty reported in \cite{VERBEKE2022}. 
$L_\mathrm{geo}$ was computed for each of these realizations. The mean and standard deviation of the distribution of $L_\mathrm{geo}$ values so obtained is given in Table \ref{table:los_depth}.
This exercise was performed for each of the PSF sized regions.

It is important to note that the geometrical value of LoS angle and depth can differ from those for the best fit GS model.
This is because $\theta$ and $L$ describe the GS source along a given LoS, while the geometric parameters are derived from the estimated CME morphology.
Also note that while the angle with the sky plane can not provide any constraint on $\theta$, $L$ on the other hand, is tightly constrained to be smaller than $L_\mathrm{geo}$. 
This constraint on $L$ has not been used in earlier studies. 
It is evident from Figure \ref{fig:param_sensitivity} that the peak flux density and peak frequency of the Stokes I spectrum and the Stokes V fraction in the optically thick part are all sensitive to $L$. 
Hence to constrain $L$ using GS models, it is important that the spectral peak be included in the observed spectrum and it has at least seven measurements.
For this reason, $L$ is used as a free parameter for the regions 2 and 3, but not for other regions. The maximum value of $L$ for a given region is chosen to be $L_\mathrm{max}=L_\mathrm{geo}+\sigma(L_\mathrm{geo})$ as listed in Table \ref{table:los_depth}. We have calculated an average fraction, $f=L_\mathrm{fit}/L_\mathrm{max}$ for these two regions, where $L_\mathrm{fit}$ is the estimated value of $L$ from GS modeling. For region 2 and 3 the values of $f$ are 0.29 and 0.23 respectively and have a mean of $\sim$0.26. Assuming the filling fraction of the GS sources from different regions to lie in the same ballpark, we keep $L$ fixed at 0.26 times $L_\mathrm{geo}+\sigma(L_\mathrm{geo})$ for all other regions. 

\begin{table}
\centering
    \renewcommand{\arraystretch}{1.4}
    \begin{tabular}{|p{1cm}|p{1cm}|p{1cm}|p{1cm}|p{1cm}|p{1cm}|}
    \hline
       Region No. & $L_\mathrm{geo} \newline{(R_\odot)}$ & $\sigma(L_\mathrm{geo}) \newline{(R_\odot)}$ & Region No. & $L_\mathrm{geo} \newline{(R_\odot)}$  & $\sigma(L_\mathrm{geo}) \newline{(R_\odot)}$\\ \hline \hline 
        1 & 2.1 & 0.9 &  5 & 2.6 & 1.3\\
        \hline
        2 & 3.8 & 1.4 &  6 & 4.1 & 1.5\\
        \hline
        3 & 2.7 & 1.4 &  7 & 2.3 & 1.3\\
        \hline
        4 & 1.8 & 1.1 & 8 & 4.0 & 1.4\\
       \hline
    \end{tabular}
    \caption{\textbf{Estimated geometric LoS depth from GCS modeling.} The geometric LoS depths are obtained for different PSF-sized regions using ray tracing from Earth through that region. Geometric LoS depths are given in units of solar radius.}
    \label{table:los_depth}
\end{table}

\subsection{Joint Spectral Fitting of Stokes I and V}\label{subsec:joint_fitting}
We perform a joint spectral fit using the Stokes I and V spectra for the red and cyan regions marked in the right panel of Figure \ref{fig:circular_pol}. We followed the mathematical framework described in Section \ref{subsec:upperlimtit_mathframe}. 
Combing the Stokes I detection and Stokes V upper limits, following Equation \ref{eq:join_likelihood} we define the joint likelihood function as,
\begin{equation}
    \mathcal{L}(\mathcal{D}|\lambda)=\mathcal{L}_\mathrm{I}(\mathcal{D}_\mathrm{I}|\lambda)\ \mathcal{L}_\mathrm{V}(\mathcal{D}_\mathrm{V}|\lambda),
\end{equation}
where $\mathcal{D}_\mathrm{V,i}$, $m_\mathrm{V,i}(\lambda)$, and $\sigma_\mathrm{V,i}$ are the upper limits of absolute Stokes V flux density, GS model Stokes V flux density and uncertainties in the Stokes V. $\mathcal{L}_\mathrm{I}(\mathcal{D}_\mathrm{I}|\lambda)$ is the likelihood function for the Stokes I detection following the Equation \ref{eq:likelihood_1} and $\mathcal{L}_\mathrm{V}(\mathcal{D}_\mathrm{V}|\lambda)$ is the likelihood function for the Stokes V upper limits defined in Equation \ref{eq:likelihood_2}. 

We sample the posterior distribution using the Metropolis Hastings algorithm \citep{Metropolis1953} of MCMC method. We use publicly available {\em python} package {\em lmfit} \citep{lmfit} for this purpose, which runs the MCMC chains using another {\em python} package {\em emcee} \citep{emcee2013}. We have run total 10,000 MCMC chains per spectrum. 

\begin{table*}
\centering
    \renewcommand{\arraystretch}{1.5}
    \begin{tabular}{|p{1cm}|p{1.7cm}|p{1.3cm}|p{1.3cm}|p{1.5cm}|p{1.65cm}|p{1.8cm}|p{1.5cm}|p{1.2cm}|p{1.2cm}|}
    \hline
       Region No. & Heliocentric \newline{Distance} & B (G) & $\delta$ & $ A \times 10^{20}$\newline{$(cm^{2})$} & $E_\mathrm{min}$ (keV) & $\theta$ (degrees) & $L\ (R_\odot)$ & $n_\mathrm{thermal}$ \newline{$\times 10^6$}\newline{$(cm^{-3})^*$}  & $n_\mathrm{nonth}$\newline{$\times 10^4$} \newline{$(cm^{-3})^*$} \\ \hline \hline 
        1 & 2.5 & $4.55_{-0.99}^{+1.14}$ & $5.34_{-1.22}^{+2.28}$ & $8.01_{-3.09}^{+6.80}$ & $19.97_{-10.78}^{+22.31}$ &  $71.47_{-11.97}^{+10.92}$  & $0.78^*$ &1.5 & 1.5\\
        \hline
        2 & 2.5 & $1.28_{-0.34}^{+0.40}$  & $6.40_{-1.13}^{+1.31}$  & $2.43_{-0.61}^{+0.87}$  & $178.72_{-72.49}^{+10.94}$  & $59.23_{-12.60}^{+15.81}$ & $1.52_{-0.93}^{+1.70}$ & 1.5 & 1.5\\
        \hline
        3 & 2.5 & $1.44_{-0.28}^{+0.30}$ & $4.92_{-0.63}^{+0.77}$ & $2.94_{-0.56}^{+0.67}$ & $104.66_{-34.50}^{+50.86}$ & $78.64_{-5.69}^{+4.61}$ & $0.97_{-0.47}^{+0.91}$ & 1.5 & 1.5\\
        \hline
        4 & 2.5 & $2.04_{-0.59}^{+0.76}$ & $4.33_{-0.53}^{+0.72}$ & $6.94_{-2.30}^{+4.79}$ & $38.90_{-18.19}^{+32.01}$ & $68.81_{-12.24}^{+12.00}$ & $0.75^*$ &1.5 & 1.5\\
       \hline
       5 & 3.0 & $<1.28$ & $1.68$ & $9.57^*$ & $122.45^*$ & $65.94^*$ & $1.01^*$ & 0.7 & 0.7\\
       \hline
        6 & 3.0 & $1.05_{-0.27}^{+0.37}$ & $5.43_{-0.55}^{+0.69}$ & $8.55_{-2.99}^{+7.10}$ & $122.45_{-44.84}^{+64.30}$ & $65.94_{-10.69}^{+13.19}$ & $1.69^*$ & 0.7 & 0.7\\
       \hline
       7 & 3.0 & $1.99_{-0.87}^{+0.71}$ & $6.76_{-1.82}^{+2.16}$ & $5.80_{-2.39}^{+3.24}$ & $122.45^*$ & $54.03_{-15.03}^{+22.35}$ & $0.95^*$ & 0.7 & 0.7\\
       \hline
        8 & 3.0 & $<1.42$ & $2.14$ & $9.57^*$ & $122.45^*$ & $65.94^*$ & $1.40^*$ & 0.7 & 0.7\\
       \hline
    \end{tabular}
    \caption{\textbf{Estimated plasma and GS source parameters of CME-1.} These parameters are estimated for 01:24:55 UTC. Parameters marked by $^*$ are kept fixed during the fitting.}
    \label{table:north_params}
\end{table*}

\begin{figure*}[!htbp]
    \centering
    \includegraphics[trim={0cm 0.6cm 0cm 0cm},clip,scale=0.5]{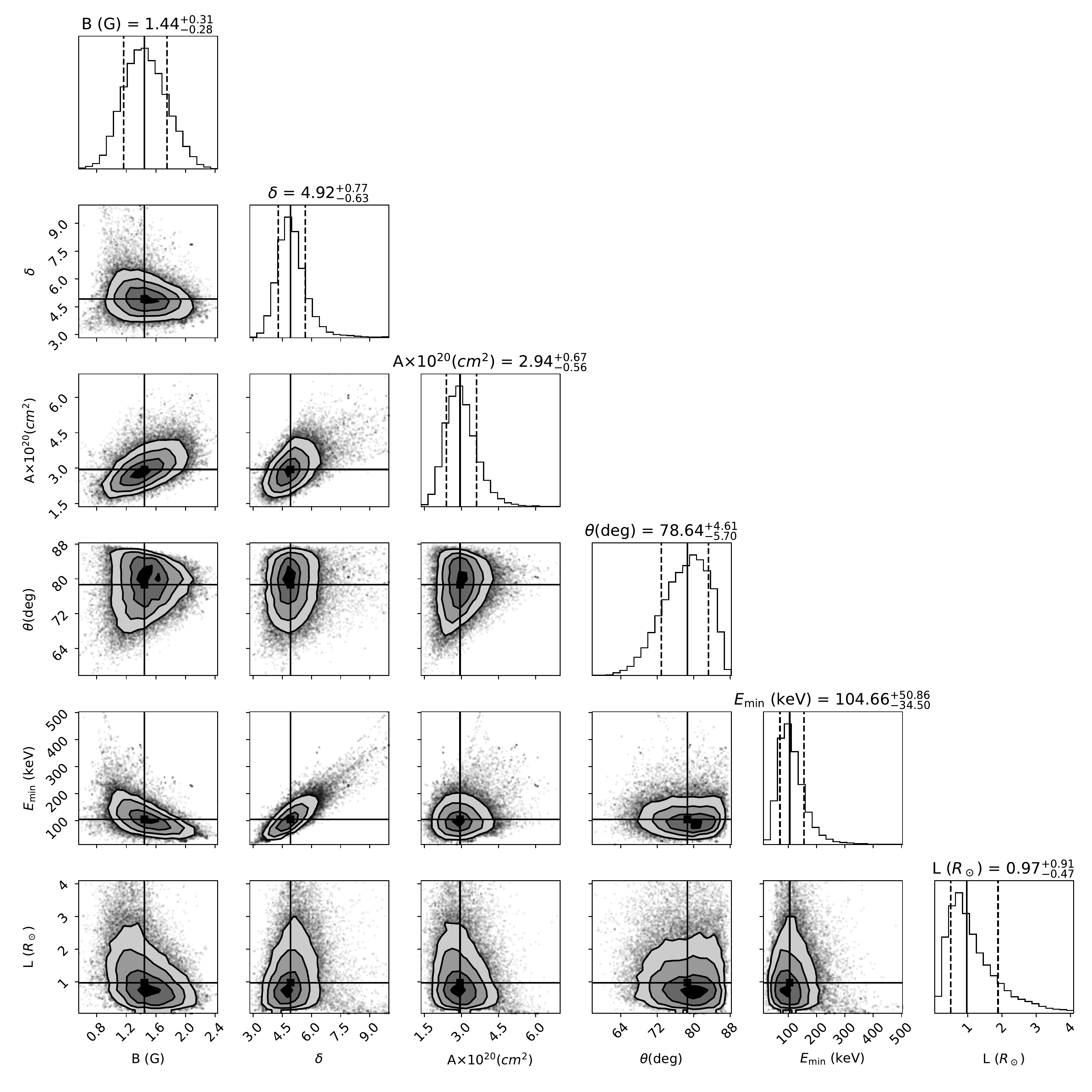}
    \caption{\textbf{Correlation of model parameters for region 3.} 2-dimensional plots show the joint probability distribution of any two parameters. The contours are at 0.5, 1, 2, and 3$\sigma$. The solid lines in the 1-dimensional histogram of posterior distributions mark the median values, and the vertical dashed lines mark the 16$^\mathrm{th}$ and 84$^\mathrm{th}$ percentiles.}
    \label{fig:corner_northern}
\end{figure*}
\begin{figure*}[!htbp]
    \centering
    \includegraphics[trim={0.5cm 0.6cm 0cm 0cm},clip,scale=0.6]{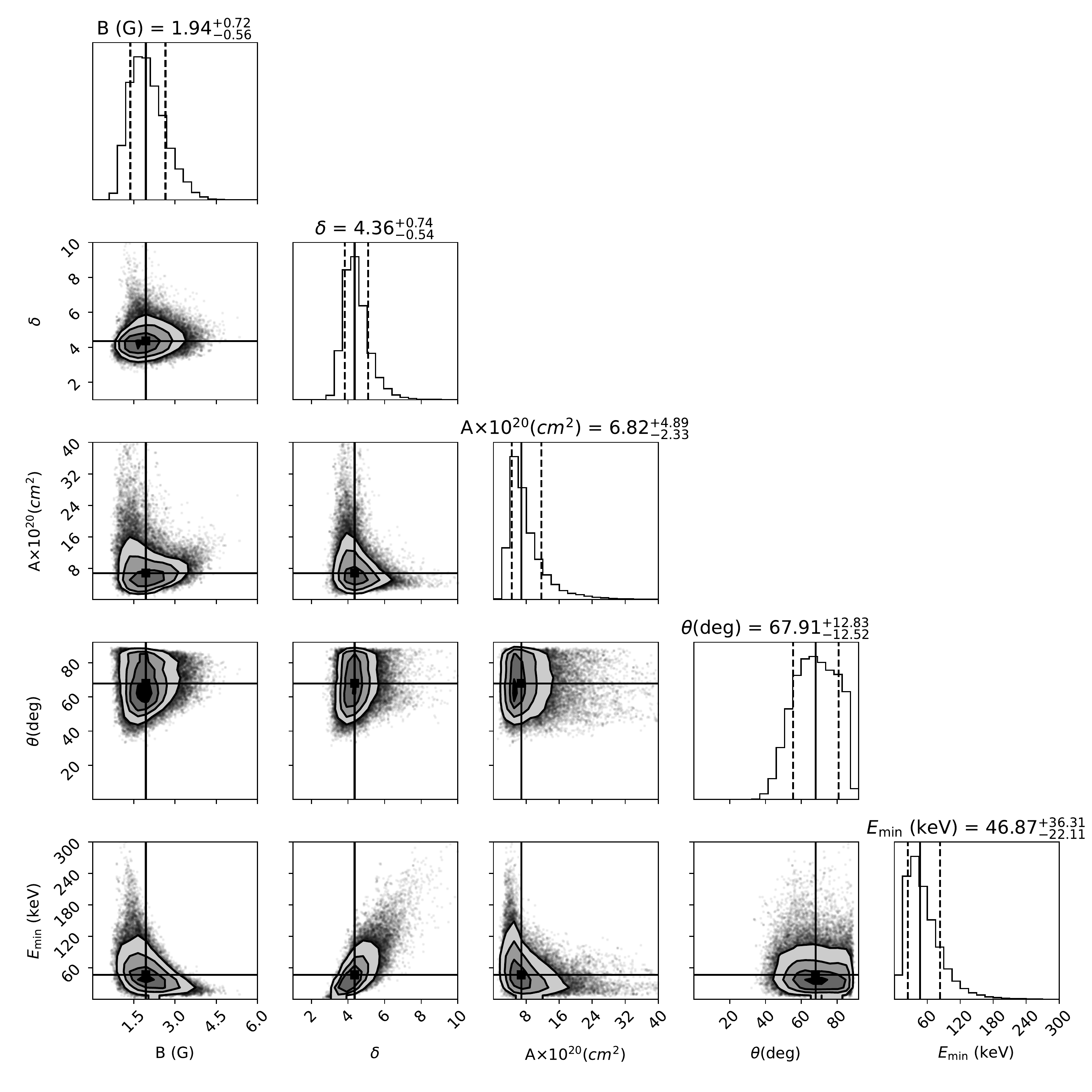}
    \caption{\textbf{Correlation of model parameters for region 4.} 2-dimensional plots show the joint probability distribution of any two parameters. The contours are at 0.5, 1, 2, and 3$\sigma$. The contours are at 0.5, 1, 2, and 3$\sigma$. The solid lines in the 1-dimensional histogram of posterior distributions mark the median values, and the vertical dashed lines mark the 16$^\mathrm{th}$ and 84$^\mathrm{th}$ percentiles.}
    \label{fig:corner_northern_4}
\end{figure*}

\begin{figure}
    \centering
    \includegraphics[trim={0.3cm 0.4cm 0cm 0cm},clip,scale=0.53]{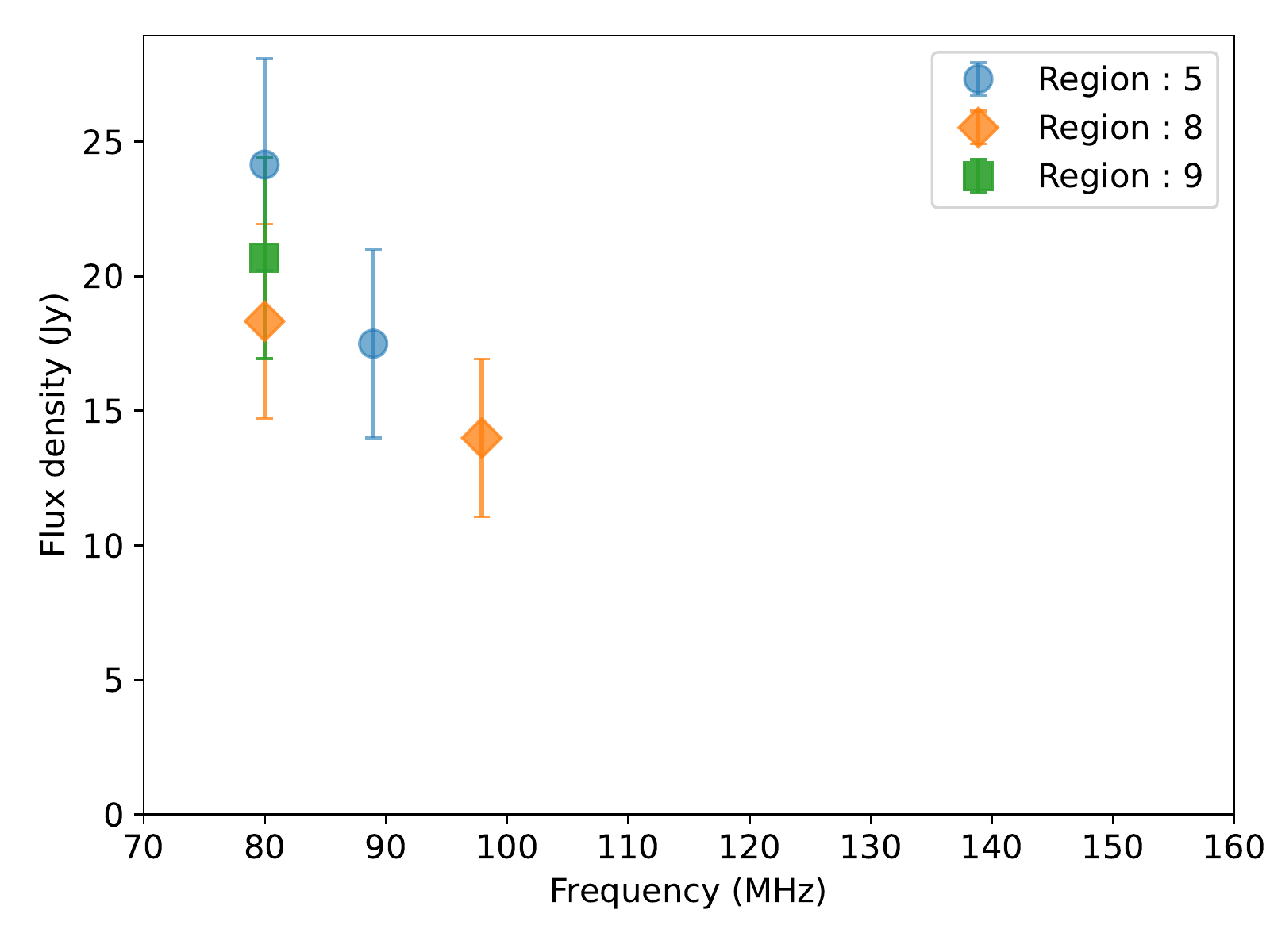}
    \caption{\textbf{Observed spectra of the green PSF-sized regions for the CME-1.} We have not done the fitting to these spectra because the number of spectral points is less than four.}
    \label{fig:green_region_spectra}
\end{figure}

\begin{figure*}[!htbp]
    \centering
    \includegraphics[trim={0.5cm 0.6cm 0cm 0cm},clip,scale=0.6]{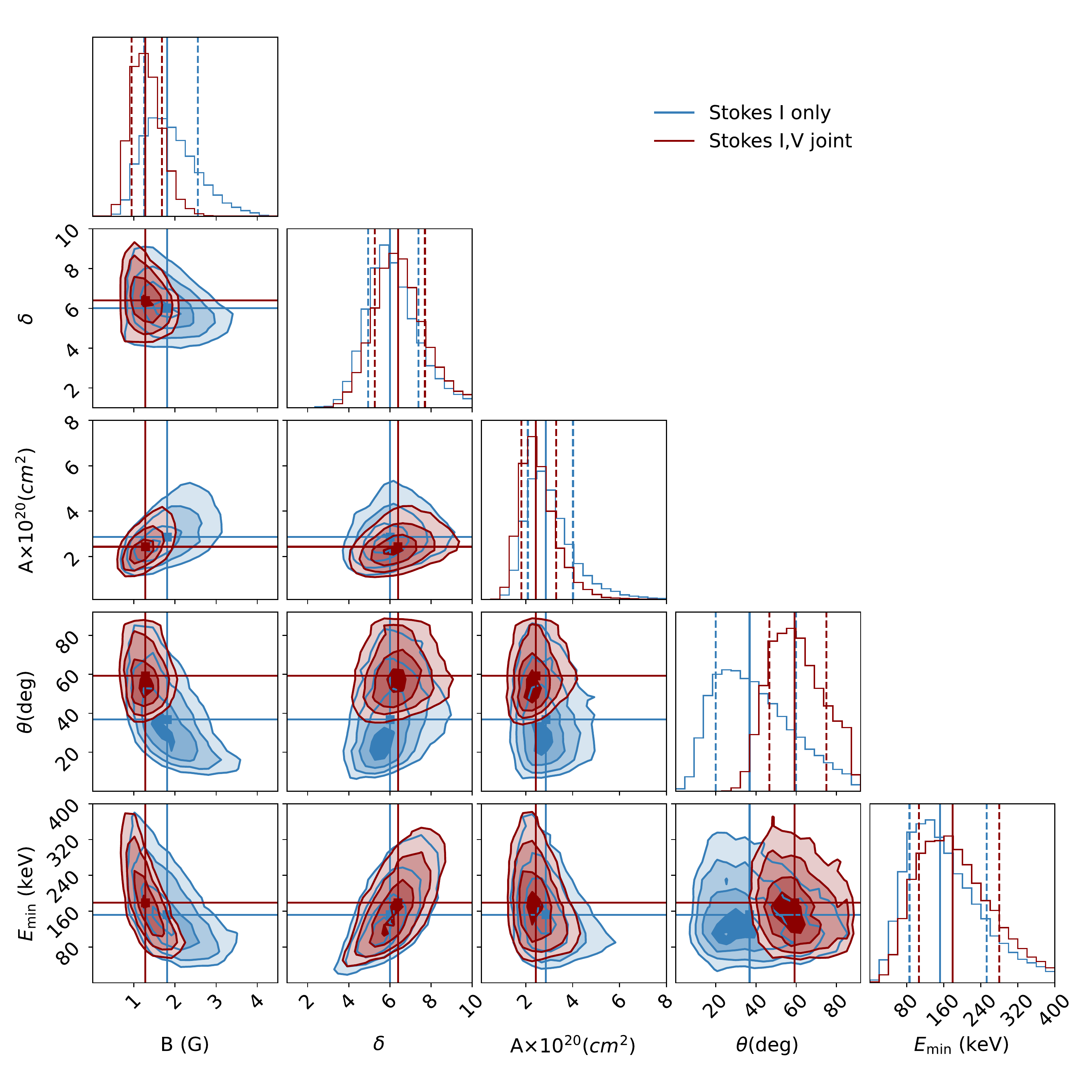}
    \caption{\textbf{Comparison between Stokes I only and Stokes I,V joint modeling for region 2.} 2-dimensional plots show the joint probability distribution of any two parameters. The contours are at 0.5, 1, 2, and 3$\sigma$. Blue contours represent the posterior distribution of parameters using only Stokes I spectrum. Maroon contours represent the posterior distribution using the joint Stokes I and V spectrum. $B$ and $\theta$ are better constrained for Stokes I and V joint fitting. The vertical dashed lines mark the 16$^{th}$ and 84$^{th}$ percentiles.}
    \label{fig:combine_corner}
\end{figure*}

We have used uniform priors, $\pi(\lambda)$, for the model parameters as follows,
\begin{enumerate}
    \item $B\ (\mathrm{G})$ : $(0,\ 20]$
    \item $\theta\ (\mathrm{degree})$ : $(0,\ 90)$
    \item $\delta$ : $(1,\ 10]$
    \item $A \times10^{20}\ (\mathrm{cm^{2}})$ : $[0.0001,\ 100]$
    \item $E_\mathrm{min}\ (\mathrm{keV})$ : $(0.1,\ 100]$
    \item $L\ (R_\odot)$ : $(0.01,L_\mathrm{max}]$ 
\end{enumerate}
The range of $B$ is guided by the choices made in previous works \citep{Vourlidas2020}. $\delta$ is also chosen based on previous studies and direct X-ray imaging observations \citep{Carley2017}. In principle, $\theta$ can take values ranging from 0$^{\circ}$ to 180$^{\circ}$. The value of $\theta$ and $180^{\circ}-\theta$ produce similar Stokes I spectra and their Stokes V spectra are inverted with respected to each other, as shown in the third and fourth panels of the second row of Figure \ref{fig:param_sensitivity}. Since, we only have upper limits on the absolute Stokes V, we can not break this degeneracy between $\theta$ and $180^{\circ}-\theta$. However, that does not impact the estimated value of $B$. Hence, we have chosen the $\theta$ to lie in the range 0$^{\circ}$--90$^{\circ}$. The minimum value of $A$ is chosen at a similar order of magnitude to that reported in M20, and the maximum value is chosen to be equal to the PSF area.  Given that there are no direct measurements of non-thermal electron distributions at these heights, the minimum value of $E_\mathrm{min}$ is chosen to be slightly higher than the energy of thermal electrons at $10^6\ \mathrm{K}$, close to the minimum value of $E_\mathrm{min}$ found by M20 . 

\subsection{Estimation of GS Model Parameters}\label{sec:plasma_parameters}

The regions for which good spectral sampling was obtained are marked in red and cyan in the right panel of Figure \ref{fig:circular_pol} are their spectra are shown in Figure \ref{fig:spectra}. The first and third rows show the Stokes I spectra and the second and fourth the Stokes V spectra. The black lines represent the GS spectra corresponding to the median values of the posterior distributions of GS model parameters. 
The reduced $\chi^2$ ($\chi^2_\mathrm{red}$) for each spectrum is listed in the corresponding Stokes I panels. Stokes V model spectra always lie below the the upper limits. 


Estimated plasma parameters are listed in Table \ref{table:north_params}. The parameters marked by stars are kept fixed to the mentioned values. 
As an example, posterior distributions of fitted parameters for region 1 are shown in Figure \ref{fig:corner_northern}. 
Distributions of all of the physical parameters show unimodal and sharply peaked clusters.
Those for regions 2 and 3 also show a similar behaviour.

For regions 4 and 6, the observations do not sample the spectral peak. As discussed in Section \ref{sec:spectrum_sensitivity},the variation in $A$ impacts only the Stokes I peak flux density, while other free parameters impact the location of the spectral peak as well as the spectral shape for both Stokes I and V. The spectral peak depends on several parameters; $B,\ \theta,\ A, E_\mathrm{min}$ and $n_\mathrm{nonth}$ as evident from Figure \ref{fig:param_sensitivity}. Among these, $A,\ E_\mathrm{min}$ and $n_\mathrm{nonth}$ do not have any significant impact on the optically thin part of the Stokes V spectra, while $B$ and $\theta$ have significant impacts on the optically thin part of the Stokes V spectra. Hence, the upper limits of Stokes V provide constraints on the $B$ and $\theta$. 
As the peak flux density is not known for these two regions, $A$ remains poorly constrained. This is evident from the posterior distribution of the parameters for region 4 shown in Figure \ref{fig:corner_northern_4}, though the ability to constrain the other parameters is not compromised much. Similar is the case for region 6.   
For the region 7 marked by cyan in the right panel of Figure \ref{fig:circular_pol}, although the spectrum samples the peak, it only has four spectral points. It is, hence, not reasonable to fit five free parameters to this spectrum. For this reason, we have kept the $E_\mathrm{min}$ fixed to a value close to that obtained for the adjacent region 6 and only fitted the other four parameters, $B$, $\delta$, $A$ and $\theta$. 

\section{Estimation of Plasma Parameters for Poorly Sampled Spectra}\label{subsec:magnetic_green}
The spectra for regions 5, 8 and 9 are shown in the Figure \ref{fig:green_region_spectra}. 
The emission from region 9 is detected only at one frequency, 80 MHz, and cannot be modelled.
Regions 5 and 8 have only two spectral points, too few for the GS spectral modeling approach adopted in the earlier section. For these regions we follow a different approach for GS. 

As for other regions, the thermal electron densities are available independently from the inversion of LASCO-C2 coronagraph image (Figure \ref{fig:electron_density}). For the non-thermal electron distributions, we use values estimated from the adjacent red regions, which lies at same heliocentric height. Values of $n_\mathrm{nonth}$ is set to 1\% of $n_\mathrm{thermal}$, while $E_\mathrm{min}$, and $\theta$ are set to those determined for the nearby region 6. 
From the two spectral points for regions 5 and 8 shown in Figure \ref{fig:green_region_spectra}, it is evident that the spectra are in optically thin part. 
Following \citet{Dulk1982,Carley2017}, $\delta$ for these regions is estimated using the spectral index ($\alpha_\mathrm{thin}$) of the optically thin part of spectrum as
\begin{equation}
    \delta=|-1.1(\alpha_\mathrm{thin}-1.2)|,
    \label{eq:delta_from_alpha}
\end{equation}
where the optically thin part of the spectrum is given by $S(\nu)=\ S_\mathrm{peak}(\nu/\nu_\mathrm{peak})^{\alpha_\mathrm{thin}}$ and $S_\mathrm{peak}$ is the peak flux density. Thus estimated values of $\delta$ for region 5 and 8 are $1.68$ and $2.14$, respectively. 

The simplified expression for $\nu_\mathrm{peak}$ \citep{Dulk1982} is valid under the assumptions we have already been making -- a power-law distribution of non-thermal electron and a homogeneous GS source.
This expression is accurate for limited ranges of GS model parameters -- $\theta$ between $\sim$20$^\circ$ and $\sim$80$^\circ$, $\delta$ between $\sim$2 and $\sim$7, and $E_\mathrm{min}$ between $\sim10$ keV to $\sim1$ MeV. The estimated values of $\theta$, $E_\mathrm{min}$ and $\delta$ from regions 1 through 6 lie in the permissible ranges. Assuming identical values for regions 5 and 8 suggests that $B$ can be estimated using the simplified expression for $\nu_\mathrm{peak}$ given as \citet{Dulk1982},
\begin{equation}
\begin{split}
    \nu_\mathrm{peak}=&2.73\times10^3\ 10^{0.27\delta}\ (sin\theta)^{0.41+0.03\delta}\times\\&(n_\mathrm{nonth}L)^{0.32-0.03\delta}\ B^{0.68+0.03\delta},
\end{split}
\label{eq:dulk_and_marsh}
\end{equation}
To determine $B$ using this analytical expression, one needs to provide the values of $n_\mathrm{nonth}$, $L$ and $\theta$. These parameters are set to their values estimated for adjacent regions. 
Since the peak of the spectra are not sampled for these regions, we can only estimate an upper limit on $B$, which are listed in Table \ref{table:north_params}.

\section{Discussion}\label{sec:discussion}
This work presents the very first spatially resolved spectro-polarimetric modeling of GS emission from a CME. 
As discussed in Section \ref{subsec:joint_fitting}, the GS model parameter phase space explored here has been motivated by physical arguments and earlier studies.
Even though the Stokes V emission lies below the detection threshold, the robust polarization calibration, high fidelity and high dynamic range imaging capabilities of P-AIRCARS and excellent snapshot PSF allows us to provide a sensitive upper limit on the Stokes V emission. 
Including these upper limits significantly reduces the spread in the distribution function of 
the model parameters and breaks some of the degeneracies in the GS model.
This section quantifies this improvement and also the benefits from the improved methodology used here.

\subsection{Advantages of Using Stokes V Spectra}\label{subsec:importance_stoks_V}
The sensitivity of the Stokes V spectra to the physical parameters of the GS model has already been demonstrated in Section \ref{subsec:stokesV_sensitivity}.
Even though the present work only uses upper limits on Stokes V emission, it already leads to a better constrained determination of GS model parameters. Use of stringent Stokes V upper limits enable us to exclude the part of the parameter space of GS models, which is consistent with the Stokes I spectra but not with the Stokes V upper limits.
To substantiate this, we compare the posterior distribution of parameters obtained using only Stokes I constraints (shown in blue in Figure \ref{fig:combine_corner}) with those obtained using joint constraints from Stokes I and V measurements (shown in maroon in Figure \ref{fig:combine_corner}).
In order to keep the number of free parameters below the number of constraints available for Stokes I only modeling and do a apples-to-apples comparison, $L$ was fixed to the value mentioned in Table \ref{table:north_params}.
The significant improvement in the ability to constrain $\theta$, $B$ and $E_\mathrm{min}$ is self evident in Figure \ref{fig:combine_corner}.
Examining the ranges spanned by the vertical dashed lines marking the 16$^{th}$ and 84$^{th}$ percentiles shows that the uncertainties in the estimates of $\theta$ has reduced by $\sim$44\% each and that in $B$ by $\sim$30\% on using joint Stokes I and V modeling.

\subsection{Need to Sample the Spectral Peak}\label{subsec:importance_peak}
A crucial feature of the spectrum is its peak, an accurate determination of which robustly constrains 
several GS model parameters. The spectral peak depends on several GS model parameters; $B,\ \theta,\ A, E_\mathrm{min},\ L$ and $n_\mathrm{nonth}$. As discussed in Section \ref{sec:spectrum_sensitivity}, changes in $A$ only impact the peak flux density and leave the fractional Stokes V spectra unchanged (Figure \ref{fig:param_sensitivity}).
The observed spectra for regions 4 and 6 do not sample the spectral peak, and this leads $A$ to be poorly constrained (Figure \ref{fig:corner_northern_4}). 
By contrast in the cases where the spectral peak has been sampled (region 2 and 3), the uncertainty in $A$ is lower by about an order of magnitude. Not only $A$, the unavailability of the spectral peak also affects the uncertainty of other parameters as well.   


\subsection{Filling Factor of GS Source}\label{subsec:filling_factor}
Earlier studies did not use any physically motivated constraints on $L$. 
For lack of a better estimate, typically $L$ was fixed to the value of the PSF diameter \citep{Mondal2020a, Vourlidas2020} assuming a spherical symmetry. The validity of this assumption was never tested . 
The CME studied here was chosen specifically to have coronagraph observations from multiple vantage points. This enabled us to building a detailed and well constrained three-dimensional model for it and use it to estimate an upper limit on $L$ (Section \ref{subsec:estimate_gcs}).
The major axis of the PSF for the current observation is $\sim9\times10^{10}$ cm, while the estimated values of $L$ vary between $\sim3\times10^{10}$ and $10\times10^{10}$ cm (Table \ref{table:north_params}). These values are close to that taken by M20, but smaller or equivalent to the PSF size.


Estimated values of $A$ are of order $10^{20}-10^{21}\ \mathrm{cm^{2}}$. The area of PSF at the lowest observing frequency is $\sim10^{22}\ \mathrm{cm^{2}}$. This leads to an areal filling fraction of 0.01$-$0.1. Assuming that the filling fractions in the sky plane and along the LoS are similar, M20 concluded that either the non-thermal electrons have a small filling fraction and/or the emission comes from regions of concentrated magnetic field. The presence of such regions has been suggested under the names of magnetic knots in the literature \citep{Karpen2012}. 
Having an independent estimate of $L$ from GS modeling and the $L_{geo}$ from geometric modeling of the CME enables us to compute volumetric filling factor, $f$, without relying on the assumption of the filling factor in the plane of the sky and along LoS being the same.
$f$ is defined as
\begin{equation}
    f=\frac{AL}{A_\mathrm{PSF}L_\mathrm{geo}}
    \label{eq:filling_frac}
\end{equation}
where, $A_\mathrm{PSF}$ is the area of the PSF. The average volumetric filling factor of GS source for the CME under study turns out to be $\sim$0.1$-$1\%. The low value of $f$ obtained here is consistent with the that arrived at by M20.

\section{Conclusion}\label{sec:conclusion}
Since the first detection 
of the GS emission from CME by \cite{bastian2001}, radio emissions from CME plasma have been detected only for a handful of fast CMEs. 
M20 presented the first detection of GS radio emissions from a slow CME.
The flux densities of radio emission reported by M20 and the present work are among the lowest reported. 
These works furnish further evidence that the earlier non-detections of GS emission from slow CMEs can be attributed to the limited dynamic range achieved in those attempts, and that these limitations can now be overcome with the high dynamic range imaging yielded by the combination of data from instruments like the MWA and imaging pipelines like P-AIRCARS \citep{Kansabanik_principle_AIRCARS,Kansabanik2022_paircarsI, Kansabanik_paircars_2}. 

Even with routine detection of CME GS emissions, the limited number of spectral points at which measurements are typically available, in contrast with the large numbers of GS model parameters and the degeneracies between some of them pose significant complications. 
These issues force one to seek independent estimates for some of the model parameters and assume physically motivated values for others. This has, in the past, limited the robustness of the GS model parameter estimates and, hence, the usefulness of this approach. This work uses a homogeneous source model and the GS model parameter phase space explored here has been motivated by physical arguments and earlier studies (Section \ref{subsec:joint_fitting}).
Under these assumptions, it presents a detailed quantitative analysis of the sensitivity of the observed Stokes I GS spectra to the various model parameters and the degeneracies present.
It also demonstrates that Stokes V spectra have a different dependence on GS model parameters than Stokes I spectra and can be used effectively to break many of these degeneracies.


For the first time, this work uses both Stokes I and V spectra for constraining the GS model parameters. Even though only sensitive upper limits on Stokes V spectra are available, their use already reduces the uncertainty in the model parameters of most interest ($B$ and $\theta$) by as much as $\sim$40\%.
We have also found that for the GS model parameters to be well constrained, it is essential that the peak of the GS spectrum be included in the observed part of the spectrum.

Another novel aspect of this work is demonstration of the usefulness of a good geometric model of the CME for determining the volume filling factor of GS emission and estimates it to be $\sim$0.1$-$1\%.. Constraining the geometric model parameters requires coronagraph observations from multiple vantage points. Work is already underway to extend this approach to other well observed CMEs for which MWA data are also available.


Based on the results from the present day instruments like the MWA, we have no doubt that the even more sensitive and wider bandwidth spectro-polarimetric imaging from the
upcoming instruments, like the Square Kilometre Array \citep[SKA;][]{Hall2005}, the \textit{Next Generation Very Large Array} \citep[ngVLA:][]{ngVLA2019}, and the \textit{Frequency Agile Solar Radiotelescope} \citep[FASR:][]{Gary2003,Bastian2005,Bastian2019}; aided by the multi-vantage point coronagraph observations will provide a routine and a robust remote sensing technique for estimating CME plasma parameters even at large coronal heights.

\facilities{Murchison Widefield Array \citep[MWA;][]{lonsdale2009,Tingay2013},Solar and Heliospheric Observatory \citep[SOHO;][]{Domingo1995}, Solar Terrestrial Relations Observatory \citep[STEREO,][]{Kaiser2008}}

\software{astropy \citep{price2018astropy}, matplotlib \citep{Hunter:2007}, Numpy \citep{Harris2020}, CASA \citep{mcmullin2007,CASA2022}, P-AIRCARS \citep{paircars_zenodo}, GCS-python \citep{gcs_python}, GScode \citep{GS_code2021}, Solar-MACH \citep{solar_mach}}

\vspace{0.5cm}
\noindent This scientific work makes use of the Murchison Radio-astronomy Observatory (MRO), operated by the Commonwealth Scientific and Industrial Research Organisation (CSIRO). We acknowledge the Wajarri Yamatji people as the traditional owners of the Observatory site.  Support for the operation of the MWA is provided by the Australian Government's National Collaborative Research Infrastructure Strategy (NCRIS), under a contract to Curtin University administered by Astronomy Australia Limited. We acknowledge the Pawsey Supercomputing Centre, which is supported by the Western Australian and Australian Governments. D.K. gratefully acknowledges Barun Maity (NCRA-TIFR) for useful discussions. We also thank the anonymous referee for the comments and suggestions, which have helped improve the clarity and the presentation of this work. D.K. and D.O. acknowledge support of the Department of Atomic Energy, Government of India, under the project no. 12-R\&D-TFR-5.02-0700. S.M. acknowledges the
partial support from USA NSF grant AGS-1654382 to the New Jersey Institute of Technology. This research has also made use of NASA's Astrophysics Data System (ADS).

\bibliography{sample631}{}
\bibliographystyle{aasjournal}


\end{document}